\documentclass[10pt]{iopart}

\usepackage{iopams}
\usepackage{cite}
\usepackage{graphicx}
\usepackage{subfigure}
\usepackage{bbm}
\usepackage{booktabs}
\usepackage[colorlinks=true,linkcolor=blue,citecolor=blue]{hyperref}
\usepackage{multicol}
\usepackage[titletoc]{appendix}

\begin{document}

\title{Global phase diagram of Coulomb-interacting anisotropic Weyl semimetal with disorder}

\author{Jing-Rong Wang$^{1}$ Wei Li$^{2}$, Gang Wang$^{3}$, and Chang-Jin Zhang$^{1,4,5}$
\footnote[0]{$^{5}$Author to whom any correspondence should be
addressed.}}

\address{$^{1}$Anhui Province Key Laboratory of Condensed Matter
Physics at Extreme Conditions, High Magnetic Field Laboratory of Anhui Province,
Chinese Academy of Sciences, Hefei 230031, China}
\address{$^{2}$Key Laboratory of Materials Physics,
Institute of Solid State Physics, Chinese Academy
of Sciences, Hefei 230031, China }
\address{$^{3}$College of Physics, Optoelectronics and Energy,
Soochow University, Suzhou 215006, China}
\address{$^{4}$Institute of Physical Science and Information
Technology, Anhui University, Hefei 230601, China}
\ead{zhangcj@hmfl.ac.cn}

\begin{abstract}
Taking into account the interplay between the disorder and Coulomb
interaction, the phase diagram of three-dimensional anisotropic Weyl
semimetal is studied by renormalization group theory.  Weak disorder is irrelevant in
anisotropic Weyl semimetal, while the disorder becomes relevant
and drives a quantum phase transition from semimetal to
compressible diffusive metal phases if the disorder strength is larger than a critical value. The long-range Coulomb interaction
is irrelevant in clean anisotropic Weyl semimetal. However, interestingly, we find that
the long-range Coulomb interaction exerts a dramatic influence on the
critical disorder strength for phase transition to compressible
diffusive metal. Specifically, the critical disorder strength can receive a
prominent change even though an arbitrarily weak Coulomb
interaction is included. This novel behavior is closely related to
the anisotropic screening effect of Coulomb interaction,
and essentially results from the specifical energy dispersion
of the fermion excitations in anisotropic Weyl semimetal.
The theoretical results are helpful for understanding
the physical properties of the candidates of anisotropic Weyl semimetal, such as pressured BiTeI, and
some other related materials.
\end{abstract}


\submitto{\JPCM}

\maketitle

\ioptwocol

\section{Introduction}

In the fast-expanding field of topological phases of matter, various
semimetals (SMs) have been attracting particular research interest.
Major results consist of, among others, the predictions and/or
realizations of  three-dimensional Dirac SM (3D DSM) \cite{Vafek14, Wehling14,
Armitage17}, 3D Weyl SM (WSM) \cite{Armitage17, Wan11,
Huang15, WengHM15, Xu15A, Lv15A, Burkov16, Yan17, Hasan17}, nodal
line SM (NLSM)~\cite{Armitage17, Weng16, FangChen16} \emph{etc}. A
rich diversity of novel physics emerges in these materials. For
instance, the chiral anomaly of WSM gives rise to interesting features
and phenomena, such as the negative
magnetoresistance~\cite{HuangChenGFGroup15, ZhangChengLong16}, planar Hall effect \cite{Burkov17, Nandy17},
anomalous thermoelectric response~\cite{Panfilov14}, just to mention a few.

For SMs, the dimension of Fermi surface is at least two less than the
dimension of system \cite{Vafek14, Wehling14, Armitage17}, in
contrast to conventional metals in which the dimension of Fermi surface
is one less than the dimension of system. In many SMs including DSM, WSM, Luttinger SM,
the conduction and valence bands touch each other at discrete nodal points in the
Brillouin zone. In NLSM, the conduction and valence bands touch at
one dimension lines. Therefore, the density of states (DOS)
vanishes at the Fermi level in SMs, which is  different from conventional metals
with a large finite DOS at the Fermi level.

Due to the abovementioned difference between SMs and conventional metals,
the effects of interaction and disorder in SMs may be obviously different
from the one in conventional metals. The Coulomb interaction in conventional
metals acquires static screening and becomes short-ranged \cite{Shankar94, ColemanBook}.
However, in SMs, the Coulomb interaction only receives dynamical screening and is still long-ranged.
The renormalization group (RG) studies revealed that the
long-range Coulomb interaction can be marginally irrelevant
\cite{Kotov12, Gonzalez94, Son07, Hofmann14, Goswami11, Hosur12,
Gonzalez14, Hofmann15, Throckmorton15, Lai15, Jian15, ZhangShiXin17, Cho16,
Isobe16, WangLiuZhang17A}, irrelevant \cite{Abrikosov72, Yang14A,
Huh16}, or relevant \cite{Abrikosov74, Moon13, Herbut14, Janssen15,
Dumitrescu15, Janssen16, Janssen17}, depending on the energy
dispersion of the fermion excitations in SMs. Additionally, the disorder
effects in SMs and conventional metals may exhibit different behaviors.
In 3D conventional metals, arbitrarily weak random scalar potential (RSP) results
in finite disorder scattering rate and drives the system to compressible diffusive
metal (CDM) state \cite{PALee85}. However, for 3D DSM, many studies showed that finite disorder scattering
rate is generated and the system is driven to CDM phase if the strength of RSP is larger a critical
value~\cite{SyzranovReview}. In 2D conventional metals, arbitrarily weak RSP drives
the system to Anderson insulator \cite{PALee85, Evers08},
but arbitrarily weak RSP induces 2D DSM to CDM \cite{Evers08, DasSarma11}.

Interplay of correlated electron
interactions and disorder is an interesting question which attracted extensive studies~\cite{Finkelstein84, Castellani84,
PunnooseFinkelstein05, Abrahams01, Kravchenko04, Spivak10,
WangJing11, WangLiuZhang16, WangJing17, Ye98, Ye99, Stauber05, Herbut08, Vafek08,
Foster08, WangLiu14, Moon14, Roy16, Gonzalez17, Zhao16, Nandkishore17,
WangYuXuan17, Mandal18, WangLiuZhang17B}. It is well known that the celebrated metal-insulator transition in
2D system probably results from the interplay of Coulomb interaction and disorder \cite{Finkelstein84, Castellani84,
PunnooseFinkelstein05, Abrahams01, Kravchenko04, Spivak10}.
The interplay of long-range Coulomb interaction and disorder in various SMs acquired  particular
interest \cite{Ye98, Ye99, Stauber05, Herbut08, Vafek08, Foster08,
WangLiu14, Goswami11, Moon14, Roy16, Gonzalez17, Zhao16, Nandkishore17, Mandal18, WangYuXuan17,
WangLiuZhang17B, YangZK18, ZhaoPL19, Sikkenk19}. Actually, the interplay of
long-range Coulomb interaction and disorder has been studied in 2D
DSM \cite{Ye98, Ye99, Stauber05, Herbut08, Vafek08, Foster08,
WangLiu14}, 3D DSM/WSM \cite{Goswami11, Moon14, Roy16, Gonzalez17}, 3D
Luttinger SM \cite{Nandkishore17, Mandal18}, 3D NLSM
\cite{WangYuXuan17}, and 3D multi-WSMs \cite{WangLiuZhang17B}.
According to the concrete conditions, the Coulomb interaction and disorder could
compete or promote each other. It was shown that interplay of Coulomb interaction and
disorder may result in quantum phase transitions (QPTs) from SM to CDM, excitonic insulator,
Anderson insulator, or a stable quantum  critical state \cite{Ye98, Ye99, Stauber05, Herbut08, Vafek08, Foster08,
WangLiu14, Goswami11, Moon14, Roy16, Gonzalez17, Zhao16, Nandkishore17, Mandal18, WangYuXuan17,
WangLiuZhang17B, YangZK18, ZhaoPL19, Sikkenk19}.

Recently, 3D anisotropic WSM (AWSM), which is a novel SM, whose fermion dispersion
 is linear along two directions and quadratic along the third one \cite{Yang14A, Yang13,
Yang14B}. 3D AWSM state can be obtained through fine tuning to the
topological phase transition point between 3D WSM and normal band
insulator \cite{Yang13, Yang14B}. Recent studies showed that 3D AWSM exhibits intriguing
properties. Yang \emph{et al.} found that
long-range Coulomb interaction becomes irrelevant in a clean 3D AWSM
\cite{Yang14A}, which is consistent with the pioneering work of
Abrikosov \cite{Abrikosov72}. The studies given by Roy \emph{et al.} \cite{Roy18}
and Luo \emph{et al.} \cite{Luo18A, Luo18B} pointed out that the weak disorder
is irrelevant in 3D AWSM, but a QPT from
SM to  CDM phases will
be induced at a critical disorder strength. Li \emph{et al.} studied the quantum critical behaviors
in the vicinity of a quantum critical point (QCP) from AWSM to superconducting phase \cite{LiXin18}.
The study by Copetti and Landsteiner  unveiled that anomalous Hall viscosity is developed in AWSM \cite{Copetti19}.

In this article, we study the interplay of  long-range Coulomb
interaction and disorder in 3D AWSM.
Due to the long-range Coulomb interaction becomes irrelevant in clean 3D AWSM,
it is intuitively expected that the Coulomb interaction would not play an important
role in the phase transitions occurred in disordered AWSM. However,
after performing concrete RG calculations, we find
that the Coulomb interaction has dramatic influence on the
phase diagram of disordered AWSM. Remarkably, we show that even in
the presence of arbitrarily weak Coulomb interaction, the critical disorder strength can receive a prominent modification in
some conditions. We indicate that this novel behavior is closely related to the anisotropic screening effect
of Coulomb interaction, and essentially results from the special energy dispersion of fermion excitations in AWSM.

The rest paper is structured as follows. The model is described in
section~\ref{Sec:Model}. In section~\ref{Sec:ExpansionParameter}, we discuss the expansion
parameters for the RG analysis of interactions and disorder in SMs, and explain the method
which we use to analyze the interplay of Coulomb interaction and disorder in 3D AWSM. In  section~\ref{Sec:Results}, we show the RG
equations of model parameters and numerical results of the RG
equations. In this section, we illustrate the phase diagrams of
disordered AWSM with Coulomb interaction. We also calculate the dynamical exponents,
correlation length exponent, and anomalous dimension of the fermion field
at the phase boundary. After presenting these
results, in section~\ref{Sec:ObservableQuantities}, we compare the
observable quantities  in SM phase, CDM phase, and at the phase boundary. In
section~\ref{Sec:PhysicalImplications}, we discuss the physical
implications of our results for candidate materials of 3D AWSM, and some
other related materials. In section~\ref{Sec:Comparison}, we discuss previous studies about the
interplay of Coulomb interaction and disorder in other SMs, and compare with the results  for AWSM. The
main results are summarized in section~\ref{Sec:Summary}. The details of
calculation and derivation are  presented in the Appendices.

\section{Effective action \label{Sec:Model}}

The Hamiltonian of free 3D anisotropic Weyl fermions is given by \cite{Yang14A, Yang13, Yang14B}
\begin{eqnarray}
H_{f}=\int d^3\mathbf{x}\psi^{\dag}(\mathbf{x})\left(-iv\partial_{x}\sigma_{1}-iv\partial_{y}\sigma_{2}-A\partial_{z}^{2}\sigma_{3}\right)\psi(\mathbf{x}),\nonumber
\\
\end{eqnarray}
where $\psi$ represents two-component spinor, and $\sigma_{1,2,3}$ are the Pauli matrices. $v$ and $A$ are model parameters.
The energy dispersion of fermions  takes the form
\begin{eqnarray}
E(\mathbf{k})=\pm\sqrt{v^2k_{\bot}^{2}+A^{2}k_{z}^{4}},
\end{eqnarray}
where $k_{\bot}^{2}=k_{x}^{2}+k_{y}^{2}$.
The Hamiltonian for long-range Coulomb interaction between the fermions is expressed as
\begin{eqnarray}
H_{C}=\frac{1}{4\pi}\int d^3\mathbf{x}d^3\mathbf{x}'\rho(\mathbf{x})\frac{e^{2}}{\epsilon\left|\mathbf{x}-\mathbf{x}'\right|}
\rho(\mathbf{x}'),
\end{eqnarray}
where $\rho(\mathbf{x})=\psi^{\dag}(\mathbf{x})\psi(\mathbf{x})$ is the fermion density operator, $e$ the electric charge,
and $\epsilon$ the dielectric constant. The effective strength of long-range Coulomb interaction is defined as $\alpha=e^{2}/v\epsilon$.
The action for the fermion-disorder coupling is
\begin{eqnarray}
S_{dis}=\sum_{j=0}^{3}\int d\tau d^3\mathbf{x}V_{j}\psi^{\dag}(\mathbf{x})\Gamma_{j}\psi(\mathbf{x}).
\end{eqnarray}
The quenched random field $V_{j}$ is taken as a Gaussian white noise distribution which satisfies $\left\langle V_{j}(\mathbf{x})\right\rangle=0$
and $\left\langle V_{j}(\mathbf{x})V_{j}(\mathbf{x}')\right\rangle=\Delta_{j}\delta^{3}\left(\mathbf{x}-\mathbf{x}'\right)$. The Coulomb interaction
can be treated by introducing a boson field $\phi$ through Hubbard-Stratonovich transformation \cite{Goswami11,Yang14A, Lai15, Jian15, ZhangShiXin17, Huh16, WangLiuZhang17B}.
The disorder potential is averaged by adopting the replica method \cite{Goswami11, Ye99, Herbut08, Foster08, Moon14, Roy16, Gonzalez17, Nandkishore17,
Mandal18, WangYuXuan17, WangLiuZhang17B, YangZK18, ZhaoPL19, Sikkenk19, Roy18, Luo18A, Roy14}.

Accordingly, the total effective action can be written as
\begin{eqnarray}
S&=&S_{f}+S_{b}+S_{fb}+S_{dis}, \label{Eq:TotalAction}
\\
S_{f}&=&\int\frac{d\omega}{2\pi}\frac{d^3\mathbf{k}}{(2\pi)^{3}}\psi_{a}^{\dag}\left(i\omega-vk_{x}\sigma_{1}-vk_{y}\sigma_{2}\right.\nonumber
\\
&&\left.-Ak_{z}^{2}\sigma_{3}\right) \label{Eq:ActionFermion}
\psi_{a},
\\
S_{b}&=&\int\frac{d\omega}{2\pi}\frac{d^{3}\mathbf{k}}{(2\pi)^{3}}\phi\left(\frac{1}{\sqrt{\eta}}k_{\bot}^{2}
+\sqrt{\eta}k_{z}^{2}\right)\phi, \label{Eq:ActionBoson}
\\
S_{fb}&=&ig\int d\tau d^3\mathbf{x}\psi_{a}^{\dag}\psi_{a}\phi, \label{Eq:ActionFBCoupling}
\\
S_{dis}&=&\sum_{j=0}^{3}\frac{\Delta_{j}}{2}\int d\tau d\tau'd^3\mathbf{x}\left(\psi_{a}^{\dag}\Gamma_{j}\psi_{a}\right)_{\tau}
\left(\psi_{b}^{\dag}\Gamma_{j}\psi_{b}\right)_{\tau'}, \label{Eq:ActionDisorder}
\end{eqnarray}
where $g=\frac{\sqrt{4\pi}e}{\sqrt{\epsilon}}$. $\eta$ is utilized to parametrize the anisotropy of the Coulomb interaction. $a,b=1,2,...,m$ are the
replica indices. The limit $m\rightarrow0$ will be taken at the end of calculation. The type of disorder depends on the concrete expression of $\Gamma_{j}$.
The matrix $\Gamma_{0}=\mathbbm{1}$ corresponds to RSP.  For $\Gamma_{1,2,3}=\sigma_{1,2,3}$, the types of disorder are the
three components of random vector potential (RVP), which are dubbed as $x$-, $y$-, and $z$-RVP respectively in the following.
The parameter $\Delta_{j}$ with $j=0,1,2,3$ represents the strength of disorder. It should be indicated that single component
of RVP could be initially introduced separately  \cite{Sbierski16}.

\section{Expanding parameters for the RG analysis \label{Sec:ExpansionParameter}}

For a fermion system, if the disorder or interaction is marginal  at tree-level, it is not
necessary to employ a formal controlled parameter. We can perform the RG analysis directly for the physical
fermion dispersion and physical dimension of the system. If the disorder or interaction is irrelevant or relevant
at tree-level, we should utilize a formal controlled parameter in the RG analysis. Usually, we can generalize
the dimension of the system to $d$ and perform the $\epsilon$ expansion. The RG equation for the coupling parameter $X$
takes the general form
\begin{eqnarray}
\frac{dX}{d\ell}=\epsilon X+FX^{2},
\end{eqnarray}
where $\epsilon X$ stands for the tree-level contribution and $FX^{2}$ represents the one-loop order contribution.
If the disorder or interaction is irrelevant at tree-level for the physical  dimension, $\epsilon$ is negative
for the physical dimension, and becomes zero for the corresponding lower critical dimension. If the disorder or interaction is relevant at tree-level for the physical dimension, $\epsilon$ is positive for the physical dimension, and  becomes zero for the corresponding upper critical dimension.

Recently, another method was used  in the RG analysis for the interaction or disorder effect in some SMs \cite{Roy18, Luo18A, Roy17A, RoyFoster18}.
Namely, generalizing the dispersion of fermions to employ a controlled parameter $1/n$.
This method was firstly proposed by Roy, Goswami, and Juri\v{c}i\'{c} to study the influence of short-range four-fermion interactions on multi-WSMs \cite{Roy17A}.

In SMs, the disorder may be marginal, irrelevant, or relevant at tree-level, which is determined by the
dispersion of fermions and dimension of the system \cite{Ye98, Ye99, Stauber05, Herbut08, Vafek08, Foster08,
WangLiu14, Goswami11, Moon14, Roy16, Gonzalez17, Zhao16, Nandkishore17, Mandal18, WangYuXuan17,
WangLiuZhang17B, YangZK18, ZhaoPL19, Sikkenk19, Roy18, Luo18A, Luo18B, Ostrovsky06, Foster12, Syzranov16, Sbierski16}. In 3D AWSM, the disorder is irrelevant at tree-level.
Roy \emph{et al.} studied the disorder effects in 3D AWSM by generalizing the fermion dispersion
$E=\pm\sqrt{v^2k_{\bot}^{2}+A^{2}k_{z}^{4}}$ to $E=\pm\sqrt{v^{2}k_{\bot}^{2}+A^{2}k_{z}^{2n}}$ where $n$ is
an even integer, and performing $1/n$ expansion \cite{Roy18}. They also studied the disorder effects in 3D AWSM
by generalizing the dimension of system to  $d$ and performing $\epsilon=\frac{5}{2}-d$ expansion.

In many SMs, including 2D DSM \cite{Gonzalez94, Son07, Hofmann14}, 3D DSM/WSM \cite{Roy16, Goswami11, Hosur12, Throckmorton15}, multi-WSMs \cite{Lai15, Jian15, WangLiuZhang17B, ZhangShiXin17}, NLSM \cite{Huh16, WangYuXuan17}, and 3D AWSM \cite{Yang14A}, the long-range Coulomb interaction is marginal
at tree-level. Thus, it is not necessary to employ a formal controlled expansion parameter in the
RG analysis of influence of Coulomb interaction on these SMs. In Luttinger SM, the long-rang Coulomb interaction is relevant at tree-level. $\epsilon=4-d$ expansion was employed to
study the effects of long-range Coulomb interaction in Luttinger SM \cite{Nandkishore17, Mandal18, Moon13, Herbut14, Janssen17}.

In SMs, short-range four-fermion interaction is irrelevant at tree-level, since the DOS vanishes at the Fermi level \cite{Roy16, Roy17A, RoyFoster18}.
Thus, a formal expansion parameter should be employed to study the influence of four-fermion interactions on
SMs by the RG theory. Roy, Goswami, and Juri\v{c}i\'{c} generalized dispersions of multi-Weyl fermions to the
expression $E=\sqrt{A^{2}k_{\bot}^{2n}+v^{2}k_{z}^{2}}$ and employed the expansion parameter
 $1/n$ to study the impact of four-fermion interactions in multi-WSMs \cite{Roy17A}. For 2D semi-DSM, Roy and
Foster generalized the fermion dispersion from $E=\pm\sqrt{v^2k_{x}^{2}+A^{2}k_{y}^{4}}$ to
$E=\pm\sqrt{v^2k_{x}^{2}+A^{2}k_{y}^{2n}}$ with $n$ being an even integer and used the expansion parameter $1/n$ to study
the influence of four-fermion interactions on 2D semi-DSM \cite{RoyFoster18}.

When performing RG analysis of the interplay of interaction and disorder, or the interplay of different interactions in SMs,
different controlled parameters may be needed to employ.

In 3D DSM, the disorder and four-fermion interaction are irrelevant at tree-level, but the long-range Coulomb interaction is marginal at tree-level.
Additionally, the disorder and four-fermion interaction take different scalings
at tree-level. In reference~\cite{Roy16}, Roy and Das Sarma studied the interplay of disorder and four-fermion interactions in 3D DSM.
They employed $\epsilon_{1}=1-d$ as the formal controlled expansion parameter for
the four-fermion interactions, but $\epsilon_{2}=2-d$ as the formal controlled expansion parameter for the
disorder. In reference~\cite{Roy16}, Roy and Das Sarma also studied the interplay of four-fermion interactions
and long-range Coulomb interaction. They generalized the dimension of the system to $d$ and carried out
the $\epsilon_{1}=1-d$ expansion to calculate the corrections induced by four-fermion
interactions. When calculating the corrections induced by
long-range Coulomb interaction, the dimension of system was directly taken as the physical value $d=3$, and none
formal expansion parameter was employed. Goswami and Chakravarty analyzed the interplay of disorder and
long-range Coulomb interaction in 3D DSM \cite{Goswami11}. In reference~\cite{Goswami11},
we also notice that none expansion parameter was employed for the
RG analysis of the corrections induced by Coulomb interaction.

In this article, we study the interplay of Coulomb interaction and disorder in 3D AWSM.
In 3D AWSM, long-range Coulomb interaction is marginal at tree-level, and disorder is irrelevant at tree-level. It is not necessary to employ a formal expansion parameter for the long-range Coulomb interaction. Thus, in the Appendices, when calculating the corrections induced  by long-range Coulomb interaction, we take the physical Hamiltonian density
\begin{eqnarray}
\mathcal{H}_{f}=v\left(k_{x}\sigma_{1}+k_{y}\sigma_{2}\right)+Ak_{z}^{2}\sigma_{3},
\label{Eq:HamitonianGeneral}
\end{eqnarray}
directly. In the calculation, we utilize the momentum shell $b\Lambda<\sqrt{v^2k_{\bot}^{2}+A^{2}k_{z}^{4}}<\Lambda$, where $b=e^{-\ell}$ with $\ell$ being the
running parameter.

Disorder is irrelevant in 3D AWSM at tree-level.
In order to supply an explicit controlled parameter for the RG analysis of the disorder couplings, similar to references~\cite{Roy18, Luo18A},
we generalize the Hamiltonian density of fermions to
\begin{eqnarray}
\mathcal{H}_{f}=v\left(k_{x}\sigma_{1}+k_{y}\sigma_{2}\right)+Ak_{z}^{n}\sigma_{3},
\end{eqnarray}
where $n$ is an even integer. In this formalism, $1/n$ serves the controlled parameter in terms of disorder coupling.
In addition, it is easy to find that the disorder coupling is marginal at tree-level in the limit $n\rightarrow\infty$. We adopt the momentum shell $b\Lambda<\sqrt{v^2k_{\bot}^{2}+A^{2}k_{z}^{2n}}<\Lambda$
when calculating the corrections induced by disorder.

In \ref{App:Propagators} to \ref{App:RGEquations}, we calculate all of the corrections to one-loop order, and derive the RG equations for the related parameters.

\section{Renormalization group analysis \label{Sec:Results}}

In this section,
we analyze the influence of Coulomb interaction and disorder on 3D AWSM by RG theory \cite{Shankar94}.

Considering the influence of Coulomb interaction and disorder, the dynamical exponents within the $x$-$y$ plane and along the $z$ axis can be generally written as
\begin{eqnarray}
z_{1}&=&1+\frac{d\ln(v/v_{0})}{d\ell}, \label{Eq:Definitionz1}
\\
z_{2}&=&n+\frac{d\ln(A/A_{0})}{d\ell}. \label{Eq:Definitionz2}
\end{eqnarray}
The first terms of equations~(\ref{Eq:Definitionz1}) and (\ref{Eq:Definitionz2}) represent
the tree-level contribution. The second terms stand for the contribution from
renormalization of $v$ and $A$ induced by Coulomb interaction and disorder.

In subsection~\ref{SubSec:OnlyCoulomb}, we show the results only considering Coulomb interaction.
In this subsection, $n=2$ in equation~(\ref{Eq:Definitionz2}) is taken directly.
In subsection~\ref{SubSec:Expansion}, we carry out
$1/n$ expansion by discarding the subleading terms of one-loop order contribution
induced by disorder scattering, and then take the physical value $n=2$ for the tree-level contribution finally.
In subsection~\ref{SubSec:DisNoExpansion},  we directly take the physical value $n=2$  and present the numerical results
incorporating subleading terms contributed by disorder scattering.

Yang \emph{et al.} has studied the influence of Coulomb interaction on AWSM by RG theory \cite{Yang14A}.
The disorder effects in AWSM has been analyzed by Roy \emph{et al.} \cite{Roy18} and Luo \emph{et al.} \cite{Luo18A}.
Thus, the results shown in subsection~\ref{SubSec:OnlyCoulomb} has been obtained in reference~\cite{Yang14A}, and the results
shown in \ref{SubSec:OnlyDisorderLeading} and \ref{SubSec:OnlyDisorderSubLeading}
have been got in Refs.~\cite{Roy18, Luo18A}.  The results about interplay of Coulomb interaction and disorder in AWSM are new, and presented
in subsections~\ref{SubSec:InterplayLeading} and \ref{SubSec:InterplaySubLeading}.
In this article, we retain subsections~\ref{SubSec:OnlyCoulomb}, \ref{SubSec:OnlyDisorderLeading} and \ref{SubSec:OnlyDisorderSubLeading},
for the convenience of  comparing and analyzing the results.

\subsection{Only Coulomb interaction \label{SubSec:OnlyCoulomb}}

For 3D AWSM, if only Coulomb interaction is considered, the RG equations reduce to
\begin{eqnarray}
\frac{dv}{d\ell}&=&C_{1}v,
\\
\frac{dA}{d\ell}&=&C_{2}A,
\\
\frac{d\alpha}{d\ell}
&=&\left(-C_{1}-\frac{1}{2}\beta-\frac{1}{2}\gamma\right)\alpha,
\\
\frac{d\beta}{d\ell}
&=&\left(\frac{1}{2}-\frac{1}{2}C_{2}-\beta\right)\beta,
\\
\frac{d\gamma}{d\ell}
&=&\left(-\frac{1}{2}+\frac{1}{2}C_{2}-2C_{1}-\gamma\right)\gamma.
\end{eqnarray}
The parameters $\beta$ and $\gamma$ are defined as
$\beta=\frac{3\alpha}{10\pi\bar{A}}$ and
$\gamma=\frac{8}{21\pi}\alpha\bar{A}$ respectively, where $\bar{A}=\frac{\sqrt{A}\sqrt{\Lambda}}{v\sqrt{\eta}}$. The parameter can be expressed as $\gamma=\frac{8\alpha^{2}}{70\pi^{2}\beta}$, which is a combination of $\alpha$ and $\beta$. The parameter $\gamma$ is defined to make the RG equations to look compact.
The coefficients $C_{i}\equiv C_{i}(\alpha,\zeta)$ where $\zeta=\bar{A}^{2}$. The expressions of $C_{1}$ and $C_{2}$
are given by
\begin{eqnarray}
C_{1}&=&\frac{\alpha\zeta^{3/2}}{2\pi }
\int_{0}^{+\infty}d\chi\frac{1}{\sqrt{\chi}\left(1+\chi^2\right)^{1/4}}\nonumber
\\
&&\times\frac{1}{\left[\zeta+
\chi \left(1+\chi^2\right)^{1/2}\right]^{2}}, \label{Eq:C1Expression}
\\
C_{2}&=&\frac{\alpha\sqrt{\zeta}}{2\pi }
\int_{0}^{+\infty}d\chi\sqrt{\chi}\left(1+\chi^2\right)^{1/4}\nonumber
\\
&&\times\frac{-\zeta+3\chi \left(1+\chi^2\right)^{1/2}}{\left[\zeta+
\chi\left(1+\chi^2\right)^{1/2}\right]^{3}}. \label{Eq:C2Expression}
\end{eqnarray}

\begin{figure}[htbp]
\center
\includegraphics[width=3.2in]{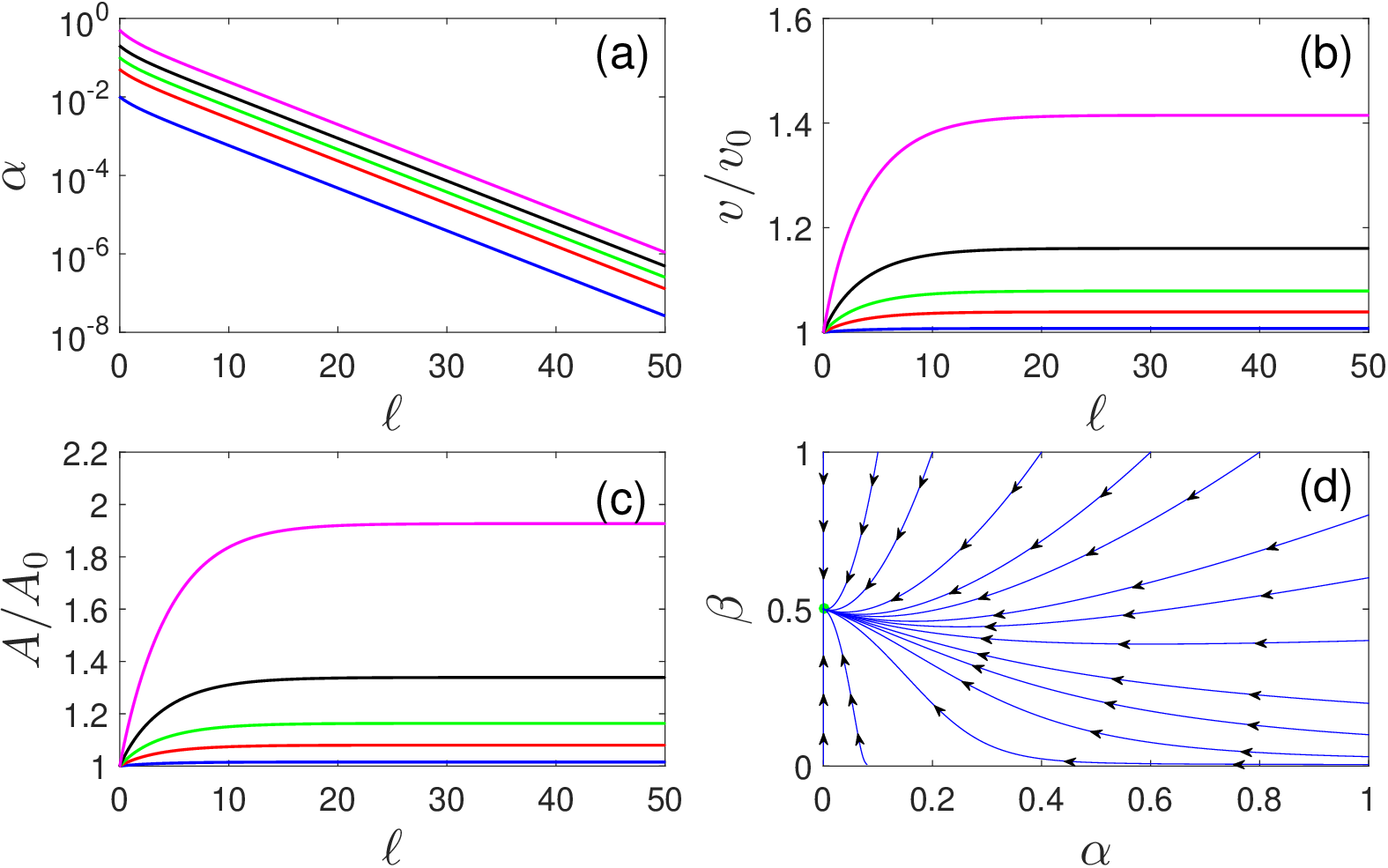}
\caption{ (a)-(c) Flows of $\alpha$, $v$, and $A$ considering only long-range Coulomb interaction.
Blue, red, green, black, magenta curves represent the initial values $\alpha_{0}=0.01, 0.05, 0.1, 0.2, 0.5$.
$\beta_{0}=1$ is taken in (a)-(c).
(d) The flow diagram on the $\alpha$-$\beta$ plane.
\label{Fig:VRGClean}}
\end{figure}

Dependence of $\alpha$ on the running parameter $\ell$ is shown in figure~\ref{Fig:VRGClean}(a). We can easily find
that $\alpha$ approaches to zero quickly in the lowest energy limit, which represents that the
Coulomb interaction is irrelevant.  According to figures~\ref{Fig:VRGClean}(b) and \ref{Fig:VRGClean}(c), $v$ and $A$
flow from the bare values $v_{0}$ and $A_{0}$ to two new constants, and only
acquire quantitative increments. Thus, the behaviors of the observable quantities
including DOS, specific heat \emph{etc.} do not change qualitatively, but only receive
quantitative corrections. This is a standard characteristic of Fermi liquid.  The flow diagram on the $\alpha$-$\beta$ plane is presented in figure~\ref{Fig:VRGClean}(d).
It is obvious that $(\alpha,\beta)$ flows to a stable fixed point $(\alpha^{*},\beta^{*})=(0,\frac{1}{2})$.
The parameter $\gamma$ satisfying $\gamma=\frac{8\alpha^{2}}{70\pi^{2}\beta}$ approaches to zero quickly in the lowest energy limit. In the low-energy regime, the asymptotical form of the parameter $\alpha$ is
given by
\begin{eqnarray}
\alpha\sim e^{-\frac{1}{4}\ell}.
\end{eqnarray}

The finite fixed value $\beta^{*}=\frac{1}{2}$ corresponds to anisotropic screening effect. Concretely, including the
correction of the polarization $\Pi(q_{\bot},q_{z})$, the dressed long-range Coulomb interaction can be written as
\begin{eqnarray}
D^{-1}(q_{\bot},q_{z})&\sim&\frac{1}{\sqrt{\eta}}q_{\bot}^{2}+\sqrt{\eta} q_{z}^{2}+\Pi(q_{\bot},q_{z})\nonumber
\\
&\sim&
\frac{1}{\sqrt{\eta}}q_{\bot}^{2}\left(1+\beta\ell\right)+\sqrt{\eta} q_{z}^{2}\left(1+\gamma\ell\right)\nonumber
\\
&\sim&\frac{1}{\sqrt{\eta}}q_{\bot}^{3/2}+\sqrt{\eta} q_{z}^{2},
\end{eqnarray}
which takes anisotropic dependence on $q_{\bot}$ and $q_{z}$.

In reference~\cite{Yang14A}, a momentum shell
is imposed to the  component $k_{z}$ in the RG analysis.
Although different RG schemes are adopted, the physical results shown in above are well consistent with reference~\cite{Yang14A}. Our results are also
in accordance with the pioneering work of Abrikosov \cite{Abrikosov72}.

For 3D DSM/WSM, previous studies exhibited that the Coulomb strength $\alpha$ flows to zero but with a slow speed \cite{Goswami11, Hosur12, Throckmorton15, Roy16}.
Accordingly, the fermion velocity acquires logarithmic-like
correction on momentum. Thus, the low-energy behaviors of the observable quantities receive
logarithmic-like corrections in their energy or temperature dependence. These results manifest that long-range
Coulomb interaction in 3D DSM/WSM is marginally irrelevant in the low-energy regime.

The obviously  different roles of Coulomb interaction in 3D AWSM and 3D DSM/WSM result from that the Coulomb field $\phi$
acquires a finite anomalous dimension $\eta_{\phi}$ in 3D AWSM, but $\eta_{\phi}$ vanishes in the
lowest energy limit in 3D DSM/WSM. In \ref{Sec:Comparing3DAWSM3DWSM}, we present the detailed
analysis of the reason for the different roles of Coulomb interaction in 3D AWSM and WSM.

\subsection{RG analysis  considering disorder scattering after performing $1/n$ expansion \label{SubSec:Expansion}}

In this subsection, for the RG equations~(\ref{Eq:VRGvGeneral})- (\ref{Eq:VRGDelta3General}),
we discard the subleading terms in the sense of $1/n$ expansion induced by disorder scattering, and then
take the physical value $n=2$.

\subsubsection{Only disorder \label{SubSec:OnlyDisorderLeading}}

Taking the large $n$ limit, the RG equations for the disorder strength parameters
can be further simplified to
\begin{eqnarray}
\frac{d\Delta_{0}}{d\ell}&=&-\frac{1}{n}\Delta_{0}+\Delta_{0}\left(\Delta_{0}+\Delta_{1}+\Delta_{2}\right),
\label{Eq:RGEDelta0Expansion}
\\
\frac{d\Delta_{1}}{d\ell}&=&-\frac{1}{n}\Delta_{1}+\Delta_{0}\Delta_{3}, \label{Eq:RGEDelta1Expansion}
\\
\frac{d\Delta_{2}}{d\ell}&=&-\frac{1}{n}\Delta_{2}+\Delta_{0}\Delta_{3}, \label{Eq:RGEDelta2Expansion}
\\
\frac{d\Delta_{3}}{d\ell}&=&-\frac{1}{n}\Delta_{3}+\Delta_{3}\left(-\Delta_{0}
+\Delta_{1}+\Delta_{2}-\Delta_{3}\right)\nonumber
\\
&&+\Delta_{0}\left(\Delta_{1}+\Delta_{2}\right). \label{Eq:RGEDelta3Expansion}
\end{eqnarray}
Under this approximation, one type of disorder could exist solely, and only RSP can drive the QPT to CDM phase.
In CDM, the fermions acquire a
finite disorder scattering $\gamma_{0}$, and DOS at the Fermi level $\rho(0)$ becomes a finite constant which is
determined by $\gamma_{0}$ \cite{Goswami11, Roy14}. $\gamma_{0}$ or $\rho(0)$ could be regarded as the order parameter for the QPT
from SM to CDM.  Both conventional metal and CDM have the characteristic that  $\rho(0)$ is finite.
The difference is that finite $\rho(0)$ in conventional metal results from finite chemical potentail, but finite $\rho(0)$ in CDM results from
finite disorder scattering rate $\gamma_{0}$.

If only considering RSP, the RG equation for strength of RSP is given by
\begin{eqnarray}
\frac{d\Delta_{0}}{d\ell}&=&-\frac{1}{n}\Delta_{0}+\Delta_{0}^{2}. \label{Eq:VRGDelta0RSPExp}
\end{eqnarray}
There is a nontrivial solution $\Delta_{0}^{*}=\frac{1}{n}$. Thus,
\begin{eqnarray}
\left(\Delta_{0}^{*},\Delta_{1}^{*},\Delta_{2}^{*},\Delta_{3}^{*}\right)&=&\left(\frac{1}{n},0,0,0\right), \label{Eq:FixedPointExpansion}
\end{eqnarray}
is an unstable fixed point.
At this fixed point, the
RG equations for $v$ and $A$ take the forms
\begin{eqnarray}
\frac{dv}{d\ell}&=&-\frac{1}{2}\Delta_{0}^{*}v,
\\
\frac{dA}{d\ell}&=&-\frac{1}{2}\Delta_{0}^{*}A.
\end{eqnarray}
The solutions of these equations are
\begin{eqnarray}
\frac{v}{v_{0}}&=&e^{-\Delta_{0}^{*}\ell/2},
\\
\frac{A}{A_{0}}&=&e^{-\Delta_{0}^{*}\ell/2}.
\end{eqnarray}
Considering the renormalization of $v$ and $A$,  the dynamical exponents become
\begin{eqnarray}
z_{1}&=&1+\frac{1}{2}\Delta_{0}^{*}=1+\frac{1}{2n}, \label{Eq:z1FixedPoint}
\\
z_{2}&=&n+\frac{1}{2}\Delta_{0}^{*}=n+\frac{1}{2n}. \label{Eq:z2FixedPoint}
\end{eqnarray}
The correlation length exponent $\nu$ can be calculated through the formula \cite{Roy14}
\begin{eqnarray}
\nu^{-1}=\left.\frac{d}{d\Delta_{0}}\left(\frac{d\Delta_{0}}{d\ell}\right)\right|_{\Delta_{0}=\Delta_{0}^{*}}.
\end{eqnarray}
We find that
\begin{eqnarray}
\nu&=&\left(-\frac{1}{n}+2\Delta_{0}^{*}\right)^{-1}
=\left(-\frac{1}{n}+2\frac{1}{n}\right)^{-1}=n.
\end{eqnarray}
We can also obtained the value of correlation length exponent by linearizing the flow equations in
the vicinity of the fixed point \cite{Goswami11}. In the vicinity of the fixed point shown in
equation~(\ref{Eq:FixedPointExpansion}), linearizing the equation~(\ref{Eq:VRGDelta0RSPExp})
and solving it yield
\begin{eqnarray}
\Delta_{0}-\Delta_{0}^{*}=\left(\Delta_{0,0}-\Delta_{0}^{*}\right)e^{\frac{1}{n}\ell}\propto e^{\nu^{-1}\ell}.
\end{eqnarray}
Thus, we get the same result $\nu=n$.
For the physical value $n=2$, we have $z_{1}=\frac{5}{4}$, $z_{2}=\frac{9}{4}$,
and $\nu=2$. The results $z_{1}=\frac{5}{4}$ and $\nu=2$ are exactly consistent with the ones given by Roy \emph{et al.} \cite{Roy18}.
The dynamical exponent along the $z$ axis at the fixed point $z_{2}=\frac{9}{4}$
is not shown in reference~\cite{Roy18}.

The first terms of equations~(\ref{Eq:RGEDelta0Expansion})-(\ref{Eq:RGEDelta3Expansion}) represent
the tree-level contribution. Taking the limit $n\rightarrow\infty$, these terms vanish and
equations~(\ref{Eq:RGEDelta0Expansion})-(\ref{Eq:RGEDelta3Expansion}) become to the RG equations
for the disorder effects in 2D DSM \cite{Evers08, Ostrovsky06, Foster12}.
Additionally, for the exponents shown in equations~(\ref{Eq:z1FixedPoint}) and (\ref{Eq:z2FixedPoint}),
taking $n\rightarrow\infty$, we can find $z_{1}\rightarrow1$ and $z_{2}\rightarrow \infty$.
This result is well consistent  with the theory for 2D Dirac fermions with a linear dispersion within the
$x$-$y$ plane \cite{Evers08, Ostrovsky06, Foster12}.

\subsubsection{Interplay of Coulomb interaction and disorder \label{SubSec:InterplayLeading}}

Considering the interplay of Coulomb interaction and disorder, the RG equations can be written as
\begin{eqnarray}
\frac{dv}{d\ell}&=&\left(C_{1}-\frac{1}{2}\sum_{j=0}^{3}\Delta_{j}\right)v, \label{Eq:VRGv}
\\
\frac{dA}{d\ell}&=&\left(C_{2}-\frac{1}{2}\sum_{j=0}^{3}\Delta_{j}\right)A, \label{Eq:VRGA}
\\
\frac{d\alpha}{d\ell}
&=&\left(-C_{1}-\frac{1}{2}\beta-\frac{1}{2}\gamma
+\frac{1}{2}\sum_{j=0}^{3}\Delta_{j}\right)\alpha, \label{Eq:VRGalpha}
\\
\frac{d\beta}{d\ell}
&=&\left(\frac{1}{2}-\frac{1}{2}C_{2}-\beta
+\frac{1}{4}\sum_{j=0}^{3}\Delta_{j}\right)\beta, \label{Eq:VRGbeta}
\\
\frac{d\gamma}{d\ell}
&=&\left(-\frac{1}{2}+\frac{1}{2}C_{2}-2C_{1}-\gamma
+\frac{3}{4}\sum_{j=0}^{3}\Delta_{j}\right)\gamma, \label{Eq:VRGgamma}
\\
\frac{d\Delta_{0}}{d\ell}&=&-\frac{1}{n}\Delta_{0}+\Delta_{0}\left(\Delta_{0}+\Delta_{1}+\Delta_{2}\right)\nonumber
\\
&&-\Delta_{0}\left(2C_{1}+\frac{1}{2}C_{2}+2\beta+2\gamma\right), \label{Eq:VRDelta0}
\\
\frac{d\Delta_{1}}{d\ell}&=&-\frac{1}{n}\Delta_{1}+\Delta_{0}\Delta_{3} \label{Eq:VRDelta1}
-\Delta_{1}\left(2C_{1}+\frac{1}{2}C_{2}-C_{3}\right),
\\
\frac{d\Delta_{2}}{d\ell}&=&-\frac{1}{n}\Delta_{2}+\Delta_{0}\Delta_{3}
-\Delta_{2}\left(2C_{1}+\frac{1}{2}C_{2}-C_{3}\right), \label{Eq:VRDelta2}
\\
\frac{d\Delta_{3}}{d\ell}&=&-\frac{1}{n}\Delta_{3}+\Delta_{3}\left(-\Delta_{0}
+\Delta_{1}+\Delta_{2}-\Delta_{3}\right)\nonumber
\\
&&+\Delta_{0}\left(\Delta_{1}+\Delta_{2}\right)
-\Delta_{3}\left(2C_{1}+\frac{1}{2}C_{2}-C_{4}\right), \label{Eq:VRDelta3}
\end{eqnarray}
where
\begin{eqnarray}
C_{3}&=&\frac{\alpha\sqrt{\zeta}}{2\pi }\int_{0}^{+\infty}d\chi
\frac{2+\chi^{2}}{\sqrt{\chi}\left(1+\chi^2\right)^{5/4}}\nonumber
\\
&&\times\frac{1}{\zeta+ \chi \left(1+\chi^2\right)^{1/2}},
\\
C_{4}&=&\frac{\alpha\sqrt{\zeta}}{\pi }\int_{0}^{+\infty}d\chi
\frac{\chi^{3/2}}{\left(1+\chi^2\right)^{5/4}}\nonumber
\\
&&\times\frac{1}{\zeta+ \chi \left(1+\chi^2\right)^{1/2}}.
\end{eqnarray}
In the numerical calculation, the physical value $n=2$ is taken for the
terms $\frac{1}{n}\Delta_{i}$ in equations~(\ref{Eq:VRDelta0})-(\ref{Eq:VRDelta3}).

Incorporating the renormalization of parameters $v$ and $A$, the dynamical exponents
become
\begin{eqnarray}
z_{1}(\ell)=1-C_{1}(\ell)+\frac{1}{2}\sum_{j=0}^{3}\Delta_{j}(\ell),
\\
z_{2}(\ell)=n-C_{2}(\ell)+\frac{1}{2}\sum_{j=0}^{3}\Delta_{j}(\ell).
\end{eqnarray}
The anomalous dimension of fermion field is given by
\begin{eqnarray}
\eta_{\psi}=\frac{1}{4}\sum_{j=0}^{3}\Delta_{j}.
\end{eqnarray}
If we consider the interplay of Coulomb interaction and RSP, the dynamical exponents
are
\begin{eqnarray}
z_{1}(\ell)=1-C_{1}(\ell)+\frac{1}{2}\Delta_{0}(\ell),
\\
z_{2}(\ell)=n-C_{2}(\ell)+\frac{1}{2}\Delta_{0}(\ell).
\end{eqnarray}
The anomalous dimension $\eta_{\psi}$ becomes
\begin{eqnarray}
\eta_{\psi}=\frac{1}{4}\Delta_{0}.
\end{eqnarray}

There are several fixed points. In the vicinity of the fixed point
\begin{eqnarray}
\left(\alpha^{*},\beta^{*},\Delta_{0}^{*}\right)=(0,0,0),
\end{eqnarray}
linearizing the RG equations, we get
\begin{eqnarray}
\frac{d\alpha}{d\ell}&\sim&\left(-C_{1}-\frac{1}{2}\beta-\frac{1}{2}\gamma
+\frac{1}{2}\Delta_{0}\right)\alpha,
\\
\frac{d\beta}{d\ell}&\sim&\frac{1}{2}\beta,
\\
\frac{d\Delta_{0}}{d\ell}&\sim&-\frac{1}{2}\Delta_{0}.
\end{eqnarray}
The corresponding solutions for $\beta$ and $\Delta_{0}$ are $\beta\sim\beta_{0}e^{\frac{1}{2}\ell}$,
$\Delta_{0}\sim\Delta_{0,0}e^{-\frac{1}{2}\ell}$. Additionally, $\alpha\rightarrow0$ from the initial
value $\alpha_{0}$ with growing of $\ell$.
We can find that this fixed point is unstable since $\beta$ is relevant, and
the system is robust against RSP in the vicinity of this fixed point.

\begin{figure}[htbp]
\center
\includegraphics[width=3.2in]{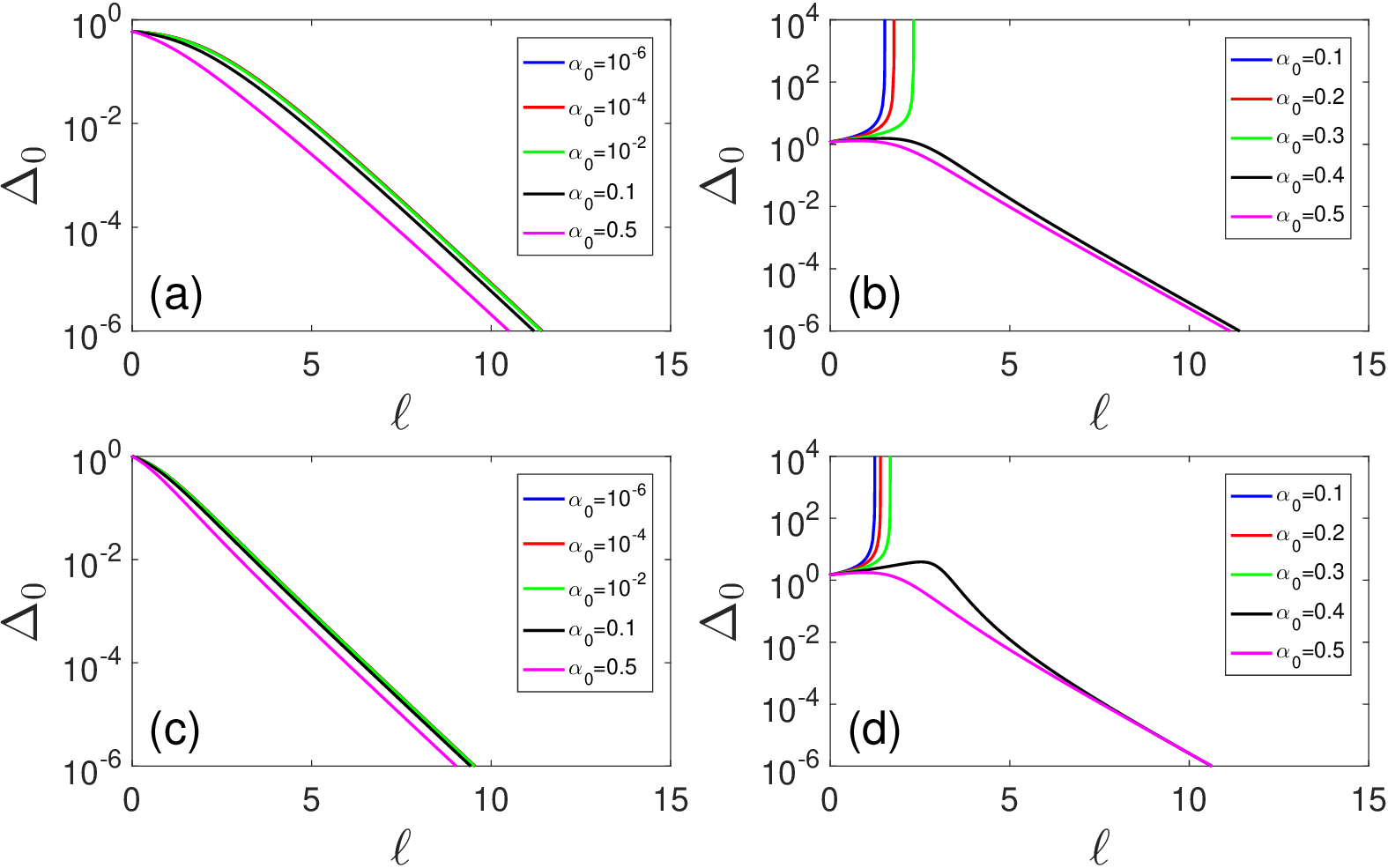}
\caption{ Flows of $\Delta_{0}$ considering initially RSP and Coulomb interaction. (a) $\Delta_{0,0}=0.6$, $\beta_{0}=0.1$;
(b) $\Delta_{0,0}=1.2$, $\beta_{0}=0.1$; (c) $\Delta_{0,0}=1.2$, $\beta_{0}=0.2$; (d) $\Delta_{0,0}=1.5$, $\beta_{0}=0.2$.
Subleading terms in the sense of $1/n$ expansion contributed by the disorder scattering
 are discarded in this figure, and figures~\ref{Fig:VRGCoulombRSPFlow}, \ref{Fig:PhaseDiagramRSPCoulombExpansion} and \ref{Fig:ExponentZExpansion}.
\label{Fig:VRGCoulombRSPExpansion}}
\end{figure}

In the vicinity of the fixed point
\begin{eqnarray}
\left(\alpha^{*},\beta^{*},\Delta_{0}^{*}\right)=\left(0,\frac{1}{2},0\right),
\end{eqnarray}
the RG equations can be linearized as
\begin{eqnarray}
\frac{d\alpha}{d\ell}&\sim&-\frac{1}{4}\alpha,
\\
\frac{d\beta}{d\ell}&\sim&-\left(\beta-\frac{1}{2}\right),
\\
\frac{d\Delta_{0}}{d\ell}&=&-\frac{3}{2}\Delta_{0}.
\end{eqnarray}
It is easy to obtain $\alpha\sim\alpha_{0}e^{-\frac{1}{4}\ell}$, $\beta-\frac{1}{2}=\left(\beta_{0}-\frac{1}{2}\right)e^{-\ell}$,
and $\Delta_{0}\sim\Delta_{0,0}e^{-\frac{3}{2}\ell}$. These results indicate that this fixed point is a stable fixed point.

In the vicinity of the fixed point
\begin{eqnarray}
\left(\alpha^{*},\beta^{*},\Delta_{0}^{*}\right)=\left(0,0,\frac{1}{2}\right),
\end{eqnarray}
through linearizing the RG equations, we get
\begin{eqnarray}
\frac{d\alpha}{d\ell}&\sim&\frac{1}{4}\alpha, \label{Eq:LinearFP3Alpha}
\\
\frac{d\beta}{d\ell}&\sim&\frac{5}{8}\beta, \label{Eq:LinearFP3Beta}
\\
\frac{d\Delta_{0}}{d\ell}&\sim&\frac{1}{2}\left(\Delta_{0}-\frac{1}{2}\right)-\frac{1}{2}
\left(2C_{1}+\frac{1}{2}C_{2}+2\beta+2\gamma\right). \label{Eq:LinearFP3Delta0}
\end{eqnarray}
Therefore, $\alpha\sim\alpha_{0}e^{\frac{1}{4}\ell}$ and $\beta\sim\beta_{0}e^{\frac{5}{8}\ell}$. These results
represent that Coulomb interaction is relevant in the vicinity of this fixed point.

\begin{figure}[htbp]
\center
\includegraphics[width=3.2in]{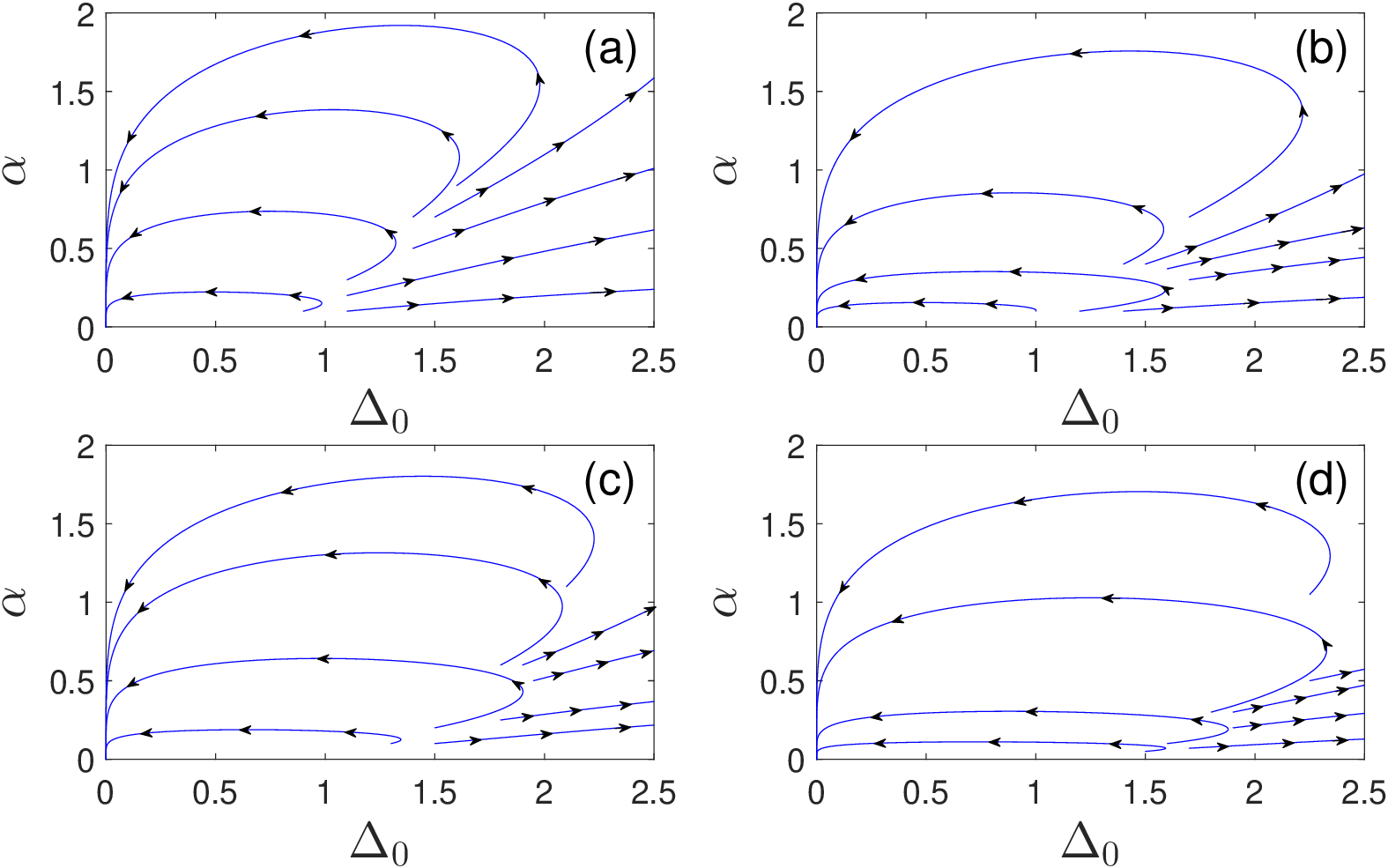}
\caption{ Projections of flow diagrams on the plane of $\Delta_{0}$ and $\alpha$ with different $\beta_{0}$. $\beta_{0}=0.1$, $0.2$, $0.3$, and $0.4$ in
(a), (b), (c), and (d) respectively.
\label{Fig:VRGCoulombRSPFlow}}
\end{figure}

From equations~(\ref{Eq:C1Expression}) and (\ref{Eq:C2Expression}),
and equations~(\ref{Eq:LinearFP3Alpha})-(\ref{Eq:LinearFP3Delta0}), we expect there should be a relation
\begin{eqnarray}
\frac{d\ln\left[\Delta_{0}-\frac{1}{2}-F(\alpha,\beta)\right]}{d\ell}\sim\frac{1}{2},
\end{eqnarray}
where $F(\alpha,\beta)$ formally is a function of $\alpha$ and $\beta$. Accordingly,
\begin{eqnarray}
\Delta_{0}-\frac{1}{2}-F(\alpha,\beta)&=&\left[\Delta_{0,0}-\frac{1}{2}-F(\alpha_{0},\beta_{0})\right]e^{\frac{1}{2}\ell}\nonumber
\\
&\propto& e^{\nu^{-1}\ell}.
\end{eqnarray}
It indicates that the correlation length exponent at the phase boundary between SM phase and CDM phase is still $\nu=2$. In the
study about interplay of Coulomb interaction and disorder in 3D DSM, Goswami \emph{et al.} also showed that the value of correlation
length exponent at the phase boundary is not changed by Coulomb interaction \cite{Goswami11}.

Dependence of $\Delta_{0}$ on $\ell$ considering initially both of RSP and long-range Coulomb interaction with different initial conditions is shown in
figure~\ref{Fig:VRGCoulombRSPExpansion}. It is clear  that RSP is suppressed by long-range Coulomb interaction.
For a given $\beta_{0}$, there are two different cases determined by the initial value $\Delta_{0,0}$. In the first case, $\Delta_{0,0}$ takes a relatively small value,
such as figures~\ref{Fig:VRGCoulombRSPExpansion}(a) and \ref{Fig:VRGCoulombRSPExpansion}(c), then $\Delta_{0}$ always flows to zero even if $\alpha_{0}$ takes arbitrarily
small value.  It represents that the system is always in SM phase once both of RSP and long-range Coulomb interaction are considered.
For a large  enough $\Delta_{0,0}$,  such as figures~\ref{Fig:VRGCoulombRSPExpansion}(b) and \ref{Fig:VRGCoulombRSPExpansion}(d), $\Delta_{0}$ flows away if the Coulomb strength takes
a smaller initial value, but approaches to zero if the initial value of Coulomb strength is larger than a critical value. It indicates that
there is a QPT from CDM phase to SM phase with increasing of Coulomb interaction.

The projections of flow diagrams on the plane of $\Delta_{0}$ and $\alpha$ are shown in figure~\ref{Fig:VRGCoulombRSPFlow}.
We can find that the system flows to the point $(0,0$) or is driven to the strong coupling regime. We notice that the flows of $\alpha$ and
$\Delta_{0}$ may take nonmonotonic behaviors under proper conditions. Accordingly, the observable quantities may exhibit
nomonotonic dependence on energy or temperature.

 \begin{figure}[htbp]
\center
\includegraphics[width=3.2in]{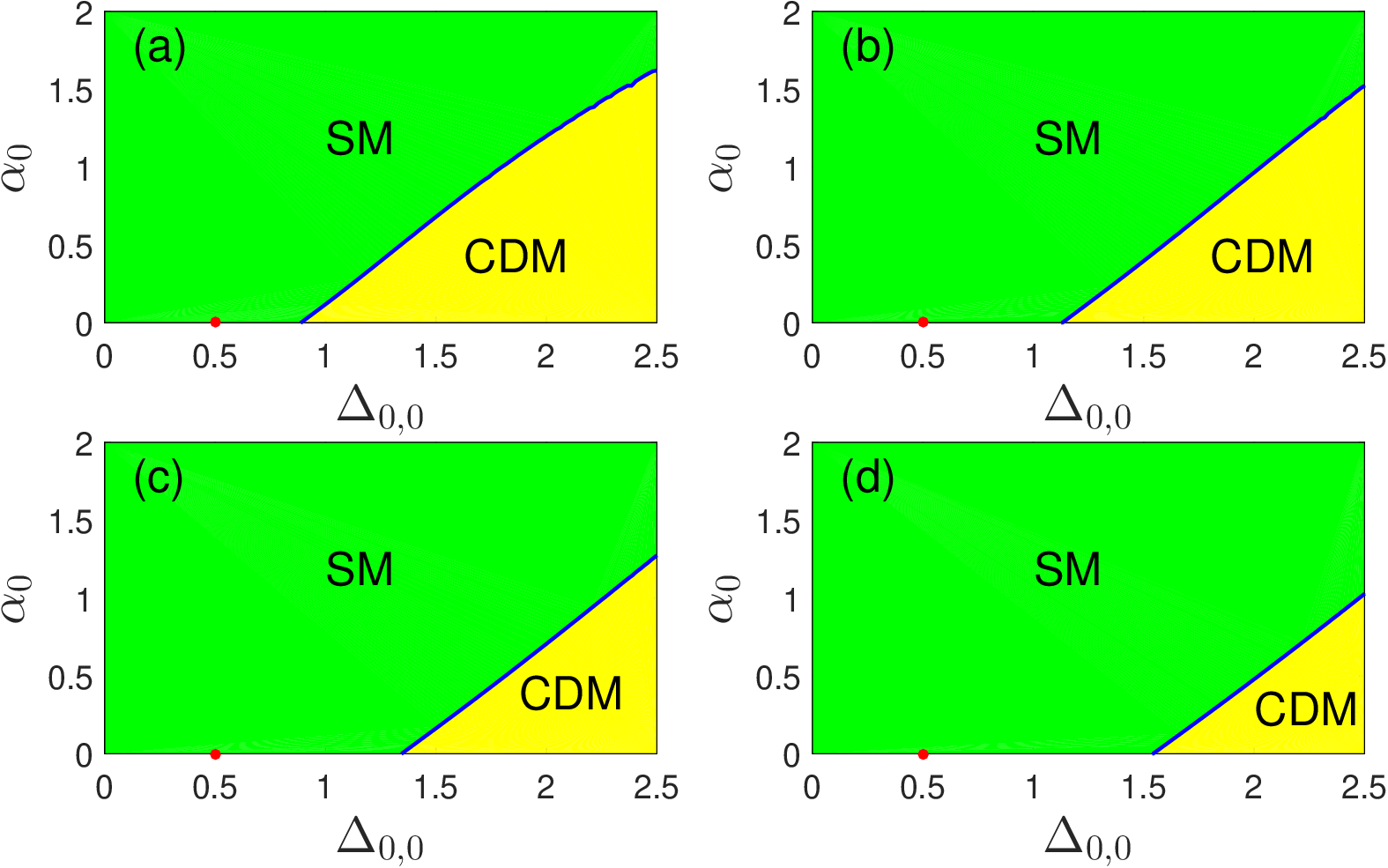}
\caption{ Phase diagrams of 3D AWSM considering initially both of RSP and Coulomb interaction.
$\beta_{0}=0.1, 0.2, 0.3, 0.4$ in (a), (b), (c), and (d) respectively. The red point represents the
critical strength of RSP corresponding to the QCP between SM and CDM phases neglecting Coulomb interaction.
\label{Fig:PhaseDiagramRSPCoulombExpansion}}
\end{figure}

Considering long-range Coulomb interaction and RSP initially, the phase diagrams  on
the plane of $\Delta_{0,0}$ and $\alpha_{0}$ with different values of $\beta_{0}$ are shown
in figures~\ref{Fig:PhaseDiagramRSPCoulombExpansion}(a)-\ref{Fig:PhaseDiagramRSPCoulombExpansion}(d). The red point represents the QCP from
SM phase to CDM phase considering only RSP initially.  According to figure~\ref{Fig:PhaseDiagramRSPCoulombExpansion},
the CDM phase is compressed, but the SM phase is enlarged by long-range Coulomb interaction.
There is a quite novel result: Once long-range Coulomb interaction
is also considered, the critical strength of RSP for the appearance of CDM receives a prominent increment, even
if the initial Coulomb strength $\alpha_{0}$ takes arbitrarily small value.
For larger $\beta_{0}$, the suppression effect for RSP by long-range Coulomb interaction is
more obvious. The parameter $\beta$ is corresponding to the anisotropic screening effect of Coulomb interaction.
It indicates that the remarkable suppression effect for RSP results essentially  from the anisotropic
screening effect of Coulomb interaction, which is contributed by the Feynman diagram shown in figure~\ref{Fig:VertexCorrection}(e).
If $\Delta_{0,0}$ is larger than a critical value, the system is in CDM phase for small
$\alpha_{0}$, but restores SM phase if $\alpha_{0}$ is large enough.

\begin{figure}[htbp]
\center
\includegraphics[width=3.2in]{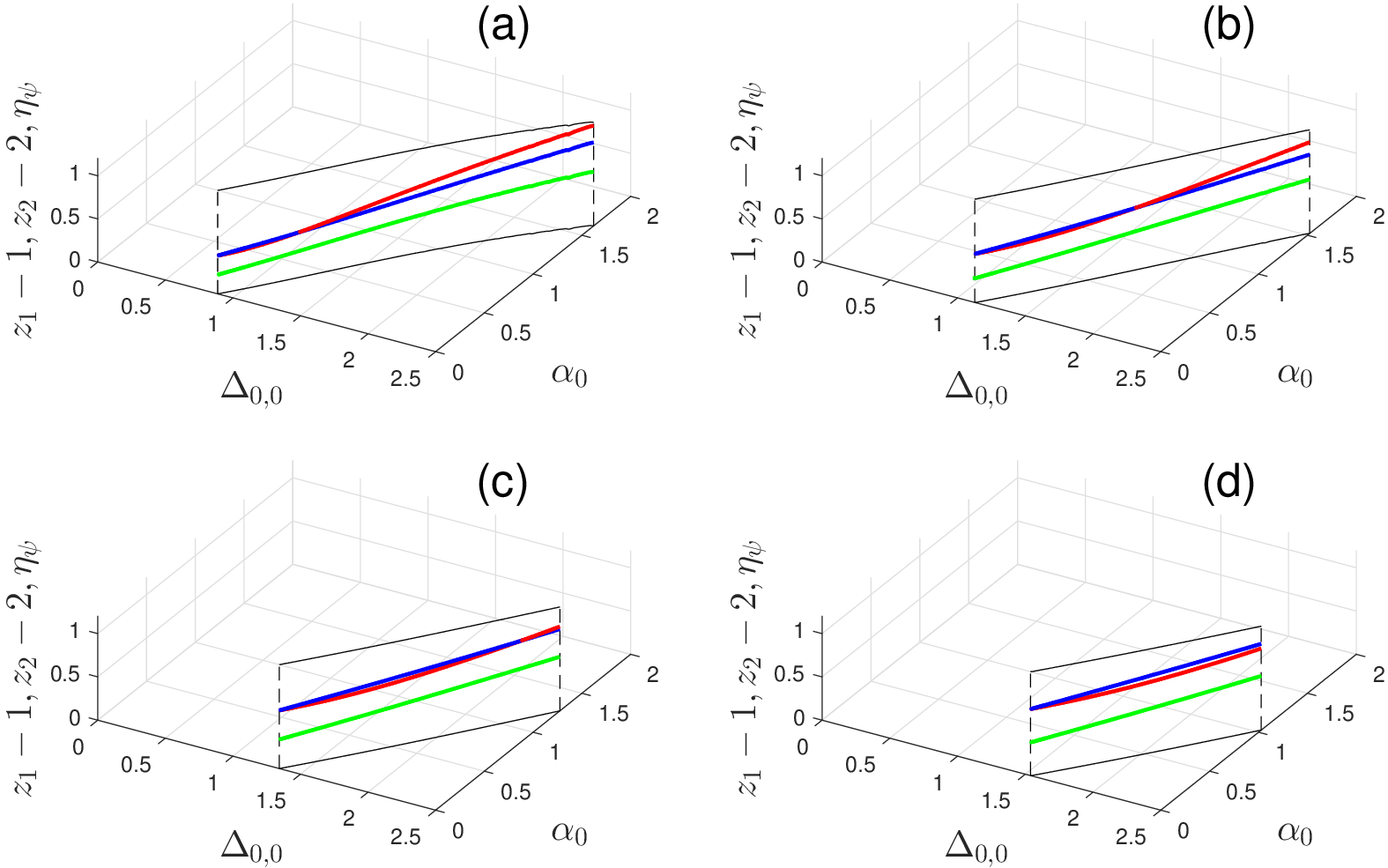}
\caption{ The dynamical exponents $z_{1}$ and $z_{2}$, and the anomalous dimension of the fermion field $\eta_{\psi}$
at the boundary between SM and CDM phases. Blue, red, and green lines are corresponding
to $z_{1}-1$, $z_{2}-2$, and $\eta_{\psi}$ respectively. $\beta_{0}=0.1$, $0.2$, $0.3$ and $0.4$ in (a), (b), (c), and (d) respectively.
\label{Fig:ExponentZExpansion}}
\end{figure}

The dynamical exponents $z_{1}$ and $z_{2}$, and the anomalous dimension of fermion field $\eta_{\psi}$ at
the phase boundary between SM phase and CDM phase are presented in figure~\ref{Fig:ExponentZExpansion}.
The blue, red, and green lines represent the values of $z_{1}$, $z_{2}$, and $\eta_{\psi}$ respectively.
Since $z_{1}$ and $z_{2}$ acquire finite corrections comparing with the free case,
the fermion dispersion receives power-law correction at the phase boundary.
The fermion damping rate at the phase boundary can be expressed as
\begin{eqnarray}
\mathrm{Im}\Sigma^{R}(\omega)\sim\omega^{1-2\eta_{\psi}},
\end{eqnarray}
which is a characteristic of non-Fermi liquid state.

\subsection{RG analysis considering influence of disorder including subleading contribution
in terms of disorder coupling \label{SubSec:DisNoExpansion}}

Taking $n=2$ directly for equations~(\ref{Eq:VRGvGeneral})-(\ref{Eq:VRGDelta3General}),
the RG equations for the
corresponding parameters are given by
\begin{eqnarray}
\frac{dv}{d\ell}&=&\left(C_{1}-\frac{1}{2}\sum_{j=0}^{3}\Delta_{j}\right)v, \label{Eq:VRGv}
\\
\frac{dA}{d\ell}&=&\left(C_{2}-\frac{1}{2}\sum_{j=0}^{3}\Delta_{j}\right)A, \label{Eq:VRGA}
\\
\frac{d\alpha}{d\ell}
&=&\left(-C_{1}-\frac{1}{2}\beta-\frac{1}{2}\gamma
+\frac{1}{2}\sum_{j=0}^{3}\Delta_{j}\right)\alpha, \label{Eq:VRGalpha}
\\
\frac{d\beta}{d\ell}
&=&\left(\frac{1}{2}-\frac{1}{2}C_{2}-\beta
+\frac{1}{4}\sum_{j=0}^{3}\Delta_{j}\right)\beta, \label{Eq:VRGbeta}
\\
\frac{d\gamma}{d\ell}
&=&\left(-\frac{1}{2}+\frac{1}{2}C_{2}-2C_{1}-\gamma
+\frac{3}{4}\sum_{j=0}^{3}\Delta_{j}\right)\gamma, \label{Eq:VRGgamma}
\\
\frac{d\Delta_{0}}{d\ell}
&=&-\frac{1}{2}\Delta_{0}+\left(\frac{5}{4}\Delta_{0}^{2}+\frac{5}{4}\Delta_{0}\Delta_{1}+\frac{5}{4}\Delta_{0}\Delta_{2}\right.\nonumber
\\
&&\left.+\frac{33}{20}\Delta_{0}\Delta_{3}
+\frac{4}{5}\Delta_{1}\Delta_{3}+\frac{4}{5}\Delta_{2}\Delta_{3}\right)\nonumber
\\
&&-\Delta_{0}\left(2C_{1}+\frac{1}{2}C_{2}+2\beta+2\gamma\right), \label{Eq:VRGDelta0}
\\
\frac{d\Delta_{1}}{d\ell}
&=&-\frac{1}{2}\Delta_{1}+\left(\frac{1}{20}\Delta_{1}\Delta_{0}+\frac{1}{20}\Delta_{1}^{2}+\frac{9}{20}\Delta_{1}\Delta_{2}\right.\nonumber
\\
&&\left.+\frac{17}{20}\Delta_{1}\Delta_{3}+\frac{4}{5}\Delta_{0}\Delta_{3}\right)\nonumber
\\
&&-\Delta_{1}\left(2C_{1}+\frac{1}{2}C_{2}-C_{3}\right), \label{Eq:VRGDelta1}
\\
\frac{d\Delta_{2}}{d\ell}
&=&-\frac{1}{2}\Delta_{2}+\left(\frac{1}{20}\Delta_{2}\Delta_{0}+\frac{9}{20}\Delta_{2}\Delta_{1}+\frac{1}{20}\Delta_{2}^{2}\right.\nonumber
\\
&&\left.+\frac{17}{20}\Delta_{2}\Delta_{3}+\frac{4}{5}\Delta_{0}\Delta_{3}\right)\nonumber
\\
&&-\Delta_{2}\left(2C_{1}+\frac{1}{2}C_{2}-C_{3}\right), \label{Eq:VRGDelta2}
\\
\frac{d\Delta_{3}}{d\ell}
&=&-\frac{1}{2}\Delta_{3}+\left(-\frac{7}{20}\Delta_{3}\Delta_{0}+\frac{17}{20}\Delta_{3}\Delta_{1}+\frac{17}{20}\Delta_{3}\Delta_{2}\right.\nonumber
\\
&&+\frac{1}{20}\Delta_{3}^{2}
+\frac{2}{5}\Delta_{0}^{2}+\frac{2}{5}\Delta_{1}^{2}+\frac{2}{5}\Delta_{2}^{2}\nonumber
\\
&&\left.+\frac{4}{5}\Delta_{0}\Delta_{1}+\frac{4}{5}\Delta_{0}\Delta_{2}\right)\nonumber
\\
&&-\Delta_{3}\left(2C_{1}+\frac{1}{2}C_{2}-C_{4}\right), \label{Eq:VRGDelta3}
\end{eqnarray}
If the Coulomb interaction is completely neglected, $\alpha$, $\beta$, $\gamma$, $C_{i}$ are all taken to be zero, which leads to
the RG equations considering only disorder. Once the Coulomb strength $\alpha$ takes arbitrarily finite initial value, $\alpha$ and $\beta$
can flow independently.

\subsubsection{Only disorder\label{SubSec:OnlyDisorderSubLeading}}

In this subsection, the numerical results considering disorder solely are displayed.

The unstable fixed point is determined by the equations
\begin{eqnarray}
&&-\frac{1}{2}\Delta_{0}+\left(\frac{5}{4}\Delta_{0}^{2}+\frac{5}{4}\Delta_{0}\Delta_{1}+\frac{5}{4}\Delta_{0}\Delta_{2}
+\frac{33}{20}\Delta_{0}\Delta_{3}\right.\nonumber
\\
&&\left.
+\frac{4}{5}\Delta_{1}\Delta_{3}+\frac{4}{5}\Delta_{2}\Delta_{3}\right)=0,
\\
&&-\frac{1}{2}\Delta_{1}+\left(\frac{1}{20}\Delta_{1}\Delta_{0}+\frac{1}{20}\Delta_{1}^{2}+\frac{9}{20}\Delta_{1}\Delta_{2}
+\frac{17}{20}\Delta_{1}\Delta_{3}\right.\nonumber
\\
&&\left.+\frac{4}{5}\Delta_{0}\Delta_{3}\right)=0,
\\
&&-\frac{1}{2}\Delta_{2}+\left(\frac{1}{20}\Delta_{2}\Delta_{0}+\frac{9}{20}\Delta_{2}\Delta_{1}+\frac{1}{20}\Delta_{2}^{2}
+\frac{17}{20}\Delta_{2}\Delta_{3}\right.\nonumber
\\
&&\left.+\frac{4}{5}\Delta_{0}\Delta_{3}\right)=0,
\\
&&-\frac{1}{2}\Delta_{3}+\left(-\frac{7}{20}\Delta_{3}\Delta_{0}+\frac{17}{20}\Delta_{3}\Delta_{1}+\frac{17}{20}\Delta_{3}\Delta_{2}
+\frac{1}{20}\Delta_{3}^{2}\right.\nonumber
\\
&&
\left.+\frac{2}{5}\Delta_{0}^{2}+\frac{2}{5}\Delta_{1}^{2}+\frac{2}{5}\Delta_{2}^{2}
+\frac{4}{5}\Delta_{0}\Delta_{1}+\frac{4}{5}\Delta_{0}\Delta_{2}\right)=0.
\end{eqnarray}
Numerical calculation gives rise to the solution
\begin{eqnarray}
&&\left(\Delta_{0}^{*},\Delta_{1}^{*},\Delta_{2}^{*},\Delta_{3}^{*}\right)\nonumber
\\
&\approx&\left(0.239358,0.0307505,0.0307505,0.0667869\right),
\end{eqnarray}
which corresponds to a nontrivial unstable fixed point.

\begin{figure}[htbp]
\center
\includegraphics[width=3.2in]{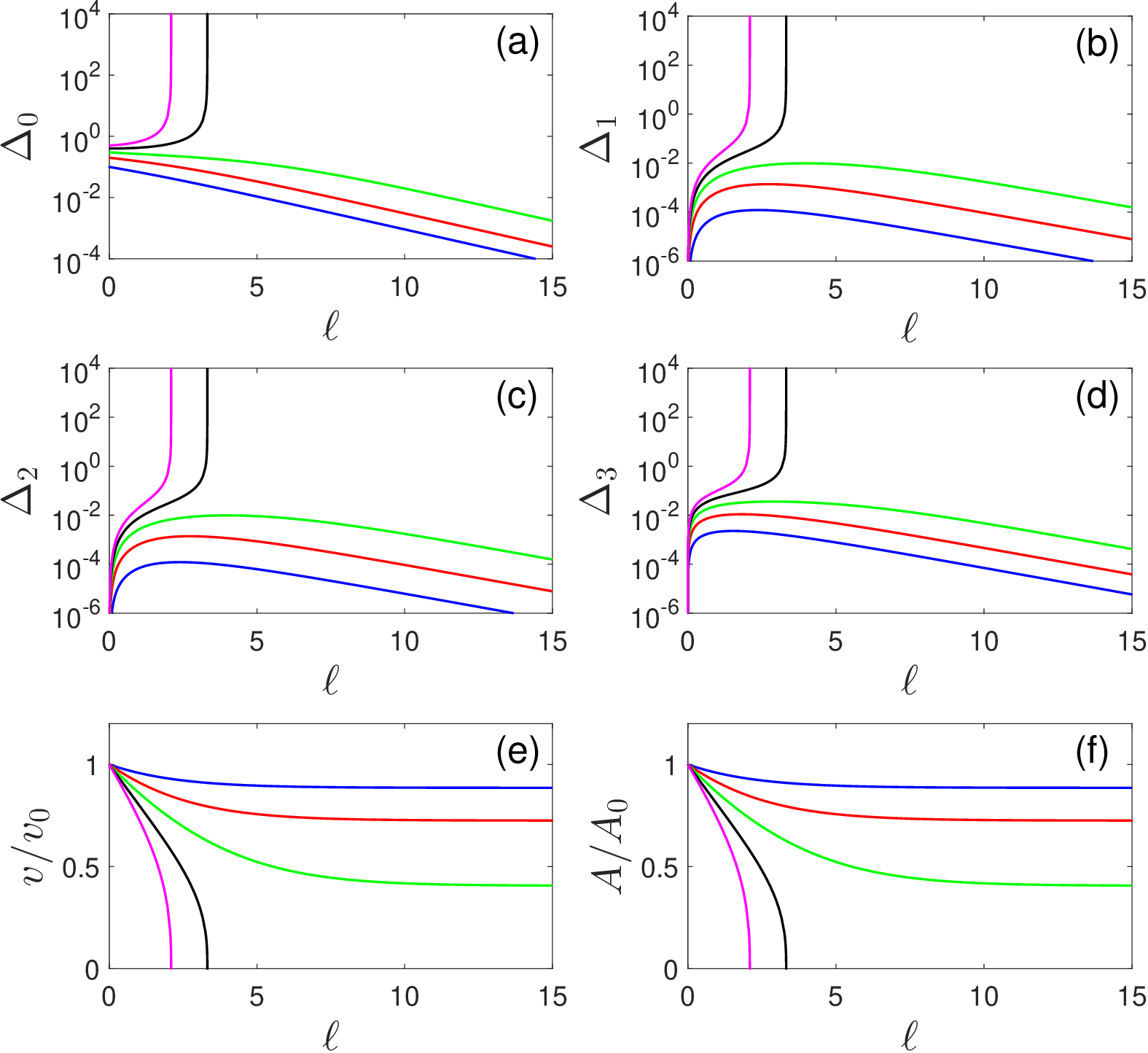}
\caption{ (a)-(f) Flows of $\Delta_{0}$, $\Delta_{1}$, $\Delta_{2}$, $\Delta_{3}$,
$v$, and $A$ only initially including RSP.
Blue, red, green, black, magenta curves represent the initial values $\Delta_{0,0}=0.1, 0.2, 0.3, 0.4, 0.5$.
\label{Fig:VRGDelta0}}
\end{figure}

At this fixed point, the dynamical exponents
are
\begin{eqnarray}
z_{1}=1+\frac{1}{2}\sum_{j=0}^{3}\Delta_{j}^{*}\approx1.1838,
\\
z_{2}=2+\frac{1}{2}\sum_{j=0}^{3}\Delta_{j}^{*}\approx2.1838.
\end{eqnarray}
According to the detailed calculation shown in \ref{App:CLEDerivation},
we find that \begin{eqnarray}
&&\left(\Delta_{0}-\Delta_{0}^{*}\right)+\sum_{j=1}^{3}c_{j}\left(\Delta_{j}-\Delta_{j}^{*}\right)\nonumber
\\
&=&\left[\left(\Delta_{0,0}-\Delta_{0}^{*}\right)+\sum_{j=1}^{3}c_{j}\left(\Delta_{j,0}-\Delta_{j}^{*}\right)
\right]e^{c_{4}\ell}\nonumber
\\
&\propto& e^{\nu^{-1}\ell}.
\end{eqnarray}
$c_{1}$, $c_{2}$, $c_{3}$ and $c_{4}$ are constants, whose values are calculated in  \ref{App:CLEDerivation}.
The concrete value of $c_{4}$ is $c_{4}\approx0.500575$.
Therefore, the correlation length exponent $\nu$ is determined by
\begin{eqnarray}
\nu=\frac{1}{c_{4}}\approx2.
\end{eqnarray}
We can find that the value of $\nu$ is not changed even if
the subleading terms induced by the disorder coupling are considered. This result is
consistent with  reference~\cite{Roy18}.

\begin{figure}[htbp]
\center
\includegraphics[width=3.2in]{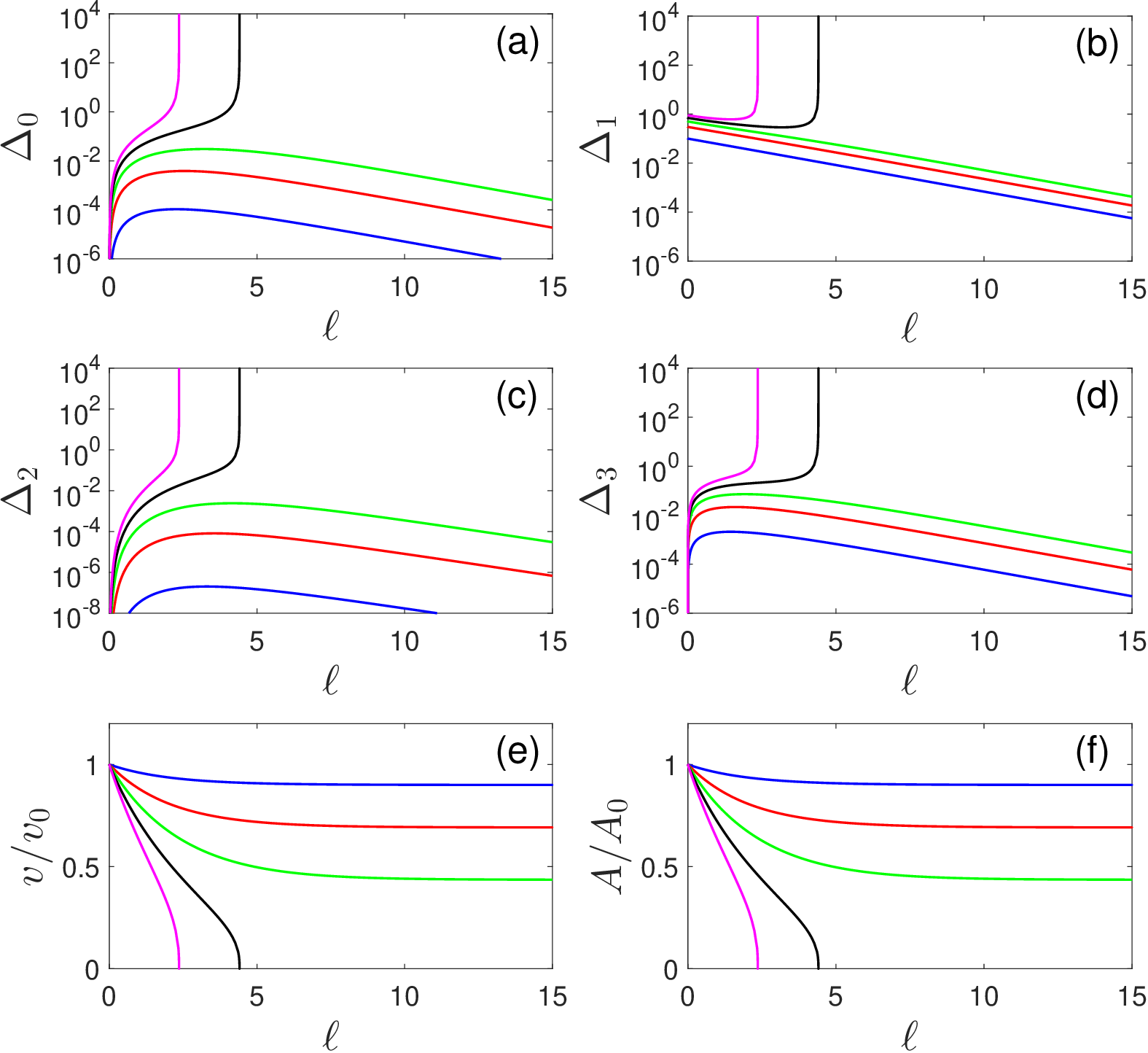}
\caption{ (a)-(f) Flows of $\Delta_{0}$, $\Delta_{1}$, $\Delta_{2}$, $\Delta_{3}$,
$v$, and $A$ only initially including $x$-RVP.
Blue, red, green, black, magenta curves represent the initial values $\Delta_{1,0}=0.1, 0.3, 0.5, 0.7, 0.9$.
\label{Fig:VRGDelta1}}
\end{figure}

If only RSP is considered initially, the flows of $\Delta_{0}$, $\Delta_{1}$, $\Delta_{2}$, $\Delta_{3}$,
$v$, and $A$ are displayed in figures~\ref{Fig:VRGDelta0}(a)-\ref{Fig:VRGDelta0}(f) respectively. If the initial strength
of RSP $\Delta_{0,0}$ is smaller than a critical value $\Delta_{0,0}^{*}$, $\Delta_{0}$ flows to
zero in the lowest energy limit, which represents that RSP is irrelevant. $\Delta_{1}$, $\Delta_{2}$, and $\Delta_{3}$ are dynamically generated
and increase with growing of $\ell$ at the beginning, but start to decrease if $\ell$ is large enough,
and approach to zero eventually. In this case, $v$ and $A$ only receive quantitative corrections and
flow to new constants which are smaller than the initial values $v_{0}$ and $A_{0}$. Accordingly,
the SM phase is stable against the weak disorder, and the observable quantities
do not acquire  qualitative modifications. If $\Delta_{0,0}$ is larger than a critical
value  $\Delta_{0,0}^{*}$,
$\Delta_{0}$ approaches to infinity at some finite energy scale. $\Delta_{1}$, $\Delta_{2}$, and $\Delta_{3}$
are dynamically generated and also flow to infinity finally. $v$ and $A$ flow
to zero at the same finite energy scale. These behaviors are generally believed to signify that the
system becomes unstable and is driven to CDM phase \cite{Goswami11, Roy16, Roy14, Syzranov16, Roy18, Luo18A, Luo18B}.
The critical value $\Delta_{0,0}^{*}$  corresponds to the QCP
between SM and CDM phases. The numerical calculation exhibits that $\Delta_{0,0}^{*}\approx0.324$.

A similar QCP was also found in 3D DSM or WSM through RG analysis \cite{Goswami11, Roy16, Roy14, Syzranov16}. RSP can solely exist in 3D DSM or WSM. However,
$x$-, $y$-, and $z$-RVP are dynamically generated in 3D AWSM although only RSP is considered initially.

The flows of $\Delta_{0}$, $\Delta_{1}$, $\Delta_{2}$, $\Delta_{3}$, $v$, and $A$ considering initially only $x$-RVP
are depicted in figures~\ref{Fig:VRGDelta1}(a)-\ref{Fig:VRGDelta1}(f) respectively. We find that there is a similar threshold value $\Delta_{1,0}^{*}\approx0.626$,
which defines a QCP.  If $\Delta_{1,0}<\Delta_{1,0}^{*}$, $\Delta_{0}$, $\Delta_{1}$, $\Delta_{2}$, and $\Delta_{3}$ all
approach to zero finally, which indicates that the disorder is irrelevant and the SM phase is stable. If $\Delta_{1,0}>\Delta_{1,0}^{*}$,
$\Delta_{0}$, $\Delta_{1}$, $\Delta_{2}$, and $\Delta_{3}$ all flow away, which represents the instability to CDM phase. If only $y$-RVP
is included initially, we obtain similar results, which are not shown here.

\begin{figure}[htbp]
\center
\includegraphics[width=3.2in]{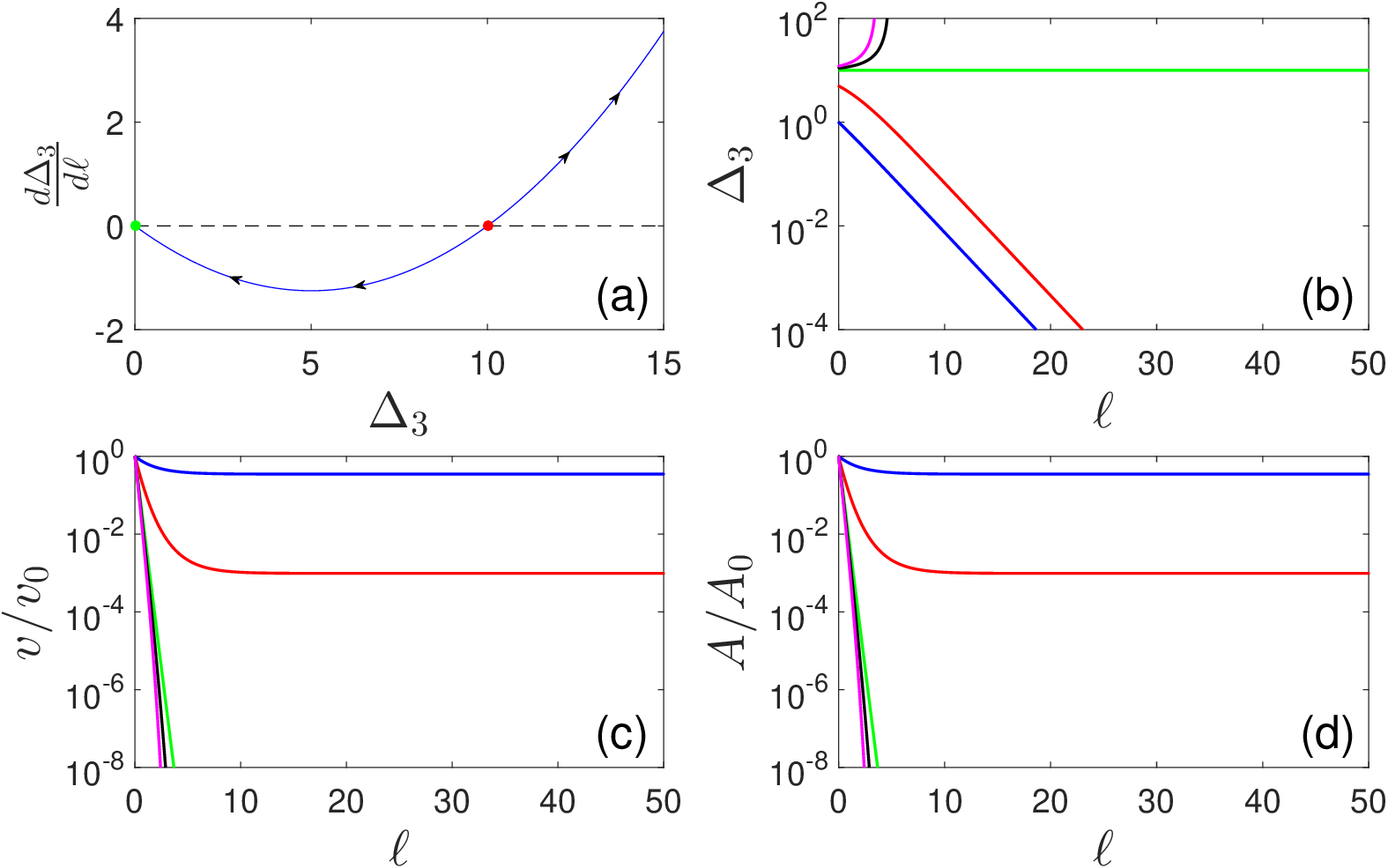}
\caption{ (a) Dependence of $\frac{d\Delta_{3}}{d\ell}$ on $\Delta_{3}$; (b)-(d) Flows of $\Delta_{3}$, $v$, and $A$.
Only $z$-RVP is initially considered. In (b)-(d), blue, red, green, black, magenta curves
represent the initial values $\Delta_{3,0}=1, 5, 10, 11, 12$.
\label{Fig:VRGDelta3}}
\end{figure}

If only $z$-RVP is included initially, it is found that RSP, $x$-RVP, and $y$-RVP are not dynamically generated, and
$z$-RVP can exist solely. The dependence of $\frac{d\Delta_{3}}{d\ell}$ on $\Delta_{3}$ is shown in figure~\ref{Fig:VRGDelta3}(a). The flows
of $\Delta_{3}$, $v$, and $A$ with different initial values $\Delta_{3,0}$ are depicted in figures~\ref{Fig:VRGDelta3}(b), \ref{Fig:VRGDelta3}(c),
and \ref{Fig:VRGDelta3}(d) respectively. If $\Delta_{3,0}$ is smaller than $\Delta_{3,0}^{*}=10$, $\Delta_{3}$ approaches to zero quickly.
If $\Delta_{3,0}>\Delta_{3,0}^{*}$, $z$-RVP becomes relevant and flows away. Thus, there is a QCP from SM to CDM phases
at $\Delta_{3,0}=\Delta_{3,0}^{*}$.

For usual WSM,  if only single component of RVP  exists initially, the RG calculations given by Sbierski \emph{et al.}
exhibit that other types of disorder will not be generated dynamically \cite{Sbierski16}. Additionally, there is not a QPT to CDM phase for any strength of
single component of RVP. The numerical simulations performed by Sbierski \emph{et al.} reveal consistent results comparing with their RG calculations \cite{Sbierski16}.
We can find that usual WSM and AWSM exhibit obviously different behaviors if single component of RVP exists initially. These differences
are closely related to the different properties of Hamiltonian in usual WSM and AWSM.
The Hamiltonian of WSM satisfies $\mathcal{H}_{f}(-\mathbf{k})=-\mathcal{H}_{f}(\mathbf{k})$, but
\begin{eqnarray}
\mathcal{H}_{f}(-\mathbf{k})\neq-\mathcal{H}_{f}(\mathbf{k}),\label{Eq:HamiltonianPR}
\end{eqnarray}
for AWSM. Accordingly, the fermion propagator of WSM
satisfies
\begin{eqnarray}
G_{0}(\omega,\mathbf{k})+G_{0}(-\omega,-\mathbf{k})=0.
\end{eqnarray}
However, the fermion propagator of AWSM has the characteristic
\begin{eqnarray}
G_{0}(\omega,\mathbf{k})+G_{0}(-\omega,-\mathbf{k})\neq0. \label{Eq:PropagatorAniWSMPro}
\end{eqnarray}
Therefore, the two Feynman diagrams shown in figures~\ref{Fig:VertexCorrection}(b) and \ref{Fig:VertexCorrection}(c) lead to zero correction for
the fermion-disorder coupling in WSM, but induce finite nontrivial correction for the fermion-disorder
coupling in AWSM.

Thus, we expect that there should be a QPT to CDM qualitatively if the initial strength of $z$-RVP
is large enough, although the quantitative value for the critical strength of $z$-RVP given by our RG
study may be not  accurate.  This should be an intrinsic property for AWSM resulting from  equation~(\ref{Eq:PropagatorAniWSMPro}).
Numerical simulation methods, including kernel polynomial method \cite{Pixley16}, Lanzos method \cite{Pixley16, FuBo17}, may provide more reliable results for this question.

\begin{figure}[htbp]
\center
\includegraphics[width=3.2in]{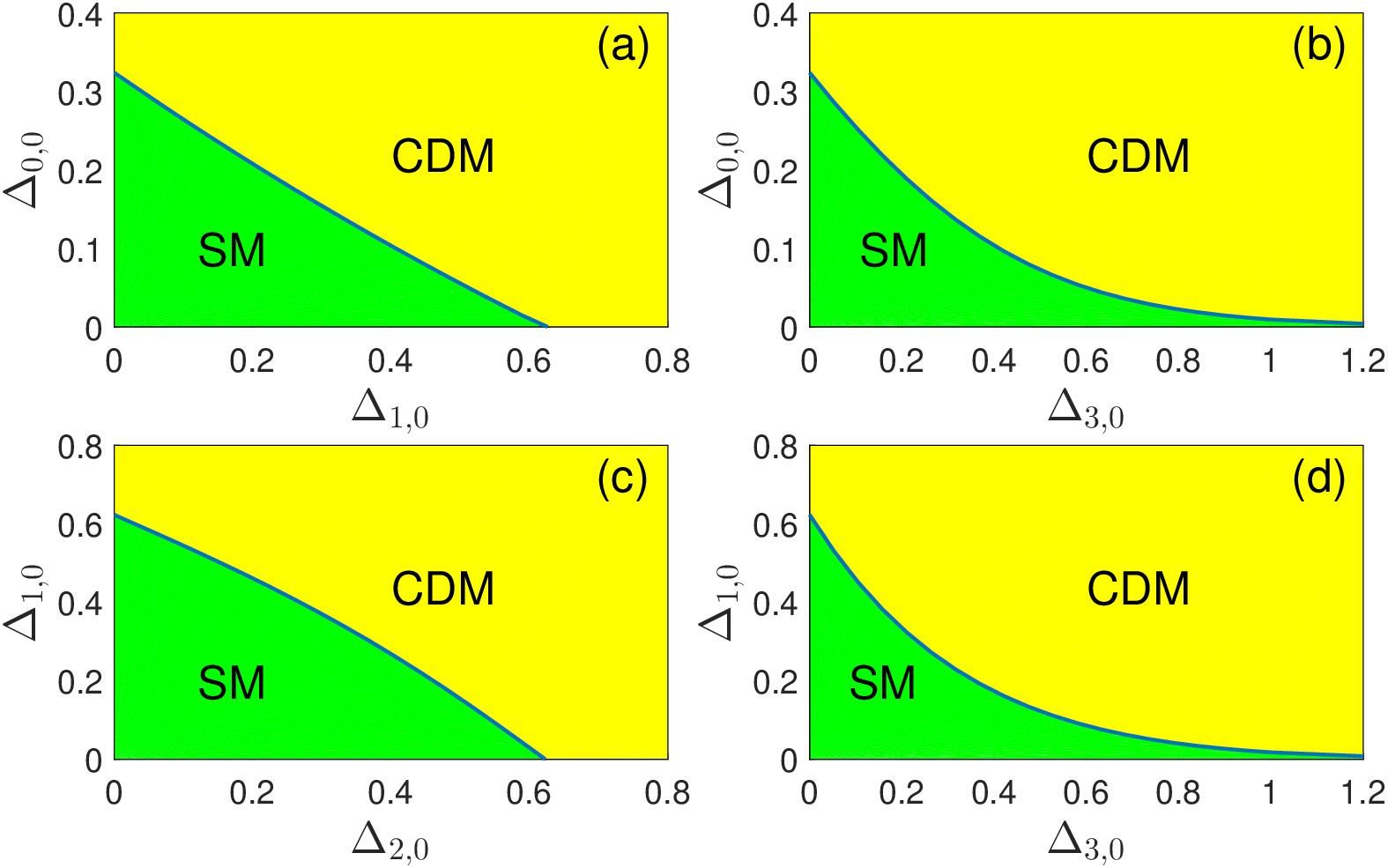}
\caption{ Phase diagrams of 3D AWSM considering initially two types of disorder. (a) RSP and $x$-RVP; (b) RSP and $z$-RVP;
(c) $x$-RVP and $y$-RVP; (d) $x$-RVP and $z$-RVP.
\label{Fig:MixtureDiagram}}
\end{figure}

\begin{figure}[htbp]
\center
\includegraphics[width=3.2in]{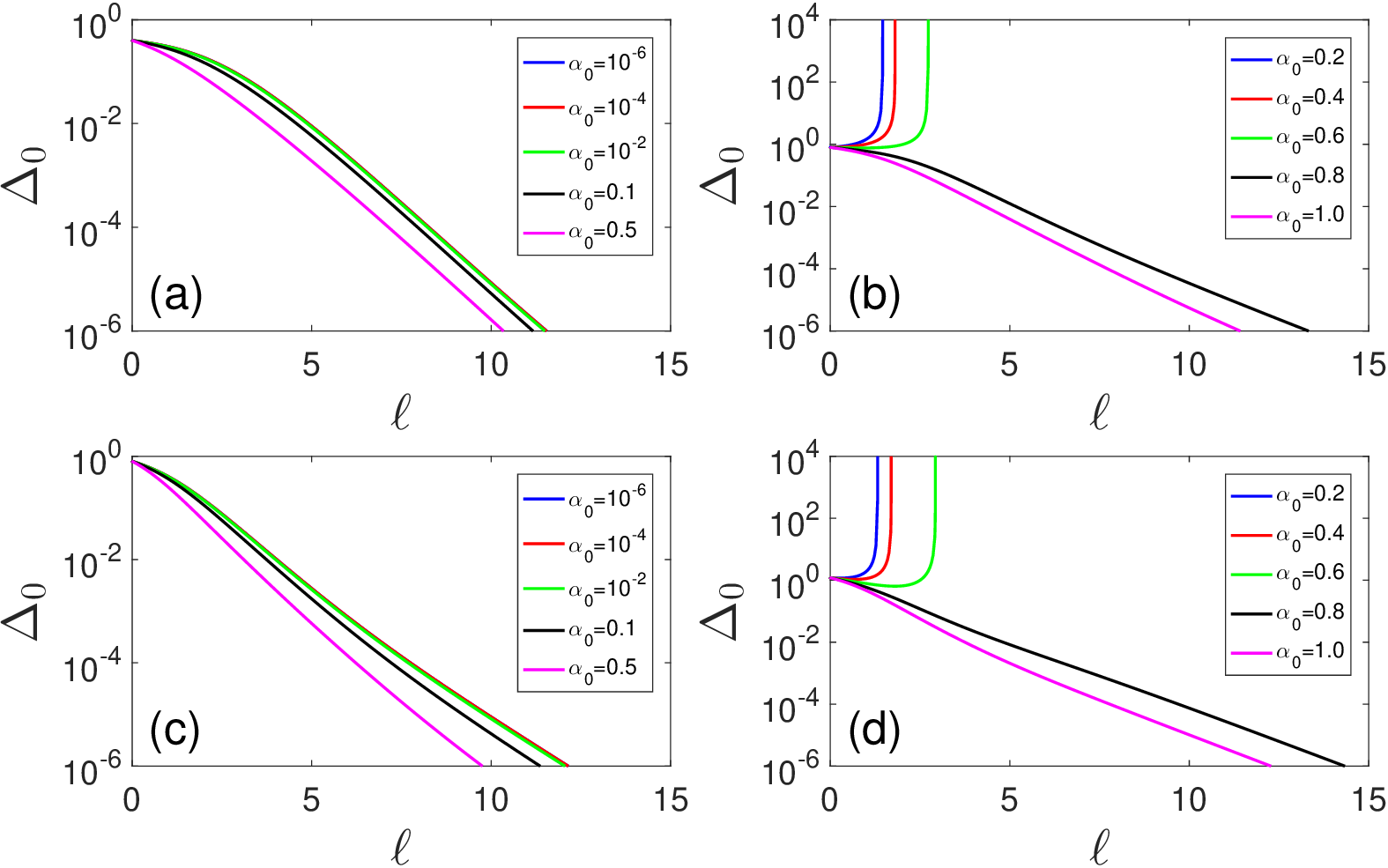}
\caption{ Flows of $\Delta_{0}$ considering initially both of RSP and Coulomb interaction. (a) $\Delta_{0,0}=0.4$, $\beta_{0}=0.1$;
(b) $\Delta_{0,0}=0.8$, $\beta_{0}=0.1$; (c) $\Delta_{0,0}=0.8$, $\beta_{0}=0.5$; (d) $\Delta_{0,0}=1.2$, $\beta_{0}=0.5$.
\label{Fig:VRGCoulombRSP}}
\end{figure}

From equation~(\ref{Eq:RGEDelta3Expansion}), considering only $\Delta_{3}$, and taking physical value $n=2$ for the tree-level contribution, the RG equation for $\Delta_{3}$ can be further written as
\begin{eqnarray}
\frac{d\Delta_{3}}{d\ell}=-\frac{1}{2}\Delta_{3}-\Delta_{3}^{2}. \label{Eq:Delta3RGLargen}
\end{eqnarray}
We can find that $\Delta_{3}$ is always irrelevant. There is not a QPT to CDM phase with increasing of $\Delta_{3}$.
 We should also notice that the generalized Hamiltonian density equation~(\ref{Eq:HamitonianGeneral})
becomes $\mathcal{H}_{0}=v\left(k_{x}\sigma_{1}+k_{y}\sigma_{2}\right)$
in the limit $n\rightarrow\infty$. Therefore, the intrinsic properties shown in equations~(\ref{Eq:HamiltonianPR}) and (\ref{Eq:PropagatorAniWSMPro}) for 3D AWSM are not satisfied for the generalized Hamiltonian equation~(\ref{Eq:HamitonianGeneral}) in the limit $n\rightarrow\infty$. This may be the reason why there is not a QPT to CDM in the limit $n\rightarrow\infty$, according to equation~(\ref{Eq:Delta3RGLargen}).

The phase diagrams considering initially two types of disorder are presented in figure~\ref{Fig:MixtureDiagram}.
The green and yellow regions stand for SM and CDM phases respectively. There is a critical line separating  the SM
and CDM phases. A QPT between SM to CDM phases appears if the initial values of disorder strength are tuned to across
the critical line.

 \begin{figure}[htbp]
\center
\includegraphics[width=3.2in]{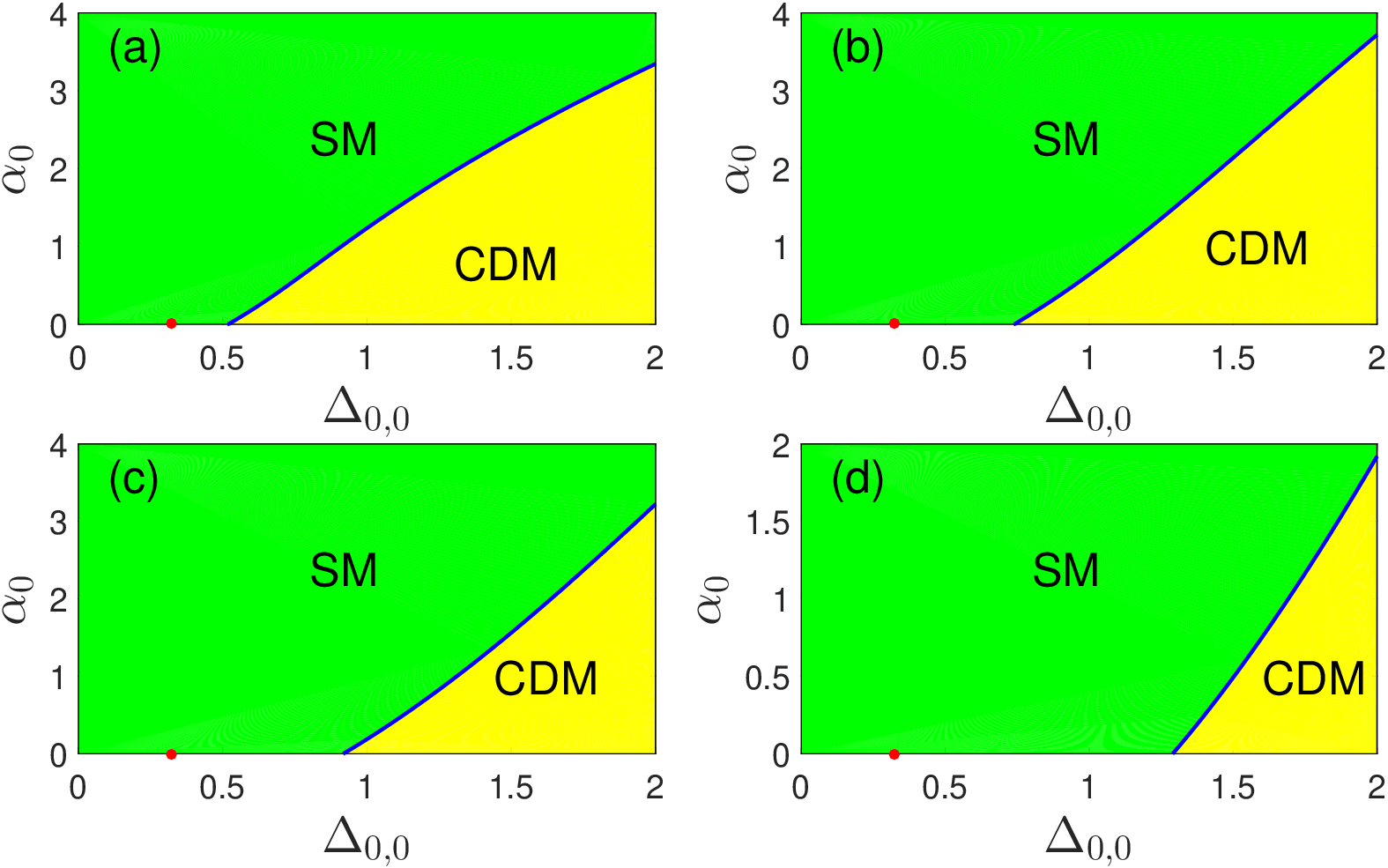}
\caption{ Phase diagrams of 3D AWSM considering initially both of RSP and Coulomb interaction.
$\beta_{0}=0.1, 0.3, 0.5, 1$ in (a), (b), (c), and (d) respectively. The red point represents the
critical strength of RSP corresponding to the QCP between SM and CDM phases neglecting Coulomb interaction.
\label{Fig:PhaseDiagramRSPCoulomb}}
\end{figure}

 \begin{figure}[htbp]
\center
\includegraphics[width=3.2in]{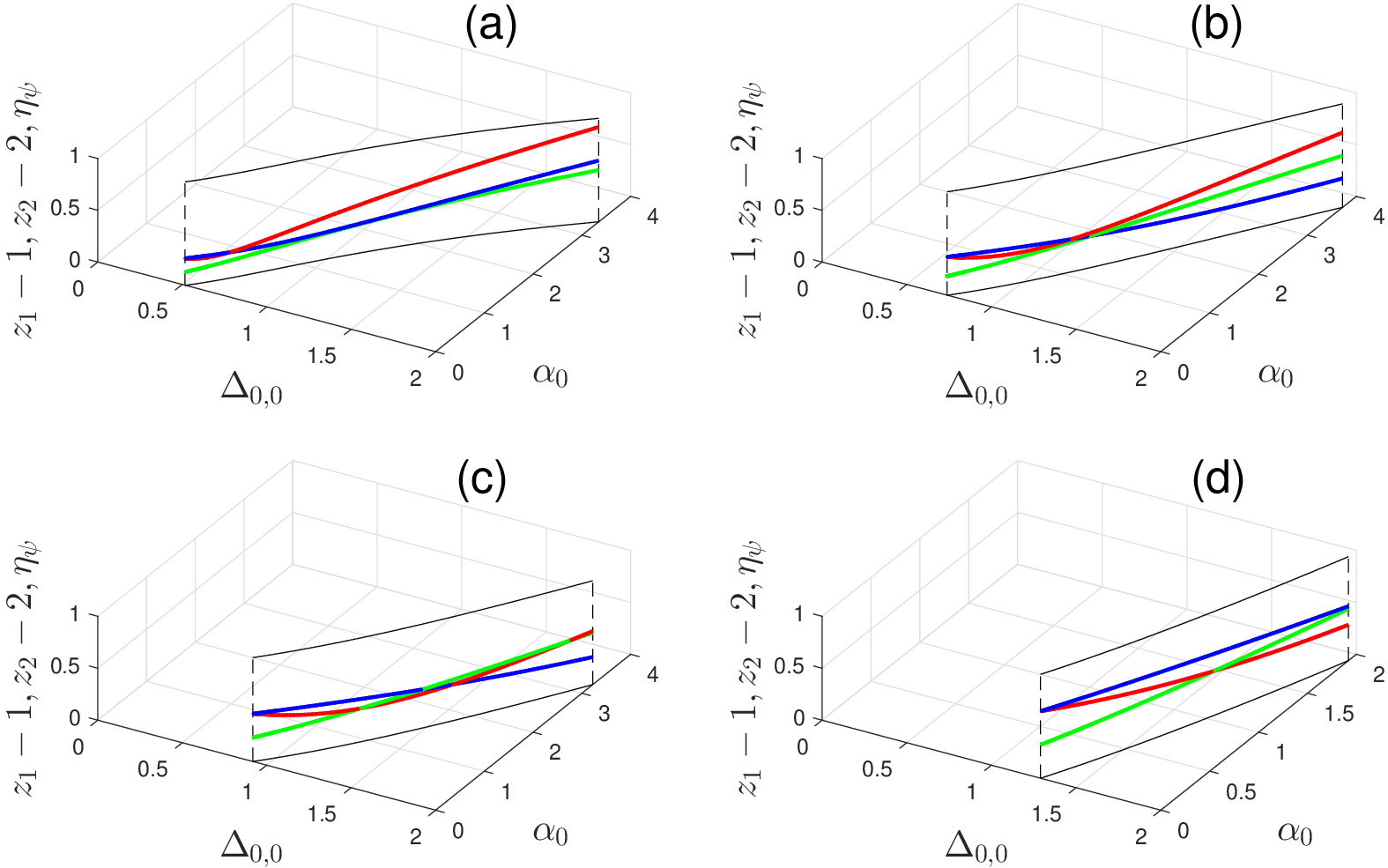}
\caption{ The dynamical exponents $z_{1}$ and $z_{2}$, and the anomalous dimension of fermion field $\eta_{\psi}$
 at the boundary between SM and CDM phases. Blue, red, and green lines are corresponding
to $z_{1}-1$, $z_{2}-2$, and $\eta_{\psi}$ respectively. $\beta_{0}=0.1$, $0.3$, $0.5$ and $1$ in (a), (b), (c), and (d) respectively.
\label{Fig:ExponentZ}}
\end{figure}

\subsubsection{Interplay of Coulomb interaction and disorder \label{SubSec:InterplaySubLeading}}

In this subsection, we analyze the interplay of Coulomb interaction
and disorder in 3D AWSM.

Considering initially both of long-range Coulomb interaction and RSP, the flows of $\Delta_{0}$ with different initial
conditions are shown in figure~\ref{Fig:VRGCoulombRSP}. We can find that RSP is suppressed by Coulomb interaction.
For small initial value $\Delta_{0,0}$, RSP always flows to zero. However, for large enough initial value $\Delta_{0,0}$, RSP  flows away
for small $\alpha_{0}$, but approaches to zero  if $\alpha_{0}$ is larger than a critical value. These results are qualitatively same as the ones
displayed in figure~\ref{Fig:VRGCoulombRSPExpansion}, in which subleading terms induced by the disorder coupling are discarded.

\begin{figure}[htbp]
\center
\includegraphics[width=3.2in]{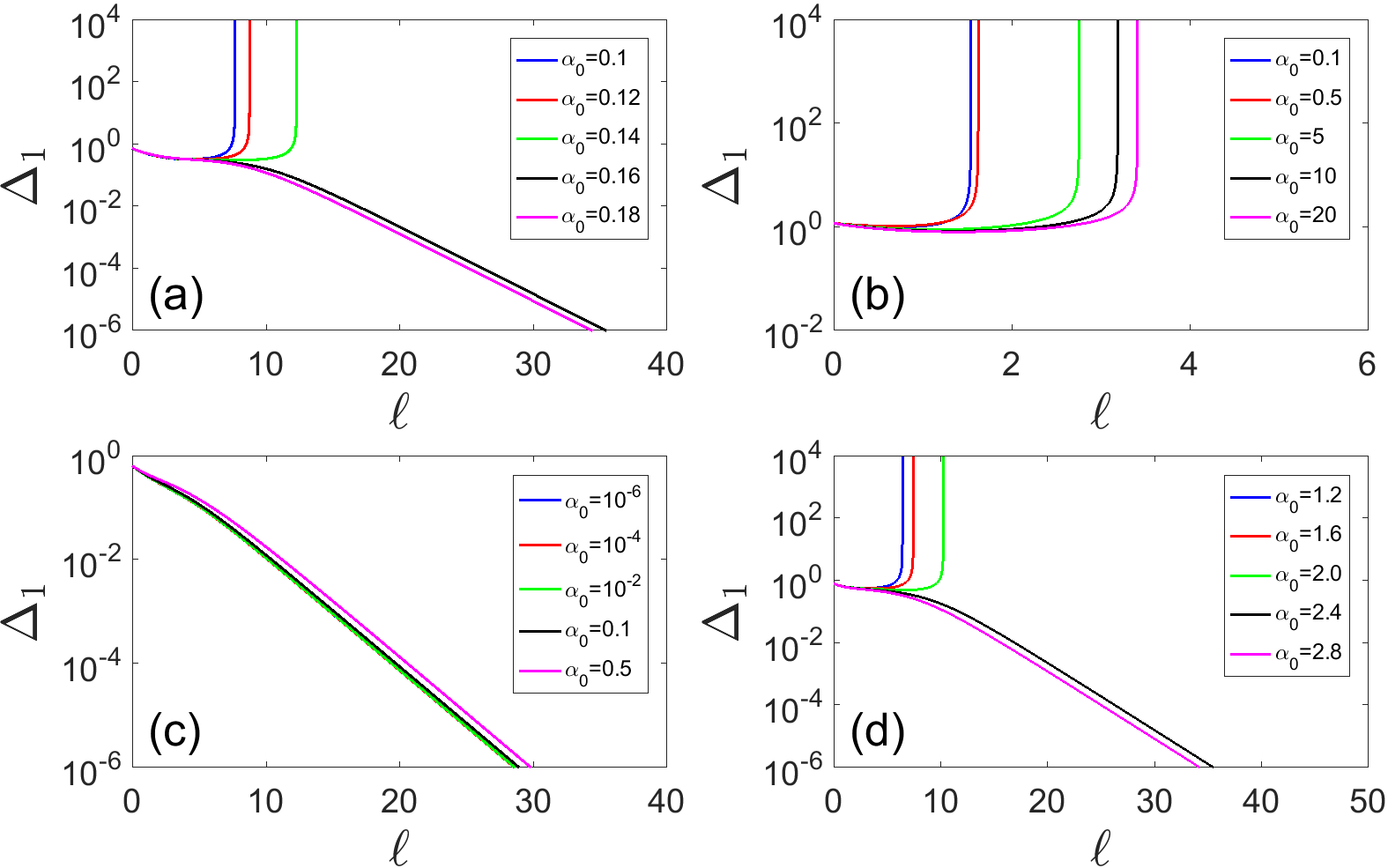}
\caption{ Flows of $\Delta_{1}$ considering initially both of $x$-RVP and Coulomb interaction. (a) $\Delta_{1,0}=0.75$, $\beta_{0}=0.1$;
(b) $\Delta_{1,0}=1.2$, $\beta_{0}=0.1$; (c) $\Delta_{1,0}=0.65$, $\beta_{0}=0.1$; (d) $\Delta_{1,0}=0.8$, $\beta_{0}=0.2$
\label{Fig:VRGCoulombRVP1}}
\end{figure}

For different parameter $\beta_{0}$, the phase diagrams on the plane of $\Delta_{0,0}$ and $\alpha_{0}$ are shown in
figure~\ref{Fig:PhaseDiagramRSPCoulomb}. For a given $\beta_{0}$, if $\Delta_{0,0}$ takes a small value, the system is always in
SM phase, if $\Delta_{0,0}$ takes a large enough value, the system is driven from CDM phase to SM phase with the increasing of $\alpha_{0}$.
Remarkably, the critical strength of RSP is changed obviously once Coulomb interaction is considered even if  $\alpha_{0}$ takes
arbitrarily small value. These characteristics are also qualitatively same as figure~\ref{Fig:PhaseDiagramRSPCoulombExpansion}, although there are
some quantitative differences.

The dynamical exponents $z_{1}$ and $z_{2}$, and the anomalous dimension of fermion field $\eta_{\psi}$
at the phase boundary between SM phase and CDM phase are shown in figure~\ref{Fig:ExponentZ} by
blue, red, and green lines respectively.

\begin{figure}[htbp]
\center
\includegraphics[width=3.2in]{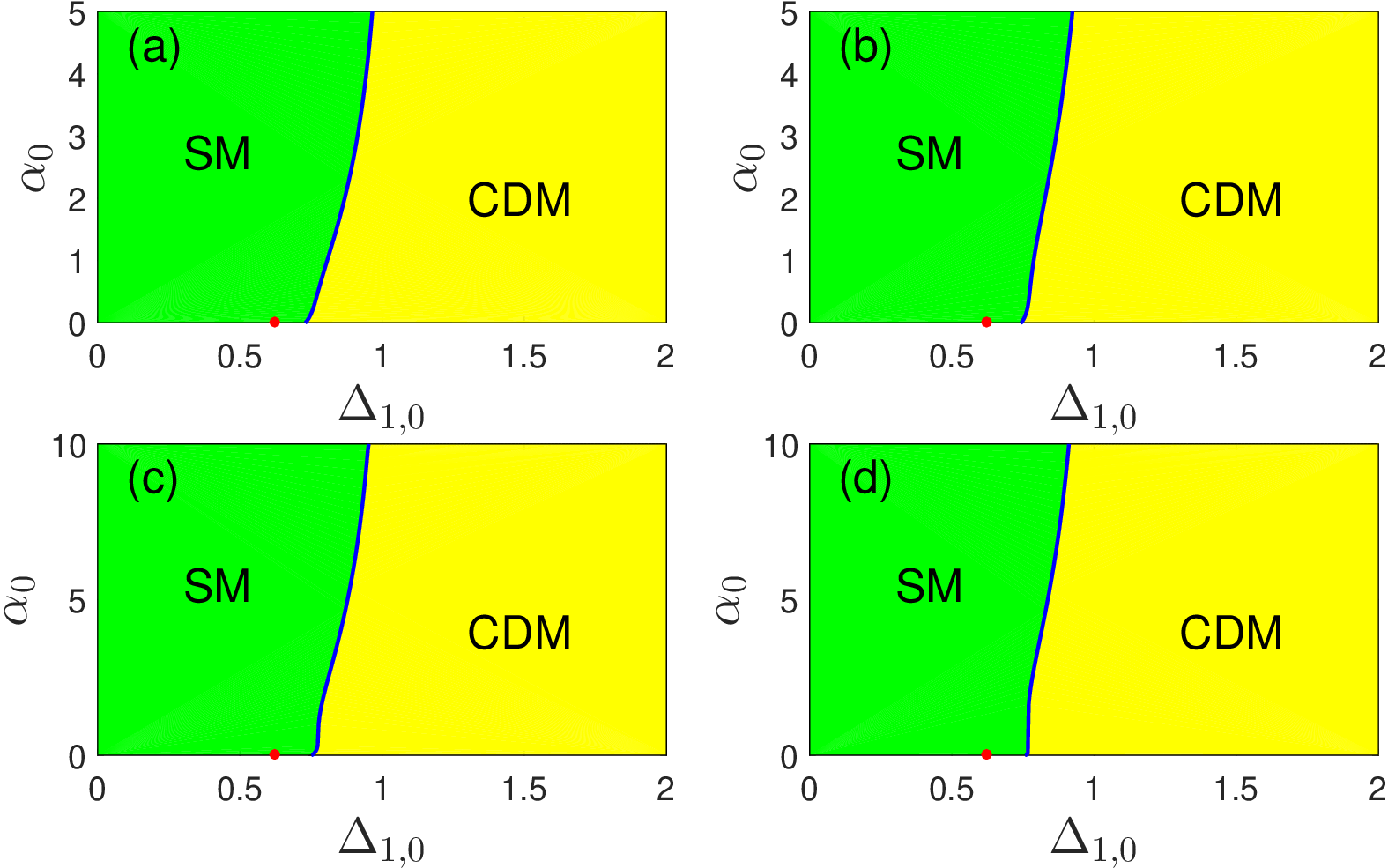}
\caption{ Phase diagrams of 3D AWSM considering initially both of $x$-RVP and Coulomb interaction.
$\beta_{0}=0.1, 0.15, 0.2, 0.25$ in (a), (b), (c), and (d) respectively. The red point represents the
critical value of $x$-RVP between SM and CDM phases neglecting the Coulomb interaction.
\label{Fig:PhaseDiagramRVP1Coulomb}}
\end{figure}

If both of long-rang Coulomb interaction and $x$-RVP are initially considered, the flows of $\Delta_{1}$ with
different initial conditions are presented in figure~\ref{Fig:VRGCoulombRVP1}. For a given $\beta_{0}$, there are three different cases.
In the first case, such as figure~\ref{Fig:VRGCoulombRVP1}(c), $\Delta_{1,0}$ takes relatively small value.
In this case, $\Delta_{1}$ always flows to zero if $\alpha_{0}$ takes arbitrarily finite value, which indicates
that the system is always in SM phase. In the second case, such as figure~\ref{Fig:VRGCoulombRVP1}(b) , $\Delta_{1,0}$ takes a
large enough value. Accordingly, $\Delta_{1}$ flows away even if the Coulomb strength takes quite large value, which represents that the system is driven to CDM phase. In the third case, as shown in figures~\ref{Fig:PhaseDiagramRVP1Coulomb}(a) and \ref{Fig:PhaseDiagramRVP1Coulomb}(d),
$\Delta_{1,0}$ takes an intermediate value. In this case, $\Delta_{1}$ flows away if $\alpha_{0}$ takes a small value,
but approaches to zero if $\alpha_{0}$ is larger than a critical value. It indicates that there is a QPT from CDM to SM phases with increasing of $\alpha_{0}$ for the third case.

\begin{figure}[htbp]
\center
\includegraphics[width=3.2in]{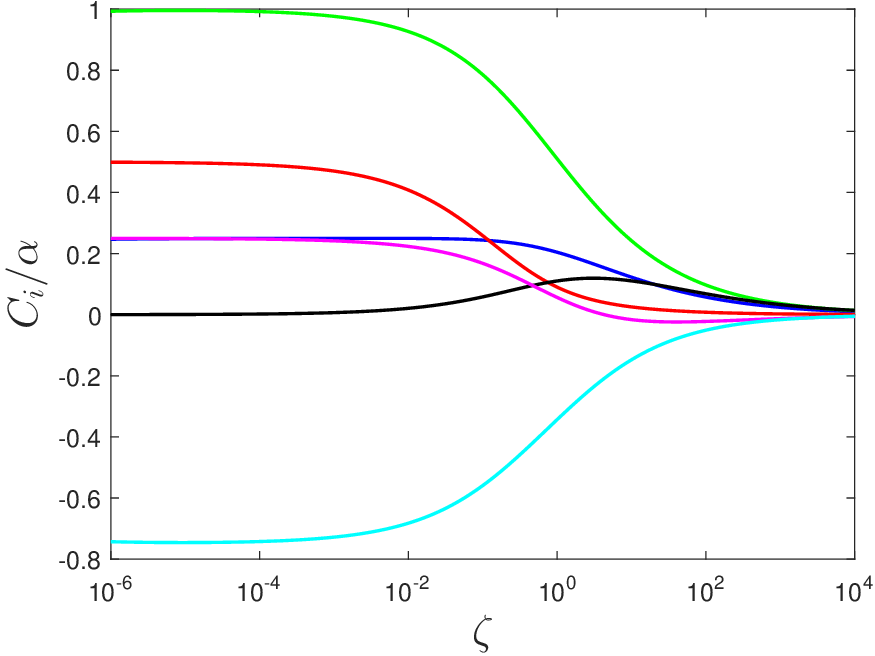}
\caption{The relation between $C_{i}/\alpha$ and $\zeta$. The blue, red, green, black, magenta, cyan lines are corresponding to
$C_{1}$, $C_{2}$, $C_{3}$, $C_{4}$, $C_{5}$, and $C_{6}$ respectively.  $C_{5}=-\left(2C_{1}+\frac{1}{2}C_{2}-C_{3}\right)$ and
$C_{6}=-\left(2C_{1}+\frac{1}{2}C_{2}-C_{4}\right)$.  \label{Fig:CiZeta}}
\end{figure}

The phase diagrams considering long-range Coulomb interaction and $x$-RVP initially with different values of $\beta_{0}$ are depicted
in figure~\ref{Fig:PhaseDiagramRVP1Coulomb}.
The red point in figure~\ref{Fig:PhaseDiagramRVP1Coulomb} represents the critical value $\Delta_{1,0}^{*}$ if only $x$-RVP is considered.
We can find that the critical strength of $x$-RVP considering infinitesimally weak Coulomb interaction also
has a finite difference with $\Delta_{1,0}^{*}$. However, for large enough $\Delta_{1,0}$, the
system seems always in CDM phase, and can not restore the SM phase by strong Coulomb interaction. For an intermediate range of
$\Delta_{1,0}$, the system is in CDM phase for weak $\alpha_{0}$, but restores the SM phase if $\alpha_{0}$ is large enough.

These behaviors are probably
due to subtle interplay of several effects. Firstly, the Feynman
diagram as shown in figure~\ref{Fig:VertexCorrection}(e) leads to the corrections
\begin{eqnarray}
\delta\Delta_{0}^{(5)}&=&-2\Delta_{0}\left(\sqrt{\eta}C_{\bot}
+\frac{C_{z}}{\sqrt{\eta}}\right)\ell,
\\
\delta\Delta_{1}^{(5)}&=&0,
\end{eqnarray}
which represent that figure~\ref{Fig:VertexCorrection}(e) induces the suppression effect for RSP, but does not result in correction for coupling of $x$-RVP and fermions.
The contribution from figure~\ref{Fig:VertexCorrection}(e) to RSP is the last term of equation~(\ref{Eq:deltaDelta0App}).
Secondly, the contributions from
Feynman diagram shown in figure~\ref{Fig:VertexCorrection}(d) to RSP and $x$-RVP are
\begin{eqnarray}
\delta\Delta_{0}^{(4)}&=&0,
\\
\delta\Delta_{1}^{(4)}&=&\Delta_{1}C_{3}\ell,
\end{eqnarray}
which indicate that figure~\ref{Fig:VertexCorrection}(d) does not lead to correction for RSP but enhances $x$-RVP.
The contribution from figure~\ref{Fig:VertexCorrection}(d) to $x$-RVP is the last term of equation~(\ref{Eq:deltaDelta1App}).
Thirdly, the fermion self-energy induces renormalization  of the parameters $v$ and $A$, which could result in
correction to $\Delta_{i}$ in the RG equations, due to that the effective disorder strength is determined by
$\frac{\Delta_{i}\sqrt{\Lambda}}{\pi^{2}v^{2}\sqrt{A}}$. It should be noticed that the replacement
$\frac{\Delta_{i}\sqrt{\Lambda}}{\pi^{2}v^{2}\sqrt{A}}\rightarrow\Delta_{i}$ has been employed
in the derivation for the RG equations. These three effects yield the term
\begin{eqnarray}
-\Delta_{0}\left(2C_{1}+\frac{1}{2}C_{2}+2\beta+2\gamma\right) \label{Eq:CoulombCorrectionRSP}
\end{eqnarray}
for the RG equation of $\Delta_{0}$ as shown in equation~(\ref{Eq:VRGDelta0}),
and the term
\begin{eqnarray}
-\Delta_{1}\left(2C_{1}+\frac{1}{2}C_{2}-C_{3}\right) \label{Eq:CoulombCorrectionRVPX}
\end{eqnarray}
for the RG equation of $\Delta_{1}$ given by equation~(\ref{Eq:VRGDelta1}). The term (\ref{Eq:CoulombCorrectionRSP}) is always negative.
Accordingly, these three effects result in that the Coulomb interaction suppresses RSP.
According to figure~\ref{Fig:CiZeta}, the term shown in equation~(\ref{Eq:CoulombCorrectionRVPX}) is positive in a wide range of $\zeta$.
It represents that the three effects mentioned above could enhance $x$-RVP in some conditions.
Fourthly, RSP and $x$-RVP  dynamically generate and enhance each other. The promotion
effect between RSP and $x$-RVP may be suppressed as the generation of RSP is prevented by long-range Coulomb interaction.
The complex behaviors considering initially both of long-range Coulomb interaction and $x$-RVP are due to the interplay of
the four effects aforementioned. The phase diagram including both of long-range Coulomb interaction and $y$-RVP has
similar characteristics.

\begin{figure}[htbp]
\center
\includegraphics[width=3.2in]{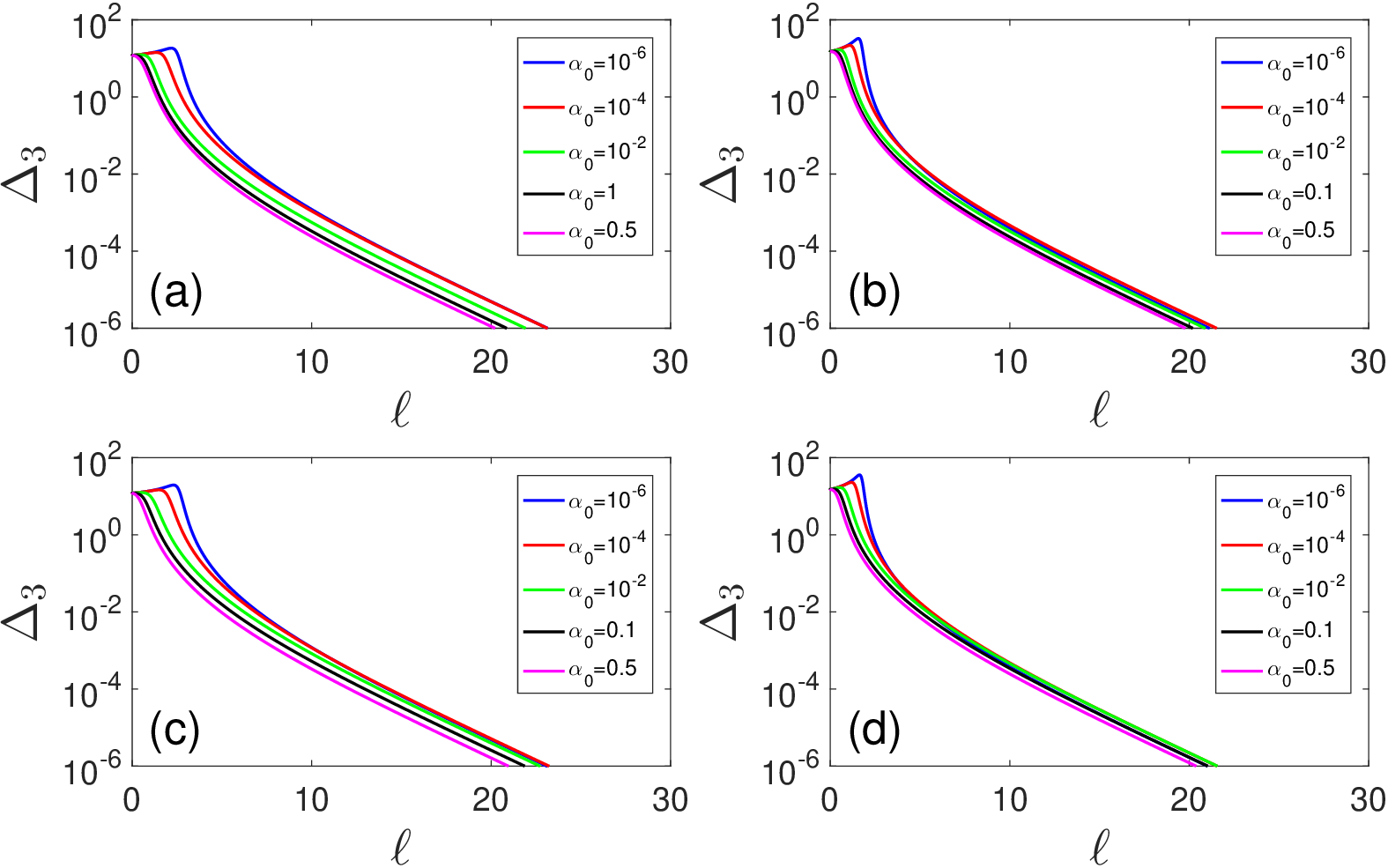}
\caption{ Flows of $\Delta_{3}$ considering initially $z$-RVP and Coulomb interaction. (a) $\Delta_{3,0}=12$, $\beta_{0}=0.1$;
(b) $\Delta_{3,0}=15$, $\beta_{0}=0.1$; (c) $\Delta_{3,0}=12$, $\beta_{0}=0.5$; (d) $\Delta_{3,0}=15$, $\beta_{0}=0.5$.
\label{Fig:VRGCoulombRVP3}}
\end{figure}

According to figure~\ref{Fig:CiZeta}, the term
\begin{eqnarray}
-\Delta_{3}\left(2C_{1}+\frac{1}{2}C_{2}-C_{4}\right) \label{Eq:CoulombCorrectionRVPZ}
\end{eqnarray}
is always negative. Thus, the long-range Coulomb interaction always tends to suppress $z$-RVP.
Considering initially both of Coulomb interaction and $z$-RVP, the flow of $\Delta_{3}$ is presented in
figure~\ref{Fig:VRGCoulombRVP3}. We find that for $\Delta_{3,0}>\Delta_{3,0}^{*}$, $\Delta_{3}$ grows with lowering of the
energy scale at first, but begins to decrease if the running parameter $\ell$ is large enough, and always approaches to zero in the
lowest energy limit. Thus, the system is always is in SM phase if both of Coulomb interaction and $z$-RVP are
considered. The remarkable suppression effect of Coulomb interaction for $z$-RVP should result from the special
energy dispersion of 3D anisotropic Weyl fermions.

\section{Observable Quantities \label{Sec:ObservableQuantities}}

In this section, in order to better understand the physical properties of 3D AWSM in the presence of disorder,
we compare the behaviors of observable quantities
including DOS, specific heat, and compressibility in SM phase, CDM phase, and at the phase boundary.

\subsection{DOS}

In SM phase, the retarded fermion propagator takes the form as
\begin{eqnarray}
G_{0}^{\mathrm{ret}}(\omega,\mathbf{k})=\frac{1}{\omega-\left(vk_{x}\sigma_{1}+vk_{y}\sigma_{2}
+Ak_{z}^{2}\sigma_{3}\right)+i\delta}.
\end{eqnarray}
The spectral function is
\begin{eqnarray}
\mathcal{A}(\omega,\mathbf{k})
&=&-\frac{1}{\pi}\mathrm{Tr}\left[\mathrm{Im}\left[G_{0}^{\mathrm{ret}}
(\omega,\mathbf{k})\right]\right]\nonumber
\\
&=&\frac{|\omega|}{E_{\mathbf{k}}}\left[\delta\left(\omega+E_{\mathbf{k}}\right)
+\delta\left(\omega-E_{\mathbf{k}}\right)\right],
\end{eqnarray}
where $E_{\mathbf{k}}=\sqrt{v^{2}k_{\bot}^{2} +A^{2}k_{z}^{4}}$. The
DOS is given by
\begin{eqnarray}
\rho(\omega)
&=&\int\frac{d^3\mathbf{k}}{(2\pi)^{3}}\mathcal{A}(\omega,\mathbf{k})
=\frac{|\omega|^{3/2}}{2\pi^{2}v^{2}\sqrt{A}},
\end{eqnarray}
which vanishes in the limit $\omega\rightarrow0$.

 \begin{figure}[htbp]
\center
\includegraphics[width=3.2in]{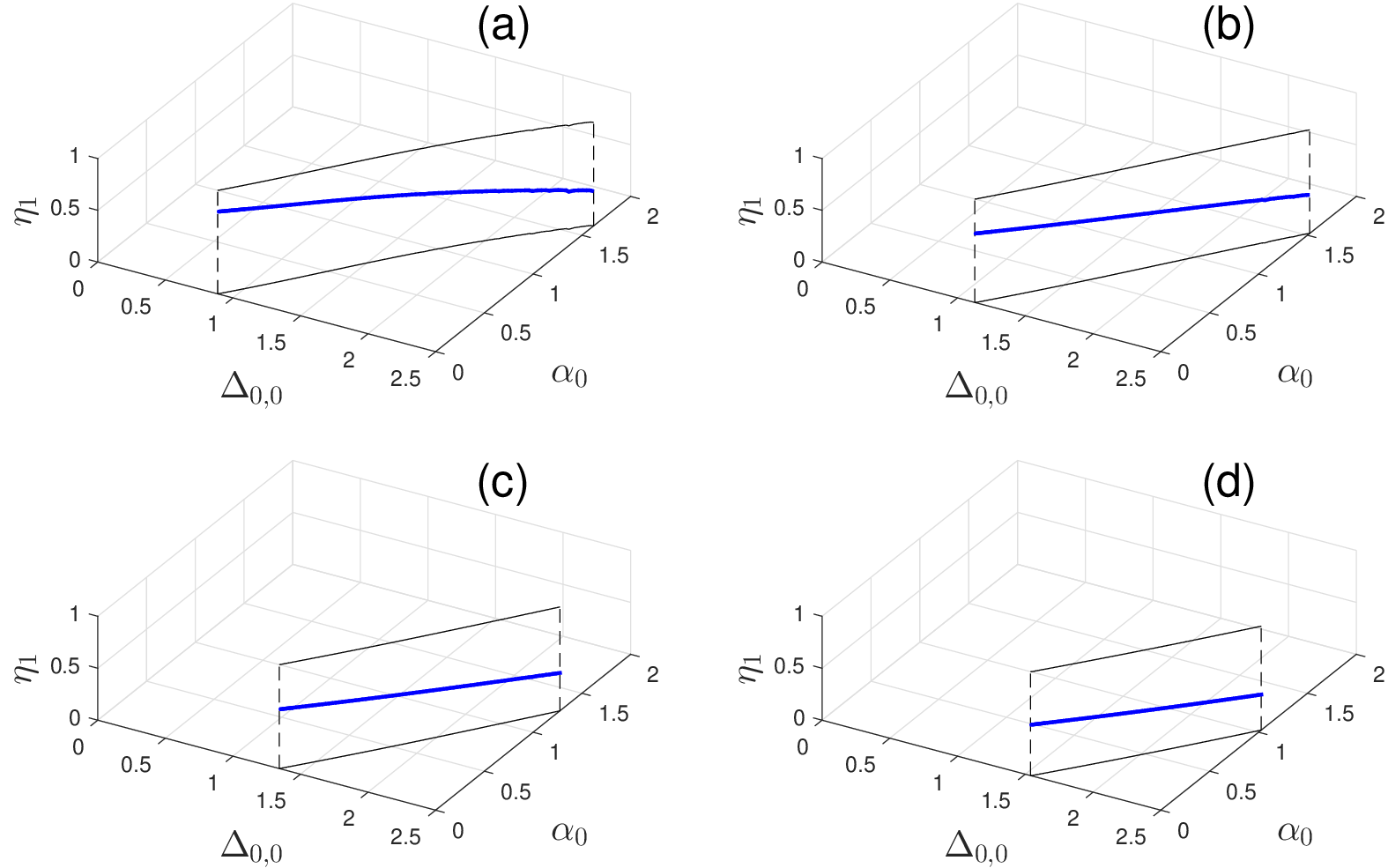}
\caption{ The parameter $\eta_{1}$ at the phase boundary between SM and CDM phases. $\beta_{0}=0.1$, $0.2$, $0.3$ and $0.4$ in (a), (b), (c), and (d) respectively.
Subleading terms in the sense of $1/n$ expansion induced by the  disorder coupling are discarded.
\label{Fig:Eta1ZExpansion}}
\end{figure}

 \begin{figure}[htbp]
\center
\includegraphics[width=3.2in]{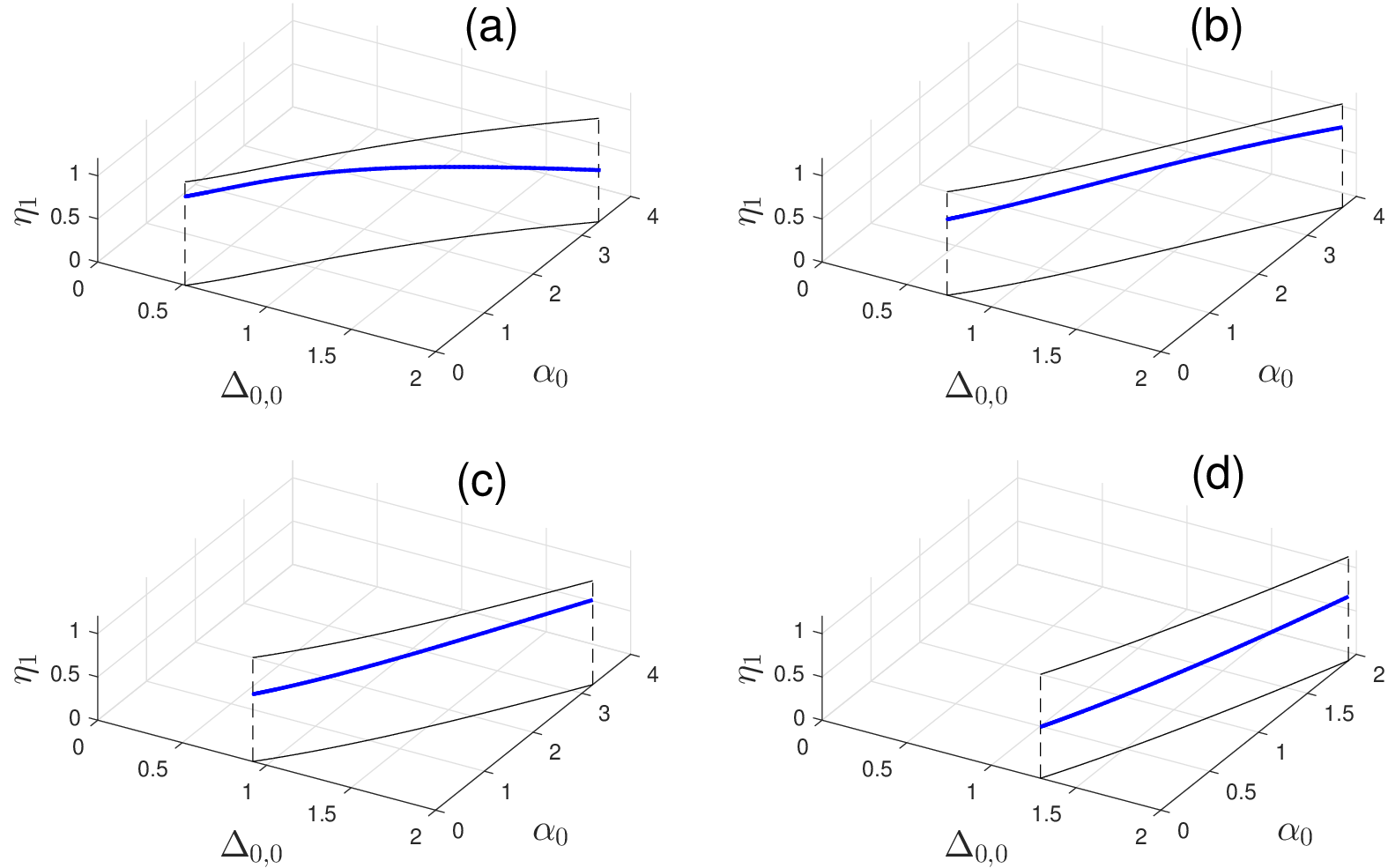}
\caption{ The parameter $\eta_{1}$ at the phase boundary between SM and CDM phases. $\beta_{0}=0.1$, $0.3$, $0.5$ and $1$ in (a), (b), (c), and (d) respectively.
Subleading terms in the sense of $1/n$ expansion induced by the disorder coupling are included.
\label{Fig:Eta1}}
\end{figure}

In CDM phase, the fermions acquire a finite disorder scattering rate
$\gamma_{0}$. Accordingly, the retarded fermion propagator becomes
\begin{eqnarray}
G^{\mathrm{ret}}(\omega,\mathbf{k})=\frac{1}{\omega+i\gamma_{0}-\left(vk_{x}\sigma_{1}+vk_{y}\sigma_{2}
+Ak_{z}^{2}\sigma_{3}\right)}.
\end{eqnarray}
The spectral function can be written as
\begin{eqnarray}
\mathcal{A}(\omega,\mathbf{k})
&=&-\frac{1}{\pi}\mathrm{Tr}\left[\mathrm{Im}\left[G^{\mathrm{ret}}(\omega,\mathbf{k})\right]\right]\nonumber
\\
&=&\frac{2}{\pi}\frac{\gamma_{0}\left(\omega^2+\gamma_{0}^2+E_{\mathbf{k}}^{2}\right)}
{\left(\omega^{2}-\gamma_{0}^{2}-E_{\mathbf{k}}^{2}\right)^{2}+4\omega^{2}\gamma_{0}^{2}}.
\end{eqnarray}
The corresponding DOS can be obtained via
\begin{eqnarray}
\rho(\omega)&=&\int\frac{d^3\mathbf{k}}{(2\pi)^{3}}\mathcal{A}(\omega,\mathbf{k}).
\end{eqnarray}
In the limit $\omega\ll\gamma_{0}\ll\Lambda$, DOS is approximated as
\begin{eqnarray}
\rho(\omega)
&\approx&\frac{4\gamma_{0}\sqrt{\Lambda}}{\pi^{3}v^{2}\sqrt{A}}.
\end{eqnarray}
It is clear that $\rho(0)$ takes a finite value in CDM phase.

At the phase boundary between SM phase and CDM phase, the DOS satisfies
\begin{eqnarray}
\rho(\omega)\sim\omega^{\frac{2}{z_{1}}+\frac{1}{z_{2}}-1}\sim\omega^{\eta_{1}},
\end{eqnarray}
where
\begin{eqnarray}
\eta_{1}=\frac{2}{z_{1}}+\frac{1}{z_{2}}-1.
\end{eqnarray}
The values of $\eta_{1}$ at the phase boundary discarding subleading terms in the sense of $1/n$ expansion
induced by the disorder coupling are shown in figure~\ref{Fig:Eta1ZExpansion}. The values of $\eta_{1}$ at the phase boundary including subleading
terms are presented in figure~\ref{Fig:Eta1}.

\subsection{Specific heat}

In SM phase, the fermion propagator in Matsubara formalism can be
expressed as
\begin{eqnarray}
G_{0}(\omega_{n},\mathbf{k})
&=&-\frac{i\omega_{n}+vk_{x}\sigma_{1}+vk_{y}\sigma_{2}
+Ak_{z}^{2}\sigma_{3}}{\omega_{n}^{2}+E_{\mathbf{k}}^{2}},
\end{eqnarray}
where $\omega_{n}=(2n+1)\pi T$ with $n$ being integers. The free
energy of fermions is
\begin{eqnarray}
F_{f}(T)
&=&-2T\sum_{\omega_{n}}\int\frac{d^3\mathbf{k}}{(2\pi)^3}
\ln\left[\left(\omega_{n}^2+E_{\mathbf{k}}^{2}\right)^{1/2}\right].
\end{eqnarray}
Carrying out the frequency summation, one gets
\begin{eqnarray}
F_{f}(T) =-2\int\frac{d^3\mathbf{k}}{(2\pi)^3}
\left[E_{\mathbf{k}}+2T\ln\left(1+e^{-\frac{E_{\mathbf{k}}}{T}}\right)\right],
\end{eqnarray}
which is divergent due to the first term in the brackets.  In order to get a finite free energy, we
redefine $F_{f}(T)-F_{f}(0)$ as $F_{f}(T)$, and obtain
\begin{eqnarray}
F_{f}(T) &=&-4T\int\frac{d^3\mathbf{k}}{(2\pi)^3}
\ln\left(1+e^{-\frac{E_{\mathbf{k}}}{T}}\right)\nonumber
\\
&=&-\frac{3(8-\sqrt{2})\zeta\left(\frac{7}{2}\right)}
{16\pi^{3/2}v^{2}\sqrt{A}}T^{7/2},
\end{eqnarray}
where $\zeta(x)$ is Riemann zeta function. Using the formula
\begin{eqnarray}
C_{v}(T)=-T\frac{\partial^2 F_{f}(T)}{\partial T^2},
\end{eqnarray}
$C_{v}$ can be written as
\begin{eqnarray}
C_{v}(T)=\frac{105(8-\sqrt{2})
\zeta\left(\frac{7}{2}\right)}{64\pi^{3/2}v^{2}\sqrt{A}}T^{5/2}\propto T^{5/2}.\label{Eq:CvSM}
\end{eqnarray}

In CDM phase, the fermion propagator in Matsubara formalism has the form as
\begin{eqnarray}
G(\omega_{n},\mathbf{k})
&=&-\frac{i\omega_{n}'+vk_{x}\sigma_{1}+vk_{y}\sigma_{2}
+Ak_{z}^{2}\sigma_{3}}{\omega_{n}'^{2}+E_{\mathbf{k}}^{2}},
\end{eqnarray}
where $\omega_{n}'=\omega_{n}+\gamma_{0}\mathrm{sgn}(\omega_{n})$.
The free energy of the fermions is given by
\begin{eqnarray}
F_{f}(T)
&=&-2T\sum_{\omega_{n}}\int\frac{d^3\mathbf{k}}{(2\pi)^3}
\ln\left[\left(\omega_{n}'^2+E_{\mathbf{k}}^{2}\right)^{1/2}\right].
\end{eqnarray}
In the low temperature regime, $F_{f}(T)$ can be approximated as
\begin{eqnarray}
F_{f}(T)
&\approx&-\frac{4}{\pi}\int\frac{d^3\mathbf{k}}{(2\pi)^3}\bigg\{E_{\mathbf{k}}\arctan\left(\frac{E_{\mathbf{k}}}{\pi T+\gamma_{0}}\right)\nonumber
\\
&&+\frac{\gamma_{0}}{2}\ln\left[\left(\pi T+\gamma_{0}\right)^{2}+E_{\mathbf{k}}^{2}\right]\bigg\}.
\end{eqnarray}
In the condition $T\ll\gamma_{0}\ll\Lambda$, we get
\begin{eqnarray}
C_{v}(T)=-T\frac{\partial^2 F_{f}(T)}{\partial T^2}
\approx\frac{4\gamma_{0}\sqrt{\Lambda}}{\pi v^{2}\sqrt{A}} T\propto T,  \label{Eq:CvCDM}
\end{eqnarray}
which is obviously different from equation~(\ref{Eq:CvSM}).

The specific heat at the phase boundary between SM phase and CDM phase takes the form
\begin{eqnarray}
C_{v}(T)\sim T^{\frac{2}{z_{1}}+\frac{1}{z_{2}}}\sim T^{\eta_{2}},
\end{eqnarray}
where
\begin{eqnarray}
\eta_{2}=\frac{2}{z_{1}}+\frac{1}{z_{2}}=\eta_{1}+1.
\end{eqnarray}

\subsection{Compressibility}

In order to calculate the compressibility, we introduce the chemical potential $\mu$ at the beginning.
Accordingly, the fermion propagator reads as
\begin{eqnarray}
G_{0}(\omega_{n},\mathbf{k})
&=&-\frac{i\omega_{n}+\mu+vk_{x}\sigma_{1}+vk_{y}\sigma_{2}
+Ak_{z}^{2}\sigma_{3}}{\left(\omega_{n}-i\mu\right)^{2}+E_{\mathbf{k}}^{2}}.\nonumber
\\
\end{eqnarray}
The free energy of fermions has the form
\begin{eqnarray}
F_{f}(T,\mu)
&=&-2T\sum_{\omega_{n}}\int\frac{d^3\mathbf{k}}{(2\pi)^3}
\ln\Big\{\big[\left(\omega_{n}-i\mu\right)^2\nonumber
\\
&&+E_{\mathbf{k}}^{2}\big]^{1/2}\Big\}.
\end{eqnarray}
Performing the summation of frequency and integration of momenta, we obtain
\begin{eqnarray}
F_{f}(T,\mu)
&=&\frac{3}{4\pi^{3/2}v^{2}\sqrt{A}}T^{7/2}
\left[\mathrm{Li}_{\frac{7}{2}}\left(-e^{\frac{\mu}{T}}\right)\right.\nonumber
\\
&&\left.+\mathrm{Li}_{\frac{7}{2}}\left(-e^{-\frac{\mu}{T}}\right)\right],
\end{eqnarray}
where $\mathrm{Li}_{x}(y)$ is the polylogarithm function.
Utilizing the formula
\begin{equation}
\kappa(T,\mu)=-\frac{\partial^2 F_{f}(T,\mu)}{\partial \mu^2},
\end{equation}
we arrive at
\begin{eqnarray}
\kappa(T,\mu)
&=&-\frac{3}{4\pi^{3/2}v^{2}\sqrt{A}}T^{3/2}
\left[\mathrm{Li}_{\frac{3}{2}}\left(-e^{\frac{\mu}{T}}\right)\right.\nonumber
\\
&&\left.+\mathrm{Li}_{\frac{3}{2}}\left(-e^{-\frac{\mu}{T}}\right)\right].
\end{eqnarray}
Taking $\mu=0$, one can get the compressibility in the SM phase as following
\begin{eqnarray}
\kappa(T)
&=&\frac{3\left(\sqrt{2}-1\right)\zeta\left(\frac{3}{2}\right)}
{2\sqrt{2}\pi^{3/2}v^{2}\sqrt{A}}T^{3/2}.
\end{eqnarray}

For calculating the compressibility in CDM phase, we incorporate a
finite disorder scattering rate $\gamma_{0}$ into the fermion propagator
\begin{eqnarray}
G_{0}(\omega_{n},\mathbf{k})
&=&-\frac{i\omega_{n}'+\mu+vk_{x}\sigma_{1}+vk_{y}\sigma_{2}
+Ak_{z}^{2}\sigma_{3}}{\left(\omega_{n}'-i\mu\right)^{2}+E_{\mathbf{k}}^{2}},\nonumber
\\
\end{eqnarray}
where $\omega_{n}'=\omega_{n}+\gamma_{0}\mathrm{sgn}(\omega_{n})$.
The corresponding free energy is expressed as
\begin{eqnarray}
F_{f}(T,\mu)
&=&-2NT\sum_{\omega_{n}}\int\frac{d^3\mathbf{k}}{(2\pi)^3}
\ln\left\{\left[\left(\omega_{n}'-i\mu\right)^2\right.\right.\nonumber
\\
&&\left.\left.+E_{\mathbf{k}}^{2}\right]^{1/2}\right\}.
\end{eqnarray}
Carrying out the frequency summation, $F_{f}(T,\mu)$ becomes
\begin{eqnarray}
F_{f}(T,\mu)
&=&-\frac{2}{\pi}\sum_{\xi=\pm1}\int\frac{d^3\mathbf{k}}{(2\pi)^3}
\left\{E_{\mathbf{k}}\arctan\left(\frac{E_{\mathbf{k}}+\xi\mu}{\pi T+\gamma_{0}}\right)\right.\nonumber
\\
&&-\frac{\gamma_{0}}{2}\ln\left[\left(\pi T+\gamma_{0}\right)^{2}+\left(E_{\mathbf{k}}+\xi\mu\right)^{2}\right]
\nonumber
\\
&&\left.-\mu\arctan\left(\frac{E_{\mathbf{k}}+\xi\mu}{\pi T+\gamma_{0}}\right)\right\}.
\end{eqnarray}
The compressibility satisfies
\begin{eqnarray}
\kappa(T,\mu)&=&-\frac{\partial^2 F_{f}(T,\mu)}{\partial \mu^2}\nonumber
\\
&=&\frac{2}{\pi}\sum_{\xi=\pm1}\int\frac{d^3\mathbf{k}}{(2\pi)^3}\nonumber
\\
&&\times\Bigg\{\frac{2\pi T\left(\pi T+\gamma_{0}\right)^{2}}{\left[\left(\pi T+\gamma_{0}\right)^{2}+\left(E_{\mathbf{k}}+\xi\mu\right)^{2}\right]^{2}}\nonumber
\\
&&+\frac{\gamma_{0}}{\left(\pi T+\gamma_{0}\right)^{2}+\left(E_{\mathbf{k}}+\xi\mu\right)^{2}}\Bigg\}.
\end{eqnarray}
Taking $\mu=0$, in the case $T\ll\gamma_{0}\ll\Lambda$, $\kappa$ reduces to
\begin{eqnarray}
\kappa(T)
&\approx&\frac{4\gamma_{0}\sqrt{\Lambda}}{\pi^{3}v^{2}\sqrt{A}}.
\end{eqnarray}

At the  phase boundary between SM phase and CDM phase, the compressibility reads as
\begin{eqnarray}
\kappa(T)\sim T^{\frac{2}{z_{1}}+\frac{1}{z_{2}}-1}\sim T^{\eta_{1}}.
\end{eqnarray}

\section{Physical implications\label{Sec:PhysicalImplications}}

In this section, we discuss the potential implications of the theoretical results in the candidate physical systems for 3D AWSM and
other related materials.

According to the study by Yang \emph{et al.} \cite{Yang14B}, 3D AWSM might be obtained at the topological QCP between normal band insulator and WSM,
or at the QCP between  normal band insulator and topological insulator in 3D noncentrosymmetric system. The theoretical
studies predicted that 3D AWSM  can be reached at the QCP between normal band insulator and topological insulator
through tuning pressure on BiTeI, in which the inversion symmetry is broken \cite{Yang13, Bahramy12}.
The subsequent experimental measurements for pressured BiTeI through x-ray powder diffraction and infrared spectroscopy
are consistent with the theoretical prediction \cite{Xi13}. The measurements of  Shubnikov-de Haas (SdH) quantum oscillations also
reveled the existence of pressure-induced topological QPT in BiTeI \cite{Park15}. The theoretical results shown in section~\ref{Sec:Results}
would be helpful for understanding the low-energy behaviors of candidate materials  for 3D AWSM.

In 3D anisotropic DSM (ADSM), the dispersion of fermion excitations is also linear along two directions and quadratic
along the third one \cite{Yang14B}. Yang \emph{et al.} showed that 3D ADSM can be obtained at the QCP between normal band insulator
and topological 3D DSM, or at the QCP between 3D DSM and weak topological insulator or topological crystalline insulator \cite{Yang14B}.
The analysis of Yuan \emph{et al.} exhibited that 3D ADSM state is possible to be realized in ZrTe$_{5}$ at the QCP between insulating and 3D DSM
phases \cite{Yuan17}. The experimental studies on pressured ZrTe$_{5}$ through SdH  quantum oscillations showed the evidence of 3D ADSM state \cite{ZhangJingLei17}. Recently,
the experimental studies based on SdH quantum oscillations and high pressure x-ray diffraction unveiled that there is a QCP
from 3D DSM to band insulator phases in Cd$_{3}$As$_{2}$ with increasing of pressure \cite{ZhangCengXiuFaXian17}. The
theoretical results shown above should also hold on in 3D ADSM, and are valuable for understanding the physical properties
 of candidate
materials of 3D ADSM.

Various unconventional fermions, including 2D Dirac fermions \cite{Ochiai09, Bittner10}, 3D Weyl fermions \cite{LuLing14, LuLing15}, 3D double- and triple-Weyl fermions
\cite{ChenCTChanGroup16, WangQiang17}, 3D nodal line fermions \cite{YanLuLi18} \emph{etc.} have been  realized in
photonic crystal. 3D anisotropic Weyl fermions could be also obtained through properly designing the photonic crystal. In photonic crystal, the disorder can be introduced and controlled by speckled beam
\cite{Schwartz17, Billy08, Roati08}. The fermions in  photonic crystal are not influenced by Coulomb interaction. In contrast, in SM materials, the
long-range Coulomb interaction is intrinsic. Therefore, the phase diagrams of 3D anisotropic Weyl fermions under the influence of disorder
in photonic crystal and SM materials could take obvious differences, which may be verified experimentally in future.

The influence of Coulomb interaction in AWSM depends on two parameters $\alpha$ and $\beta$. The parameters
$\alpha$ and $\beta$ are closely related to $\epsilon$, $v$, and $A$, which are basic parameters of the system and can be determined experimentally.
Thus, changing of the parameters $\epsilon$, $v$, and $A$ would modify the influence of Coulomb interaction.
These three parameters may be tuned by pressure, strain \emph{etc}. in a proper way.

The results for AWSM shown in former sections are obtained at zero chemical potential. In ideal SMs, the chemical potential $\mu=0$. Then the Fermi level is exactly at the touching points, and the DOS exactly  equals to zero. However, in real samples, the chemical usually does not equal to zero exactly but takes a small finite value. Accordingly, the Fermi level
is not at the touching points, and DOS takes a finite value. For the case chemical potential $\mu=0$, in AWSM, there is a QCP from SM phase to CDM phase at zero temperature with increasing
of disorder strength. At finite temperatures, the QCP becomes a quantum critical region. For the case  $\mu$ is finite, the QCP from SM phase to CDM  phase at zero temperature is avoided. However, the quantum critical region in the energy scale $T>\mu$ still exists. Thus, we believe that the results shown in
former sections could be observed in the energy scale $T>\mu$.

\section{Comparison with other SMs \label{Sec:Comparison}}

In this section, we compare with previous studies about the interplay of disorder and Coulomb interaction in
other SMs.

For 2D DSM, there are usually  three kinds of disorder including RSP, RVP, and random mass (RM) \cite{Evers08, Ostrovsky06, Foster12,
WangLiuZhang16, WangJing17}. Considering only RSP, we can find that the effective strength $\Delta_{0}$ always approaches to infinity
at a finite running parameter $\ell_{c}$ which is determined by the initial value $\Delta_{0,0}$ \cite{Evers08, Ostrovsky06, Foster12}.
It represents that 2D DSM always becomes unstable to CDM phase under RSP. If only RVP is considered,
the effective strength $\Delta_{V}$ does not flow but is fixed to the initial value $\Delta_{V,0}$ \cite{Evers08, Ostrovsky06, Foster12,
WangLiuZhang16, WangJing17}. The fermion velocity approaches to
zero in the lowest energy limit $\ell\rightarrow\infty$. Accordingly, the observable quantities including DOS, specific heat and compressibility
are enhanced by power-law corrections of energy or temperature\cite{WangLiuZhang16, WangJing17}.
If only RM is considered, the effective strength $\Delta_{M}$ flows to zero in the lowest energy limit but with a slow speed
\cite{Evers08, Ostrovsky06, Foster12, WangLiuZhang16, WangJing17}. The fermion velocity approaches to zero slowly. Accordingly, the
observable quantities including DOS, specific heat and compressibility are enhanced by logarithmic-like corrections of energy or temperature
\cite{WangLiuZhang16, WangJing17}.
It is well known that if only Coulomb interaction is considered in 2D DSM, the effective Coulomb strength $\alpha$ approaches
to zero slowly and the fermion velocity increases logarithmically with lowering of energy scale. Thus, the observable quantities receive
logarithmic-like corrections of energy or temperature.

Interplay of disorder and Coulomb interaction in 2D DSM is closely related to the kind of disorder \cite{Stauber05, Foster08, Herbut08,Vafek08, WangLiu14}.
It was shown that RSP is suppressed by Coulomb interaction. For a given initial strength of RSP $\Delta_{0,0}$,
if the initial value of Coulomb interaction $\alpha_{0}$ is small, the system is still in the CDM phase. If $\alpha_{0}$
is larger than a critical value determined by $\Delta_{0,0}$, the system restores  SM phase \cite{Stauber05, WangLiu14 }.
Under the influence of RVP and Coulomb interaction, it was unveiled that the disorder strength $\Delta_{V}$ is fixed to $\Delta_{V,0}$
and the Coulomb strength flows to a constant \cite{Stauber05, Herbut08, WangLiu14}. Additionally, the fermion velocity flows to a constant value in the lowest energy limit.
Thus, the specific heat and compressibility qualitatively take the same behaviors as the clean and free 2D Dirac fermion system \cite{WangLiu14}. Comparing with the
cases considering only RVP or Coulomb interaction, one could notice that RVP promotes Coulomb interaction.
Considering both of
RM and Coulomb interaction, we can find that the disorder strength $\Delta_{M}$ and Coulomb strength $\alpha$ both flow to constants \cite{Stauber05, WangLiu14}.
Comparing with the cases only considering RM or Coulomb interaction, we can find that RM and Coulomb interaction promote each other.

For 3D DSM, weak RSP is irrelevant, but becomes relevant if the initial strength $\Delta_{0,0}$ is larger than a critical value $\Delta_{0,0}^{c}$ \cite{Goswami11, Roy14}.
It indicates that the system is in SM phase if $\Delta_{0,0}<\Delta_{0,0}^{c}$, but becomes CDM phase if $\Delta_{0,0}>\Delta_{0,0}^{c}$.
Goswami and  Chakravarty studied the interplay of RSP and Coulomb interaction in 3D DSM \cite{Goswami11}. They showed that RSP is suppressed by the Coulomb interaction, and
the critical value $\Delta_{0,0}^{c}$ becomes larger with the increasing of Coulomb interaction. Thus, the CDM phase can be tuned to SM phase by
increasing of Coulomb interaction.

Nandkishore and Parameswaran studied the interplay of disorder and Coulomb interaction in Luttinger SM \cite{Nandkishore17}.
For Luttinger SM, arbitrarily weak disorder no matter RSP or RVP drives the system to CDM phase. Considering both of disorder and Coulomb interaction,
it was shown that disorder always dominates Coulomb interaction, and the system is still in CDM Phase.

The interplay of disorder and Coulomb interaction in NLSM was analyzed by Wang and Nandkishore \cite{WangYuXuan17}. It is shown that arbitrarily weak disorder drives
the system to CDM phase. Additionally, the SM phase can not be restored by including of the Coulomb interaction, since disorder always dominates
Coulomb interaction.

The interplay of disorder and Coulomb interaction in multi-WSMs is studied in reference~\cite{WangLiuZhang17B}.
For double-WSM, considering both of  disorder and Coulomb interaction, the system is always in CDM phase.
However, the interplay of disorder and Coulomb interaction in triple-WSM is closely related to the kind of disorder.
Arbitrarily weak RSP drives triple-WSM to CDM phase. Considering both of RSP and Coulomb interaction, we can find that
RSP is suppressed by Coulomb interaction and the system restores SM phase  if the Coulomb interaction is strong enough.
Considering $x$-RVP or $y$-RVP, the disorder strength flows to a constant. Considering both of $x$-RVP or $y$-RVP and Coulomb interaction,
the disorder strength still flows to a constant but with a larger value, and the Coulomb interaction also flows to a constant. Considering both
of $z$-RVP and Coulomb interaction, disorder strength and Coulomb strength both approach to infinity, and Coulomb interaction dominates disorder
asymptotically. It may be corresponding to Mott insulating phase.

We can find that suppression of RSP by Coulomb interaction also exhibits in 2D DSM, 3D DSM, and triple-WSM. It is qualitatively similar
to the one in AWSM. Whereas, suppression of RSP by Coulomb interaction is more obvious in AWSM, since the SM phase can be restored
by Coulomb interaction even if $\alpha_{0}$ takes arbitrarily small value, if $\beta_{0}$ is finite. $\beta$ is related to the anisotropic
screening effect of Coulomb interaction. The anisotropic screening effect of Coulomb interaction does not exist in 2D DSM and 3D DSM. In triple-WSM,
the anisotropic screening effect of Coulomb interaction also exist, but weak RSP is relevant. Therefore, more obvious suppression of RSP by Coulomb interaction
in AWSM should be due to that there is anisotropic screening effect and weak RSP is irrelevant in AWSM. Remarkable suppression effect for $z$-RVP
by Coulomb interaction is not found in other SMs. Under the interplay of disorder and Coulomb interaction, the different behaviors of AWSM  comparing with other SMs
are closely related to the special fermion dispersion of AWSM.

\section{Summary\label{Sec:Summary}}

In summary, the low-energy behaviors of 3D AWSM under the influence of long-range Coulomb interaction
and disorder are studied by RG theory. The system could be in the SM phase, CDM phase, or at the phase boundary, depending on the initial values of strength of Coulomb interaction and disorder. We find a quite
novel result: The critical disorder strength for driving the CDM phase can be remarkably increased in some conditions,
even if the Coulomb strength takes arbitrarily small value, once the interplay of Coulomb and disorder is considered.
This novel behavior is closely related to the
anisotropic screening effect of Coulomb interaction, and essentially results from the particular dispersion of the fermions in
3D AWSM.

\ack{
J.R.W. is grateful to Prof. G.-Z. Liu for the helpful discussions. We
would like to acknowledge the support by the National Key
$\mathrm{R\&D}$ Program of China under Grants 2017YFA0403600 and 2016YFA0300404,
and the support by the National Natural Science
Foundation of China under Grants 11504379, 11604231, 11674327, 11974356,
U1532267, and U1832209. G.W. is also supported by the Natural Science Foundation of
Jiangsu Province under Grant No.~BK20160303, and by the Natural
Science Foundation of the Jiangsu Higher Education Institutions of
China under Grant No.~16KJB140012. J.R.W. is also supported by the
Natural Science Foundation of Anhui Province under Grant 1608085MA19.
A portion of this work
was supported by the High Magnetic Field Laboratory of Anhui Province.}

\appendix

\section{Propagators \label{App:Propagators}}

The propagator of 3D anisotropic Weyl fermions reads as
\begin{eqnarray}
G_{0}(\omega,\mathbf{k})=\frac{1}{i\omega-\left(vk_{x}
\sigma_{1}+vk_{y}\sigma_{2}+Ak_{z}^{2}\sigma_{3}\right)}.  \label{Eq:FermionPropagatorPhysical}
\end{eqnarray}
Coulomb interaction is marginal at tree-level in 3D AWSM. Thus,
when calculating the corrections induced by long-range Coulomb interaction,
we use the physical fermion propagator equation~(\ref{Eq:FermionPropagatorPhysical}) directly.
When calculating the correction contributed by disorder, we adopt the generalized expression of
fermion propagator
\begin{eqnarray}
G_{0}(\omega,\mathbf{k})=\frac{1}{i\omega-\left(vk_{x}
\sigma_{1}+vk_{y}\sigma_{2}+Ak_{z}^{n}\sigma_{3}\right)},
\label{Eq:FermionPropagatorGeneral}
\end{eqnarray}
where $n$ is an even integer. The corresponding dispersion of fermions takes the
form \begin{eqnarray}
E_{\pm}(\mathbf{k})
&=&\pm\sqrt{v^{2}k_{\bot}^{2}+A^{2}k_{z}^{2n}}.
\end{eqnarray}
In the RG analysis, we will find that $1/n$ serves as an effective controlled expansion parameter in terms of disorder coupling.

The propagator of boson field $\phi$ which represents the influence of Coulomb interaction is given by
\begin{eqnarray}
D_{0}(\mathbf{q})=\frac{\sqrt{\eta}}{q_{x}^{2}+q_{y}^{2}+\eta q_{z}^{2}}=\frac{\sqrt{\eta}}{q_{\bot}^{2}+\eta q_{z}^{2}}.
\label{Eq:BosonPropagatorApp}
\end{eqnarray}

\section{Boson self-energy \label{App:BosonSelfEnergy}}

\begin{figure}[htbp]
\center
\includegraphics[width=1.8in]{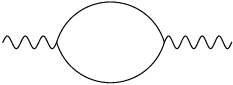}
\caption{Self-energy of bosonic field. The solid line represents the
fermion propagator, and the wavy line stands for the boson
propagator that is equivalent to the Coulomb interaction function.
\label{Fig:BosonSelfEnergy}}
\end{figure}

As shown in  figure~\ref{Fig:BosonSelfEnergy}, the self-energy of boson field $\phi$ is defined as
\begin{eqnarray}
\Pi(\Omega,\mathbf{q})&=&-g^{2}\int\frac{d\omega}{2\pi}\int'\frac{d^3\mathbf{k}}{(2\pi)^{3}}\mathrm{Tr}\left[G_{0}(\omega,\mathbf{k})\right.\nonumber
\\
&&\left.\times G_{0}(\omega+\Omega,\mathbf{k}+\mathbf{q})\right].\label{Eq:BosonSelfEnergyAppA}
\end{eqnarray}
$\int'$ represents that a momentum shell will be imposed in some proper way. Substituting equation~(\ref{Eq:FermionPropagatorPhysical}) into equation~(\ref{Eq:BosonSelfEnergyAppA}), and taking the limit $\Omega=0$, we get
\begin{eqnarray}
\Pi(0,\mathbf{q})
&=&2g^{2}\int\frac{d\omega}{2\pi}\int'\frac{d^3\mathbf{k}}{(2\pi)^{3}}
\frac{1}{\left(\omega^{2}+E_{\mathbf{k}}^{2}\right)\left(\omega^{2}+E_{\mathbf{k}+\mathbf{q}}^{2}\right)}\nonumber
\\
&&\times\Big[\omega^{2}-v^2k_{x}\left(k_{x}+q_{x}\right)
-v^{2}k_{y}\left(k_{y}+q_{y}\right)\nonumber
\\
&&-A^{2}k_{z}^{2}\left(k_{z}+q_{z}\right)^{2}\Big],
\end{eqnarray}
where $E_{\mathbf{k}}=\sqrt{v^{2}k_{\bot}^{2}+A^{2}k_{z}^{4}}$. Expanding of $q_{i}$ up
to quadratic order yields
\begin{eqnarray}
\Pi(0,\mathbf{q})
&\approx&v^{2}q_{\bot}^{2}\frac{g^{2}}{16\pi^{2}}\int' dk_{\bot}d|k_{z}|k_{\bot}\left(\frac{2}{E_{\mathbf{k}}^{3}}-\frac{v^{2}k_{\bot}^{2}}{E_{\mathbf{k}}^{5}}\right)\nonumber
\\
&&+v^{2}A^{2}q_{z}^{2}\frac{g^{2}}{2\pi^{2}}\int' dk_{\bot}d|k_{z}|k_{\bot}\frac{k_{z}^{2}k_{\bot}^{2}}{E_{\mathbf{k}}^{5}}.
\end{eqnarray}

We employ the  transformations
\begin{eqnarray}
E=\sqrt{v^2k_{\bot}^{2}+A^{2}k_{z}^{4}},\quad
\chi =\frac{Ak_{z}^{2}}{vk_{\bot}}, \label{Eq:TransforamtionsA}
\end{eqnarray}
which are equivalent to
\begin{eqnarray}
k_{\bot}=\frac{E}{v\sqrt{1+\chi^2}},\qquad
|k_{z}|=\frac{\chi^{\frac{1}{2}}E^{\frac{1}{2}}}
{A^{\frac{1}{2}}\left(1+\chi^2\right)^{\frac{1}{4}}}.  \label{Eq:TransforamtionsB}
\end{eqnarray}
One could get the relation for the integrand measures as
\begin{eqnarray}
dk_{\bot}d|k_{z}|&=&
\left|\left|
\begin{array}{cc}
\frac{\partial k_{\bot}}{\partial E} & \frac{\partial k_{\bot}}{\partial \chi}
\\
\frac{\partial |k_{z}|}{\partial E} & \frac{\partial |k_{z}|}{\partial \chi}
\end{array}
\right|\right|dEd\chi\nonumber
\\
&=&\left|\frac{\partial k_{\bot}}{\partial E}\frac{\partial |k_{z}|}
{\partial \chi}
-\frac{\partial k_{\bot}}{\partial \chi}\frac{\partial |k_{z}|}{\partial E}\right|dEd\chi\nonumber
\\
&=&\frac{E^{\frac{1}{2}}}
{2vA^{\frac{1}{2}}\chi^{\frac{1}{2}}\left(1+\chi^2\right)^{\frac{3}{4}}
}dEd\chi. \label{Eq:TransforamtionsC}
\end{eqnarray}

Performing  the integrations of $E$ and $\chi$ within the ranges $b\Lambda<E<\Lambda$
with $b=e^{-\ell}$ and $0<\chi<+\infty$, $\Pi(0,\mathbf{q})$
can be expressed as
\begin{eqnarray}
\Pi(0,\mathbf{q})
&\approx&C_{\bot}q_{\bot}^{2}\ell+C_{z}q_{z}^{2}\ell,
\end{eqnarray}
where
\begin{eqnarray}
C_{\bot}&=&\frac{3g^{2}}{40\pi^{2}A^{\frac{1}{2}}\Lambda^{\frac{1}{2}}},
\\
C_{z}&=&\frac{2g^{2}A^{\frac{1}{2}}\Lambda^{\frac{1}{2}}}{21\pi^{2}v^{2}}.
\end{eqnarray}

\begin{figure}[htbp]
\center
\includegraphics[width=3in]{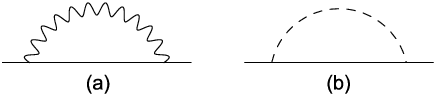}
\caption{Self-energy of fermions due to (a) Coulomb interaction and
(b) disorder. The dashed line denotes disorder scattering.
\label{Fig:FermionSelfEnergy}}
\end{figure}

\section{Fermion self-energy \label{App:FermiSelfEnergy}}

As depicted in figure~\ref{Fig:FermionSelfEnergy}(a), the self-energy of fermions induced by Coulomb interaction takes the form
\begin{eqnarray}
\Sigma_{C}(\omega,\mathbf{k})&=&-g^2\int\frac{d\Omega}{2\pi}\int'\frac{d^3\mathbf{q}}{(2\pi)^3}
 G_{0}(\Omega,\mathbf{q})\nonumber
\\
&&\times D_{0}(\omega-\Omega,\mathbf{k}-\mathbf{q}). \label{Eq:FermionSelfEnergyCoulombAppA}
\end{eqnarray}
Substituting equations~(\ref{Eq:FermionPropagatorPhysical}) and (\ref{Eq:BosonPropagatorApp}) into equation~(\ref{Eq:FermionSelfEnergyCoulombAppA}) and
retaining the leading contributions, $\Sigma_{C}$ can be approximated as
\begin{eqnarray}
\Sigma_{C}(\omega,\mathbf{k})
&\approx&v\left(k_{x}\sigma_{1}+k_{y}\sigma_{2}\right)\Sigma_{1}+Ak_{z}^{2}\sigma_{3}\Sigma_{2},
\end{eqnarray}
where
\begin{eqnarray}
\Sigma_{1}&=&\frac{g^2\sqrt{\eta}}{2}\int'\frac{d^3\mathbf{q}}{(2\pi)^3}
\frac{q_{\bot}^{2}}{E_{\mathbf{q}}\left(\mathbf{q}_{\bot}^{2}+\eta
q_{z}^{2}\right)^{2}},
\\
\Sigma_{2}&=&\frac{g^2\eta^{\frac{3}{2}}}{2}\int'\frac{d^3\mathbf{q}}{(2\pi)^3}
\frac{q_{z}^{2}\left(-q_{\bot}^{2}+3\eta q_{z}^{2}\right)}{E_{\mathbf{q}}\left(\mathbf{q}_{\bot}^{2}+\eta
q_{z}^{2}\right)^{3}}.
\end{eqnarray}
We have dropped a constant term which does not depend on external energy and momenta.
In the previous study about long-range Coulomb interaction on 3D AWSM by Yang \emph{et al.} \cite{Yang14B} and pioneer work by Abrikosov \cite{Abrikosov72}, the generated constant term  was
also discarded.

Utilizing the transformations (\ref{Eq:TransforamtionsA})-(\ref{Eq:TransforamtionsC})
and performing the integration of $E$, we arrive at
\begin{eqnarray}
\Sigma_{1}\approx C_{1}\ell,\qquad \Sigma_{2}\approx C_{2}\ell,
\end{eqnarray}
where
\begin{eqnarray}
C_{1}&=&\frac{g^2\zeta^{\frac{3}{2}}}{8\pi^{2}v}
\int_{0}^{+\infty}d\chi\frac{1}
{\chi^{\frac{1}{2}}\left(1+\chi^2\right)^{\frac{1}{4}}}\nonumber
\\
&&\times\frac{1}{\left[\zeta+\chi\left(1+\chi^2\right)^{\frac{1}{2}}\right]^{2}},
\\
C_{2}&=&\frac{g^2\zeta^{\frac{1}{2}}}{8\pi^{2}v}
\int_{0}^{+\infty}d\chi\chi^{\frac{1}{2}}\left(1+\chi^{2}\right)^{\frac{1}{4}}\nonumber
\\
&&\times\frac{-\zeta+3\chi\left(1+\chi^{2}\right)^{\frac{1}{2}}}
{\left[\zeta+\chi\left(1+\chi^2\right)^{\frac{1}{2}}\right]^{3}},
\end{eqnarray}
with $\zeta=\frac{A\Lambda}{v^{2}\eta}$.

As displayed in figure~\ref{Fig:FermionSelfEnergy}(b), the self-energy of fermions leaded by disorder scattering satisfies
\begin{eqnarray}
\Sigma_{dis}(\omega)&\approx&\sum_{j=0}^{3}\Delta_{j}\int'\frac{d^3\mathbf{k}}{(2\pi)^{3}}\Gamma_{j}
G_{0}(\omega,\mathbf{k})\Gamma_{j}. \label{Eq:FermionSelfEnergyDisAppA}
\end{eqnarray}
Substituting equation~(\ref{Eq:FermionPropagatorGeneral}) into equation~(\ref{Eq:FermionSelfEnergyDisAppA}), we obtain
\begin{eqnarray}
\Sigma_{dis}(\omega)&\approx&-\sum_{j=0}^{3}\Delta_{j}\int'\frac{d^3\mathbf{k}}{(2\pi)^{3}}
\frac{i\omega}{\omega^{2}+v^{2}k_{\bot}^{2}+A^{2}k_{z}^{2n}}. \label{Eq:FermionSelfEnergyDisMid}
\end{eqnarray}
 A constant generated term  has been discarded. We should notice that in previous studies about the
 disorder effects in 3D AWSM by Roy \emph{et al.} \cite{Roy18} and Luo \emph{et al.} \cite{Luo18A},
 the constant generated term in self-energy was also discarded. The reason, why the constant term
 generated by Coulomb interaction or disorder was discarded in previous studies
 \cite{Yang14A, Abrikosov72, Roy18, Luo18A} and also in our calculation, is assumed that the system is at
the topological QCP, although the position of the topological QCP may be moved by interaction or disorder.
In the studies about quantum critical behaviours at Landau QCP, the constant generated term of self-energy
of boson field corresponding to the order parameter is also discarded, and the system is always assumed at the QCP \cite{Huh08}.

\begin{figure}[htbp]
\center
\includegraphics[width=2.7in]{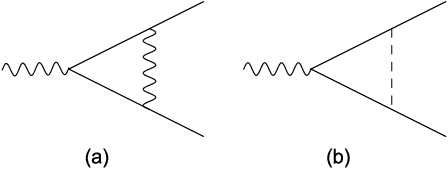}
\caption{Corrections to fermion-boson coupling due to (a) Coulomb
interaction and (b) disorder. \label{Fig:CoulombVertexCorrection}}
\end{figure}

We employ the  transformations
\begin{eqnarray}
E=\sqrt{v^2k_{\bot}^{2}+A^{2}k_{z}^{2n}},\quad
\chi =\frac{Ak_{z}^{n}}{vk_{\bot}}, \label{Eq:TransforamtionsAGeneral}
\end{eqnarray}
which are equivalent to
\begin{eqnarray}
k_{\bot}=\frac{E}{v\sqrt{1+\chi^2}},\qquad
|k_{z}|=\frac{\chi^{\frac{1}{n}}E^{\frac{1}{n}}}
{A^{\frac{1}{n}}\left(1+\chi^2\right)^{\frac{1}{2n}}}.  \label{Eq:TransforamtionsBGeneral}
\end{eqnarray}
One could get the relation for the integrand measures as
\begin{eqnarray}
dk_{\bot}d|k_{z}|&=&
\left|\left|
\begin{array}{cc}
\frac{\partial k_{\bot}}{\partial E} & \frac{\partial k_{\bot}}{\partial \chi}
\\
\frac{\partial |k_{z}|}{\partial E} & \frac{\partial |k_{z}|}{\partial \chi}
\end{array}
\right|\right|dEd\chi\nonumber
\\
&=&\left|\frac{\partial k_{\bot}}{\partial E}\frac{\partial |k_{z}|}
{\partial \chi}
-\frac{\partial k_{\bot}}{\partial \chi}\frac{\partial |k_{z}|}{\partial E}\right|dEd\chi\nonumber
\\
&=&\frac{E^{\frac{1}{n}}}
{nvA^{\frac{1}{n}}\chi^{1-\frac{1}{n}}\left(1+\chi^2\right)^{\frac{1}{2}+\frac{1}{2n}}
}dEd\chi. \label{Eq:TransforamtionsCGeneral}
\end{eqnarray}

Adopting the transformations (\ref{Eq:TransforamtionsAGeneral})-(\ref{Eq:TransforamtionsCGeneral})
for equation~(\ref{Eq:FermionSelfEnergyDisMid}), and
performing  the integrations of $E$ and $\chi$ within the ranges $b\Lambda<E<\Lambda$
with $b=e^{-\ell}$ and $0<\chi<+\infty$, $\Sigma_{dis}$ can be evaluated as following
\begin{eqnarray}
\Sigma_{dis}(\omega)
&=&-i\omega\sum_{j=0}^{3}\frac{\Delta_{j}}{2n\pi^{2}v^{2}A^{\frac{1}{n}}}
\int_{b\Lambda}^{\Lambda}dE E^{\frac{1}{n}-1}\int_{0}^{+\infty} d\chi\nonumber
\\
&&\times\frac{1}
{\chi^{1-\frac{1}{n}}\left(1+\chi^2\right)^{1+\frac{1}{2n}}
}\nonumber
\\
&\approx&-i\omega\sum_{j=0}^{3}\frac{\Delta_{j}\Lambda^{\frac{1}{n}}}
{2\pi^{2}v^{2}A^{\frac{1}{n}}}
\ell.
\end{eqnarray}

\section{Corrections to fermion-boson coupling\label{App:CoulombVertexCorrection}}

\begin{figure}[htbp]
\center
\includegraphics[width=3.3in]{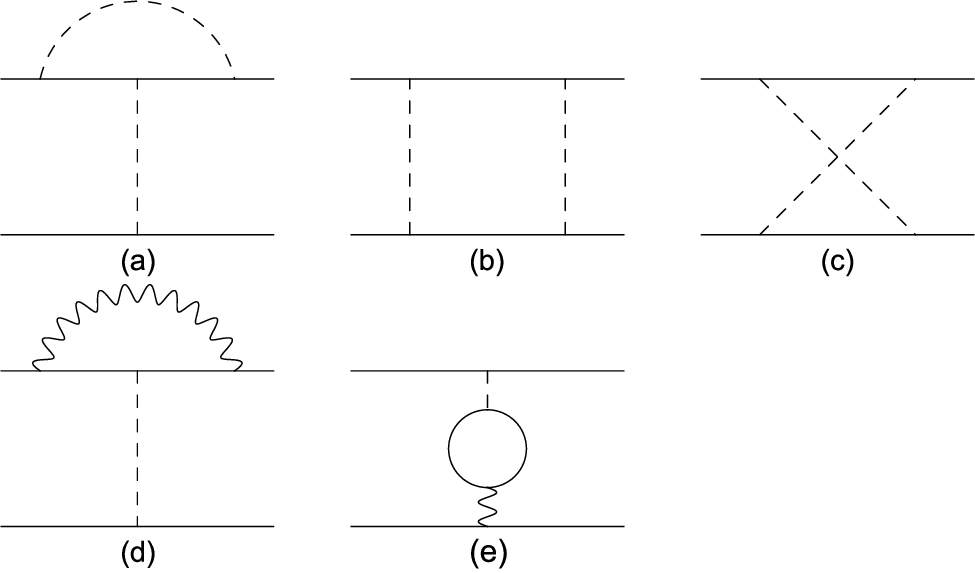}
\caption{One-loop Feynman diagrams for the corrections to the
fermion-disorder coupling. \label{Fig:VertexCorrection}}
\end{figure}

The correction to the fermion-boson coupling leaded by figure~\ref{Fig:CoulombVertexCorrection}(a)
can be written as
\begin{eqnarray}
\delta g^{(1)} &=& -g^{3}\int\frac{d\Omega}{2\pi}\int'
\frac{d^3\mathbf{q}}{(2\pi)^{3}}G_{0}(\Omega,\mathbf{q})
G_{0}(\Omega,\mathbf{q})\nonumber
\\
&&\times D_{0}(\Omega,\mathbf{q}).
\label{Eq:FermionBosonCouplingCorrectionA}
\end{eqnarray}
Substituting equations~(\ref{Eq:FermionPropagatorPhysical}) and (\ref{Eq:BosonPropagatorApp}) into equation~(\ref{Eq:FermionBosonCouplingCorrectionA}),
one can find
\begin{eqnarray}
\delta g^{(1)}=0.
\end{eqnarray}
Figure~\ref{Fig:CoulombVertexCorrection}(b) results in the correction
\begin{eqnarray}
\delta g^{(2)} &=& g\sum_{j=0}^{3}\Delta_{j}\int'
\frac{d^3\mathbf{k}}{(2\pi)^{3}}\Gamma_{j}
G_{0}(0,\mathbf{k})G_{0}(0,\mathbf{k})\Gamma_{j}.
\nonumber
\\
&\approx&g\sum_{j=0}^{3}\frac{\Delta_{j}\Lambda^{\frac{1}{n}}}
{2\pi^{2}v^{2}A^{\frac{1}{n}}}\ell.
\label{Eq:FermionBosonCouplingCorrectionB}
\end{eqnarray}
Thus the total correction is
\begin{eqnarray}
\delta g&=&\delta g^{(1)}+\delta g^{(2)}=g\sum_{j=0}^{3}\frac{\Delta_{j}\Lambda^{\frac{1}{n}}}
{2\pi^{2}v^{2}A^{\frac{1}{n}}}\ell.
\end{eqnarray}

\section{Corrections to fermion-disorder coupling
\label{App:DisVertexCorrection}}

The correction to fermion-disorder coupling from the figure~\ref{Fig:VertexCorrection}(a) is given by
\begin{eqnarray}
W^{(1)}=\sum_{i=0}^{3}W_{i}^{(1)},
\end{eqnarray}
where
\begin{eqnarray}
W_{i}^{(1)} &=&
\Delta_{i}\sum_{j=0}^{3}\Delta_{j}\left(\psi_{a}^{\dag}\Gamma_{i}\psi_{a}\right)\int'
\frac{d^3\mathbf{k}}{(2\pi)^{3}}\Big[\psi_{b}^{\dag}\Gamma_{j}
G_{0}(0,\mathbf{k})
\Gamma_{i}\nonumber
\\
&&\times G_{0}(0,\mathbf{k})\Gamma_{j}\psi_{b}\Big].
\label{Eq:VertexCorrectionA}
\end{eqnarray}
The correction from figures~\ref{Fig:VertexCorrection}(b) and \ref{Fig:VertexCorrection}(c) to the
fermion-disorder coupling reads as
\begin{eqnarray}
W^{(2)+(3)} = \sum_{i=0}^{3}\sum_{i\le j\le3}W_{ij}^{(2)+(3)},
\label{Eq:VertexCorrectionB1}
\end{eqnarray}
where
\begin{eqnarray}
W_{ij}^{(2)+(3)} &=&
\Delta_{i}\Delta_{j}\int'\frac{d^3\mathbf{k}}{(2\pi)^{3}}
\left(\psi_{a}^{\dag}\Gamma_{i} G_{0}(0,\mathbf{k})
\Gamma_{j}\psi_{a}\right)\nonumber
\\
&&\times\Big\{\psi_{b}^{\dag}
\left[\Gamma_{j}G_{0}(0,\mathbf{k})\Gamma_{i}\right.\nonumber
\\
&&\left. +
\Gamma_{i}G_{0}(0,-\mathbf{k})\Gamma_{j}\right]\psi_{b}\Big\}.
\label{Eq:VertexCorrectionB2}
\end{eqnarray}
There are ten choices for the values of $i$ and $j$. As
displayed in the figure~\ref{Fig:VertexCorrection}(d), the correction
to fermion-disorder coupling resulting from Coulomb interaction takes the form
\begin{eqnarray}
W^{(4)}=\sum_{i=0}^{3}W_{i}^{(4)},
\end{eqnarray}
where
\begin{eqnarray}
W_{i}^{(4)} &=& -\Delta_{i}g^{2}\left(\psi_{a}^{\dag}\Gamma_{i}\psi_{a}\right)
\int\frac{d\Omega}{2\pi}\int'
\frac{d^3\mathbf{q}}{(2\pi)^{3}}\Big[\psi_{b}^{\dag}G_{0}(\Omega,\mathbf{q})
\nonumber
\\
&&\times \Gamma_{i}G_{0}(\Omega,\mathbf{q})\psi_{b}\Big]D_{0}(\Omega,\mathbf{q}).
\label{Eq:VertexCorrectionC}
\end{eqnarray}
Figure~\ref{Fig:VertexCorrection}(e) yields the correction
\begin{eqnarray}
W^{(5)}=\sum_{i=0}^{3}W_{i}^{(5)},
\end{eqnarray}
where
\begin{eqnarray}
 W_{i}^{(5)} &=& 2\Delta_{i}g^{2}\left(\psi_{a}^{\dag}\Gamma_{i}\psi_{a}\right)\int
\frac{d\omega}{2\pi}\int'\frac{d^3\mathbf{k}}{(2\pi)^{3}}\nonumber
\\
&&\times\Big\{\psi_{b}^{\dag}\mathrm{Tr}\left[ G_{0}(\omega,\mathbf{k})
\Gamma_{i} G_{0}(\omega+\Omega,\mathbf{k}+\mathbf{q})\right]\nonumber
\\
&&\times D_{0}(\Omega,\mathbf{q})\psi_{b}\Big\}.
\label{Eq:VertexCorrectionD}
\end{eqnarray}
Substituting equation~(\ref{Eq:FermionPropagatorGeneral}) into
equations~(\ref{Eq:VertexCorrectionA})-(\ref{Eq:VertexCorrectionB2}),
and substituting equations~(\ref{Eq:FermionPropagatorPhysical}) and (\ref{Eq:BosonPropagatorApp}) into
equations~(\ref{Eq:VertexCorrectionC})-(\ref{Eq:VertexCorrectionD}),
 we arrive at
\begin{eqnarray}
W&=&W^{(1)}+W^{(2)+(3)}+W^{(4)}+W^{(5)}\nonumber
\\
&\approx&\frac{\delta\Delta_{0}}{2}\left(\psi_{a}^{\dag}\Gamma_{0}\psi_{a}\right)
\left(\psi_{b}^{\dag}\Gamma_{0}\psi_{b}\right)\nonumber
\\
&&+\frac{\delta\Delta_{1}}{2}\left(\psi_{a}^{\dag}\Gamma_{1}\psi_{a}\right)
\left(\psi_{b}^{\dag}\Gamma_{1}\psi_{b}\right)\nonumber
\\
&&+\frac{\delta\Delta_{2}}{2}\left(\psi_{a}^{\dag}\Gamma_{2}\psi_{a}\right)
\left(\psi_{b}^{\dag}\Gamma_{2}\psi_{b}\right)\nonumber
\\
&&+\frac{\delta\Delta_{3}}{2}\left(\psi_{a}^{\dag}\Gamma_{3}\psi_{a}\right)
\left(\psi_{b}^{\dag}\Gamma_{3}\psi_{b}\right).
\end{eqnarray}
$\delta\Delta_{i}$ with $i=0, 1, 2, 3$ are given by
\begin{eqnarray}
\delta\Delta_{0}
&=&\left[\frac{5}{2}\Delta_{0}^{2}+\frac{5}{2}\Delta_{0}\Delta_{1}
+\frac{5}{2}\Delta_{0}\Delta_{2}
+\frac{5(2n+3)}{2(2n+1)}\Delta_{0}\Delta_{3}\right.\nonumber
\\
&&\left.+\frac{5n}{2n+1}\left(\Delta_{1}\Delta_{3}+\Delta_{2}\Delta_{3}\right)\right]
\frac{2\Lambda^{\frac{1}{n}}}{5\pi^{2}v^{2}A^{\frac{1}{n}}}
\ell\nonumber
\\
&&-2\Delta_{0}\left(\sqrt{\eta}C_{\bot}
+\frac{C_{z}}{\sqrt{\eta}}\right)\ell, \label{Eq:deltaDelta0App}
\\
\delta\Delta_{1}
&=&\frac{5}{2n+1}\bigg(
-\frac{1}{2}\Delta_{1}\Delta_{0}-\frac{1}{2}\Delta_{1}^{2}
+\frac{1}{2}\Delta_{1}\Delta_{2}+\frac{3}{2}\Delta_{1}\Delta_{3}\nonumber
\\
&&+n\Delta_{0}\Delta_{3}
\bigg)
\frac{2\Lambda^{\frac{1}{n}}}{5\pi^{2}v^{2}A^{\frac{1}{n}}}
\ell+\Delta_{1}C_{3}\ell,  \label{Eq:deltaDelta1App}
\\
\delta\Delta_{2}
&=&\frac{5}{2n+1}\bigg(-\frac{1}{2}\Delta_{2}\Delta_{0}
+\frac{1}{2}\Delta_{2}\Delta_{1}-\frac{1}{2}\Delta_{2}^{2}+\frac{3}{2}\Delta_{2}\Delta_{3}\nonumber
\\
&&+n\Delta_{0}\Delta_{3}
\bigg)\frac{2\Lambda^{\frac{1}{n}}}
{5\pi^{2}v^{2}A^{\frac{1}{n}}}
\ell+\Delta_{2}C_{3}\ell, \label{Eq:deltaDelta2App}
\\
\delta\Delta_{3}
&=&\frac{5}{2n+1}\bigg[\frac{2n-1}{2}\Delta_{3}\left(-\Delta_{0}+\Delta_{1}
+\Delta_{2}\right)\nonumber
\\
&&-\frac{2n-3}{2}\Delta_{3}^{2}+\Delta_{0}^{2}+\Delta_{1}^{2}
+\Delta_{2}^{2}\nonumber
\\
&&+n\left(\Delta_{0}\Delta_{1}+\Delta_{0}\Delta_{2}\right)\bigg]
\frac{2\Lambda^{\frac{1}{n}}}{5\pi^{2}v^{2}A^{\frac{1}{n}}}\ell\nonumber
\\
&&+\Delta_{3}C_{4}\ell,  \label{Eq:deltaDelta3App}
\end{eqnarray}
where
\begin{eqnarray}
C_{3}&=&\frac{g^{2}\zeta^{\frac{1}{2}}}{8\pi^{2}v}
\int_{0}^{+\infty}d\chi
\frac{2+\chi^{2}}
{\chi^{\frac{1}{2}}\left(1+\chi^2\right)^{\frac{5}{4}}
}\nonumber
\\
&&\times\frac{1}{\zeta
+\chi\left(1+\chi^2\right)^{\frac{1}{2}}},
\\
C_{4}&=&\frac{g^{2}\zeta^{\frac{1}{2}}}{4\pi^{2}v}
\int_{0}^{+\infty} d\chi \frac{\chi^{\frac{3}{2}}}
{\left(1+\chi^2\right)^{\frac{5}{4}}
}\frac{1}
{\zeta+ \chi\left(1+\chi^2\right)^{\frac{1}{2}}}.
\end{eqnarray}

\section{Derivation of the RG equations \label{App:RGEquations}}

The action of free  fermions is
\begin{eqnarray}
S_{\psi}&=&\int\frac{d\omega}{2\pi}\frac{d^{3}\mathbf{k}}{(2\pi)^{3}}\psi_{a}^{\dag}(\omega,\mathbf{k})
\Big[i\omega-v(k_{x}\sigma_{1}+k_{y}\sigma_{2})\nonumber
\\
&&-Ak_{z}^{n}\sigma_{3}
\Big]\psi_{a}(\omega,\mathbf{k}).
\end{eqnarray}
Incorporating the self-energies of fermions induced by Coulomb interaction and disorder-scattering, the
action becomes
\begin{eqnarray}
S_{\psi}&=&\int\frac{d\omega}{2\pi}\frac{d^{3}\mathbf{k}}{(2\pi)^{3}}\psi_{a}^{\dag}(\omega,\mathbf{k})
\Big[i\omega-v(k_{x}\sigma_{1}+k_{y}\sigma_{2})\nonumber
\\
&&-Ak_{z}^{n}\sigma_{3}-\Sigma_{C}-\Sigma_{dis}
\Big]\psi_{a}(\omega,\mathbf{k})\nonumber
\\
&\approx&\int\frac{d\omega}{2\pi}\frac{d^3\mathbf{k}}{(2\pi)^{3}}\psi_{a}^{\dag}(\omega,\mathbf{k})\nonumber
\\
&&\times\Bigg[i\omega e^{\sum_{j=0}^{3}\frac{\Delta_{j}\Lambda^{\frac{1}{n}}}{2\pi^{2}v^{2}A^{\frac{1}{n}}}\ell}
-v\left(k_{x}\sigma_{1}+k_{y}\sigma_{2}\right)
e^{C_{1}\ell}\nonumber
\\
&&-Ak_{z}^{n}\sigma_{3}e^{C_{2}\ell}
\Bigg]\psi_{a}(\omega,\mathbf{k}).
\end{eqnarray}
Utilizing the transformations
\begin{eqnarray}
k_{x}&=&k_{x}'e^{-\ell}, \label{Eq:KxScaling}
\\
k_{y}&=&k_{y}'e^{-\ell}, \label{Eq:KyScaling}
\\
k_{z}&=&k_{z}'e^{-\frac{\ell}{n}}, \label{Eq:KzScaling}
\\
\omega&=&\omega'e^{-\ell}, \label{Eq:OmegaScaling}
\\
\psi_{a}&=&\psi_{a}' e^{\left(\left(2+\frac{1}{2n}\right)
-\frac{\sum_{j=0}^{3}\frac{\Delta_{j}\Lambda^{\frac{1}{n}}}
{2\pi^{2}v^{2}A^{\frac{1}{n}}}}{2}\right)\ell}, \label{Eq:PsiScaling}
\\
v&=&v'e^{\left(-C_{1}+\sum_{j=0}^{3}\frac{\Delta_{j}\Lambda^{\frac{1}{n}}}
{2\pi^{2}v^{2}A^{\frac{1}{n}}}\right)\ell},  \label{Eq:vScaling}
\\
A&=&A'e^{\left(-C_{2}+\sum_{j=0}^{3}\frac{\Delta_{j}\Lambda^{\frac{1}{n}}}
{2\pi^{2}v^{2}A^{\frac{1}{n}}}\right)\ell},  \label{Eq:AScaling}
\end{eqnarray}
the action of fermions can be written as
\begin{eqnarray}
S_{\psi'}&=&\int\frac{d\omega'}{2\pi}\frac{d^{3}\mathbf{k}'}{(2\pi)^{3}}{\psi'}_{a}^{\dag}(\omega',\mathbf{k}')
\Big[i\omega' -v'\left(k_{x}'\sigma_{1}+k_{y}'\sigma_{2}\right)\nonumber
\\
&&-A'k_{z}'^{n}\sigma_{3}
\Big]\psi'_{a}(\omega',\mathbf{k}'),
\end{eqnarray}
which has the same form as the original action of free fermions.

The action of free bosonic field $\phi$ takes the form
\begin{eqnarray}
S_{\phi}=\int\frac{d\omega}{2\pi}\frac{d^{3}\mathbf{k}}{(2\pi)^{3}}\phi(\omega,\mathbf{k})
\left(\frac{k_{\bot}^{2}}{\sqrt{\eta}}+\sqrt{\eta} k_{z}^{2}
\right)\phi(\omega,\mathbf{k}).
\end{eqnarray}
Including the correction of self-energy of boson, the action can be expressed as
\begin{eqnarray}
S_{\phi}
&=&\int\frac{d\omega}{2\pi}\frac{d^{3}\mathbf{k}}{(2\pi)^{3}}\phi(\omega,\mathbf{k})
\Big(\frac{k_{\bot}^{2}}{\sqrt{\eta}}+\sqrt{\eta} k_{z}^{2}\nonumber
\\
&&+\Pi(0,\mathbf{k})
\Big)\phi(\omega,\mathbf{k})\nonumber
\\
&\approx&\int\frac{d\omega}{2\pi}\frac{d^{3}\mathbf{k}}{(2\pi)^{3}}\phi(\omega,\mathbf{k})
\Big(\frac{k_{\bot}^{2}}{\sqrt{\eta}}e^{\sqrt{\eta}C_{\bot}\ell}\nonumber
\\
&&+\sqrt{\eta} k_{z}^{2}e^{\frac{C_{z}}{\sqrt{\eta}}\ell}
\Big)\phi(\omega,\mathbf{k}).
\end{eqnarray}
Employing the transformations equations~(\ref{Eq:KxScaling})-(\ref{Eq:OmegaScaling}), and
\begin{eqnarray}
\phi&=&\phi'e^{\left[\left(2+\frac{1}{n}\right)
-\eta_{\phi}\right]\ell} , \label{Eq:PhiScaling}
\\
\eta&=&\eta'e^{\left[-\left(2-\frac{2}{n}\right)
+\sqrt{\eta}C_{\bot}-\frac{C_{z}}{\sqrt{\eta}}\right]\ell} ,  \label{Eq:etaScaling}
\end{eqnarray}
where
\begin{eqnarray}
\eta_{\phi}=\frac{\sqrt{\eta}C_{\bot}+\frac{C_{z}}{\sqrt{\eta}}}{4},
\end{eqnarray}
we get
\begin{eqnarray}
S_{\phi'}
&=&\int\frac{d\omega'}{2\pi}\frac{d^{3}\mathbf{k}'}{(2\pi)^{3}}\phi'(\omega',\mathbf{k}')
\left(\frac{k_{\bot}'^{2}}{\sqrt{\eta'}}+\sqrt{\eta'} k_{z}'^{2}
\right)\nonumber
\\
&&\times\phi'(\omega',\mathbf{k}'),
\end{eqnarray}
which recovers the form of action of free bosons.

The action of fermion-boson coupling is
\begin{eqnarray}
S_{\psi\phi}&=&ig\int\frac{d\omega_{1}}{2\pi}\frac{d^3\mathbf{k}_{1}}{(2\pi)^{3}}\frac{d\omega_{2}}{2\pi}
\frac{d^3\mathbf{k}_{2}}{(2\pi)^{3}}\psi_{a}^{\dag}(\omega_{1},\mathbf{k}_{1})\psi_{a}(\omega_{2},\mathbf{k}_{2})\nonumber
\\
&&\times\phi(\omega_{1}-\omega_{2},\mathbf{k}_{1}-\mathbf{k}_{2}). \label{Eq:ActionFBCouplingApp}
\end{eqnarray}
Including the corrections to one-loop order, the action becomes
\begin{eqnarray}
S_{\psi\phi}&=&i\left(g+\delta g\right)\int\frac{d\omega_{1}}{2\pi}\frac{d^3\mathbf{k}_{1}}{(2\pi)^{3}}\frac{d\omega_{2}}{2\pi}
\frac{d^3\mathbf{k}_{2}}{(2\pi)^{3}}\psi_{a}^{\dag}(\omega_{1},\mathbf{k}_{1})\nonumber
\\
&&\times\psi_{a}(\omega_{2},\mathbf{k}_{2})\phi(\omega_{1}-\omega_{2},\mathbf{k}_{1}
-\mathbf{k}_{2})\nonumber
\\
&\approx&ige^{\sum_{j=0}^{3}\frac{\Delta_{j}\Lambda^{\frac{1}{n}}}{2\pi^{2}v^{2}A^{\frac{1}{n}}}\ell}\int\frac{d\omega_{1}}{2\pi}\frac{d^3\mathbf{k}_{1}}{(2\pi)^{3}}\frac{d\omega_{2}}{2\pi}
\frac{d^3\mathbf{k}_{2}}{(2\pi)^{3}}\nonumber
\\
&&\times\psi_{a}^{\dag}(\omega_{1},\mathbf{k}_{1})\psi_{a}(\omega_{2},\mathbf{k}_{2})\phi(\omega_{1}-\omega_{2},\mathbf{k}_{1}
-\mathbf{k}_{2}).\nonumber \\
\end{eqnarray}
Adopting the transformations equations~(\ref{Eq:KxScaling})-(\ref{Eq:PsiScaling}), equation~(\ref{Eq:PhiScaling}), and
\begin{eqnarray}
g=g'e^{\left(\frac{\sqrt{\eta}C_{\bot}+\frac{C_{z}}{\sqrt{\eta}}}{4}\right)\ell}, \label{Eq:gScaling}
\end{eqnarray}
we obtain
\begin{eqnarray}
S_{\psi'\phi'}
&=&ig'\int\frac{d\omega_{1}'}{2\pi}\frac{d^3\mathbf{k}_{1}'}{(2\pi)^{3}}\frac{d\omega_{2}'}{2\pi}
\frac{d^3\mathbf{k}_{2}'}{(2\pi)^{3}}{\psi'}_{a}^{\dag}(\omega_{1}',\mathbf{k}_{1}')\nonumber
\\
&&\times\psi'_{a}(\omega_{2}',\mathbf{k}_{2}')\phi'(\omega_{1}'-\omega_{2}',\mathbf{k}_{1}',
-\mathbf{k}_{2}),
\end{eqnarray}
which takes the same form as the original action.

The action of fermion-disorder coupling reads as
\begin{eqnarray}
S_{dis}&=&\sum_{j=0}^{3}\frac{\Delta_{j}}{2} \int\frac{d\omega_1
d\omega_2d^{3}\mathbf{k}_{1} d^{3}\mathbf{k}_{2}
d^{3}\mathbf{k}_{3}}{(2\pi)^{11}}
\psi^\dagger_{a}(\omega_1,\mathbf{k}_1)\Gamma_{j}\nonumber
\\
&&\times\psi_{a}(\omega_1,\mathbf{k}_2)
\psi^\dagger_{b}(\omega_2,\mathbf{k}_3)\Gamma_{j}\nonumber
\\
&&\times\psi_{b}(\omega_2,-\mathbf{k}_1-\mathbf{k}_2-\mathbf{k}_3).
\end{eqnarray}
Including the corrections to the fermion-disorder coupling, the action
is expressed as
\begin{eqnarray}
S_{dis}&=&\sum_{j=0}^{3}\frac{\left(\Delta_{j}+\delta\Delta_{j}\right)}{2} \int\frac{d\omega_1
d\omega_2d^{3}\mathbf{k}_{1} d^{3}\mathbf{k}_{2}
d^{3}\mathbf{k}_{3}}{(2\pi)^{11}}\nonumber
\\
&&\times \psi^\dagger_{a}(\omega_1,\mathbf{k}_1)\Gamma_{j}\psi_{a}(\omega_1,\mathbf{k}_2)
\psi^\dagger_{b}(\omega_2,\mathbf{k}_3)\Gamma_{j}\nonumber
\\
&&\times\psi_{b}(\omega_2,-\mathbf{k}_1-\mathbf{k}_2-\mathbf{k}_3).
\end{eqnarray}
Applying the transformations as shown in  equations~(\ref{Eq:KxScaling})-(\ref{Eq:PsiScaling}), the action can be further
written as
\begin{eqnarray}
S_{dis}
&\approx&\sum_{j=0}^{3}\frac{1}{2}\Bigg[\Delta_{j}\left(1-\frac{1}{n}\ell\right)+\delta\Delta_{j}\nonumber
\\
&&-2\Delta_{j}\sum_{j'=0}^{3}\frac{\Delta_{j'}\Lambda^{\frac{1}{n}}}{2\pi^{2}v^{2}A^{\frac{1}{n}}}\ell\Bigg]\nonumber
\\
&&\times\int\frac{d\omega_1'
d\omega_2'd^{3}\mathbf{k}_{1}' d^{3}\mathbf{k}_{2}'
d^{3}\mathbf{k}_{3}'}{(2\pi)^{11}}\nonumber
\\
&&\times\psi'^\dagger_{a}(\omega_1',\mathbf{k}_1')\Gamma_{j}
\psi'_{a}(\omega_1',\mathbf{k}_2')
\psi'^\dagger_{b}(\omega_2',\mathbf{k}_3')\Gamma_{j}\nonumber
\\
&&\times\psi'_{b}(\omega_2',-\mathbf{k}_1'-\mathbf{k}_2'-\mathbf{k}_3').
\end{eqnarray}
Let
\begin{eqnarray}
\Delta_{j}'&=&\Delta_{j}+\Delta_{j}\left(-\frac{1}{n}-2\sum_{j'=0}^{3}
\frac{\Delta_{j'}\Lambda^{\frac{1}{n}}}
{2\pi^{2}v^{2}A^{\frac{1}{n}}}\right)\ell\nonumber
\\
&&+\delta\Delta_{j},
\label{Eq:DeltaScaling}
\end{eqnarray}
we get
\begin{eqnarray}
S_{dis}
&=&\sum_{j=0}^{3}\frac{\Delta_{j}'}{2}
\int\frac{d\omega_1'
d\omega_2'd^{3}\mathbf{k}_{1}' d^{3}\mathbf{k}_{2}'
d^{3}\mathbf{k}_{3}'}{(2\pi)^{11}}\psi'^\dagger_{a}(\omega_1',\mathbf{k}_1')\Gamma_{j}\nonumber
\\
&&\times
\psi'_{a}(\omega_1',\mathbf{k}_2')
\psi'^\dagger_{b}(\omega_2',\mathbf{k}_3')\Gamma_{j}\nonumber
\\
&&\times\psi'_{b}(\omega_2',-\mathbf{k}_1'-\mathbf{k}_2'-\mathbf{k}_3').
\end{eqnarray}

Through equations~(\ref{Eq:deltaDelta0App})-(\ref{Eq:deltaDelta3App}), (\ref{Eq:PsiScaling})-(\ref{Eq:AScaling}), (\ref{Eq:etaScaling}), (\ref{Eq:gScaling}),
and (\ref{Eq:DeltaScaling}),  we finally obtain the RG equations
\begin{eqnarray}
\frac{dv}{d\ell}&=&\left(C_{1}-\frac{1}{2}\sum_{j=0}^{3}\Delta_{j}\right)v, \label{Eq:VRGvGeneral}
\\
\frac{dA}{d\ell}&=&\left(C_{2}-\frac{1}{2}\sum_{j=0}^{3}\Delta_{j}\right)A,  \label{Eq:VRGAnGeneral}
\\
\frac{d\eta}{d\ell}
&=&\left(2-\frac{2}{n}
-\frac{5}{3}\frac{1+n}{1+2n}\beta\right.\nonumber
\\
&&\left.+\frac{21}{8}\frac{n^3}{1-6n+8n^2}\gamma
\right)\eta,  \label{Eq:VRGetaGeneral}
\\
\frac{dg}{d\ell}&=&-\frac{\sqrt{\eta}C_{\bot}+\frac{C_{z}}{\sqrt{\eta}}}{4}g,  \label{Eq:VRGgGeneral}
\\
\frac{d\bar{A}}{d\ell}
&=&
\bigg[-\left(1-\frac{1}{n}\right)+\frac{1}{n}C_{2}-C_{1}
+\frac{5}{6}\frac{1+n}{1+2n}\beta\nonumber
\\
&&-\frac{21}{16}\frac{n^3}{1-6n+8n^2}\gamma\nonumber
\nonumber
\\
&&+\left(1-\frac{1}{n}\right)
\frac{1}{2}\sum_{j=0}^{3}\Delta_{j}\bigg]\bar{A},  \label{Eq:VRGAGeneral}
\\
\frac{d\alpha}{d\ell}
&=&\bigg(-C_{1}-\frac{5}{6}\frac{1+n}{1+2n}\beta
-\frac{21}{16}\frac{n^3}{1-6n+8n^2}\gamma\nonumber
\\
&&+\frac{1}{2}\sum_{j=0}^{3}\Delta_{j}\bigg)\alpha,  \label{Eq:VRGalphaGeneral}
\\
\frac{d\beta}{d\ell}
&=&
\bigg(1-\frac{1}{n}-\frac{1}{n}C_{2}
-\frac{5}{3}\frac{1+n}{1+2n}\beta\nonumber
\\
&&+\frac{1}{2n}\sum_{j=0}^{3}\Delta_{j}\bigg)\beta,  \label{Eq:VRGbetaGeneral}
\\
\frac{d\gamma}{d\ell}
&=&\bigg[-\left(1-\frac{1}{n}\right)+\frac{1}{n}C_{2}-2C_{1}\nonumber
\\
&&-\frac{21}{8}\frac{n^3}{1-6n+8n^2}\alpha\bar{A}\nonumber
\\
&&
+\frac{1}{2}\left(2-\frac{1}{n}\right)
\sum_{j=0}^{3}\Delta_{j}\bigg]\gamma,  \label{Eq:VRGgammaGeneral}
\\
\frac{d\Delta_{0}}{d\ell}&=&-\frac{1}{n}\Delta_{0}+\left[\left(1+\frac{1}{2n}\right)
\Delta_{0}\left(\Delta_{0}+\Delta_{1}
+\Delta_{2}\right)\right.\nonumber
\\
&&+\left(\frac{n+3}{2n+1}+\frac{1}{2n}\right)\Delta_{0}\Delta_{3}\nonumber
\\
&&\left.+\frac{2n}{2n+1}\left(\Delta_{1}+\Delta_{2}\right)\Delta_{3}\right]
\nonumber
\\
&&-\left(2C_{1}+\frac{1}{n}C_{2}
+\frac{10}{3}\frac{1+n}{1+2n}\beta\right.\nonumber
\\
&&\left.+\frac{21}{4}\frac{n^3}{1-6n+8n^2}\gamma\right)\Delta_{0},  \label{Eq:VRGDelta0General}
\\
\frac{d\Delta_{1}}{d\ell}&=&-\frac{1}{n}\Delta_{1}+\left[
\left(\frac{1}{2n}-\frac{1}{2n+1}\right)\left(\Delta_{1}\Delta_{0}+\Delta_{1}^{2}\right)\right.\nonumber
\\
&&+\left(\frac{1}{2n}+\frac{1}{2n+1}\right)\Delta_{1}\Delta_{2}\nonumber
\\
&&\left.+\left(\frac{1}{2n}+\frac{3}{2n+1}\right)\Delta_{1}\Delta_{3}+\frac{2n}{2n+1}
\Delta_{0}\Delta_{3}
\right]\nonumber
\\
&&
-\left(2C_{1}+\frac{1}{n}C_{2}-C_{3}\right)\Delta_{1},  \label{Eq:VRGDelta1General}
\\
\frac{d\Delta_{2}}{d\ell}&=&-\frac{1}{n}\Delta_{2}+\left[\left(\frac{1}{2n}-\frac{1}{2n+1}\right)
\left(\Delta_{2}\Delta_{0}+\Delta_{2}^{2}\right)\right.\nonumber
\\
&&+\left(\frac{1}{2n}+\frac{1}{2n+1}\right)\Delta_{2}\Delta_{1}\nonumber
\\
&&\left.+\left(\frac{1}{2n}+\frac{3}{2n+1}\right)\Delta_{2}\Delta_{3}+\frac{2n}{2n+1}\Delta_{0}\Delta_{3}
\right]\nonumber
\\
&&
-\left(2C_{1}+\frac{1}{n}C_{2}-C_{3}\right)\Delta_{2},  \label{Eq:VRGDelta2General}
\\
\frac{d\Delta_{3}}{d\ell}&=&-\frac{1}{n}\Delta_{3}+\left[\left(\frac{1}{2n}-\frac{2n-1}{2n+1}\right)\Delta_{3}\Delta_{0}\right.\nonumber
\\
&&+\left(\frac{1}{2n}+\frac{2n-1}{2n+1}\right)\Delta_{3}\left(\Delta_{1}
+\Delta_{2}\right)\nonumber
\\
&&+\left(\frac{1}{2n}-\frac{2n-3}{2n+1}\right)\Delta_{3}^{2}\nonumber
\\
&&+\frac{2}{2n+1}\left(\Delta_{0}^{2}+\Delta_{1}^{2}
+\Delta_{2}^{2}\right)\nonumber
\\
&&\left.+\frac{2n}{2n+1}\Delta_{0}\left(\Delta_{1}+\Delta_{2}\right)\right]\nonumber
\\
&&-\left(2C_{1}+\frac{1}{n}C_{2}-C_{4}\right)\Delta_{3},  \label{Eq:VRGDelta3General}
\end{eqnarray}
where $\bar{A}$, $\alpha$, $\beta$, and $\gamma$ are defined as
\begin{eqnarray}
\bar{A}&=&\frac{A^{\frac{1}{2}}\Lambda^{\frac{1}{2}}}{v\sqrt{\eta}}=\sqrt{\zeta},
\\
\alpha&=&\frac{g^{2}}{4\pi v},
\\
\beta&=&\sqrt{\eta}C_{\bot}=\frac{3}{10\pi}\frac{\alpha}{\bar{A}},
\\
\gamma&=&\frac{C_{z}}{\sqrt{\eta}}=\frac{8}{21\pi}\alpha\bar{A}.
\end{eqnarray}
It should be noticed that  re-definition
\begin{eqnarray}
\frac{\Delta_{j}\Lambda^{\frac{1}{n}}}{\pi^{2}v^{2}A^{\frac{1}{n}}}&\rightarrow&\Delta_{j}
\end{eqnarray}
has been used in the derivation of the RG equations.

\section{Different roles in 3D AWSM and WSM \label{Sec:Comparing3DAWSM3DWSM}}

In order to understand the reason for the obviously different roles of Coulomb interaction in
3D AWSM and 3D DSM/WSM, we compare the RG analysis of Coulomb interaction in 3D AWSM and 3D WSM in the following concretely.

For 3D AWSM, the scaling for the Coulomb field $\phi$ is
\begin{eqnarray}
\phi
=\phi' e^{\left(\frac{5}{2}-\eta_{\phi}\right)\ell}.
\end{eqnarray}
The term
\begin{eqnarray}
\eta_{\phi}=\frac{\sqrt{\eta}C_{\bot}+\frac{C_{z}}{\sqrt{\eta}}}{4},
\end{eqnarray}
results from the boson self-energy $\Pi(\mathbf{q})$.
The fermion-boson coupling describing the long-range Coulomb interaction is given by equation~(\ref{Eq:ActionFBCouplingApp}).
We can find that the non-trivial scaling of boson field will change the scaling of the
parameter $g$. Concretely, the scaling of $g$ is given by
\begin{eqnarray}
g=g'e^{\eta_{\phi}\ell}=g'e^{\left(\frac{\sqrt{\eta}C_{\bot}+\frac{C_{z}}{\sqrt{\eta}}}{4}\right)\ell}.
\end{eqnarray}
Thus, the RG equation of $g$ reads as
\begin{eqnarray}
\frac{dg}{d\ell}=-\frac{\sqrt{\eta}C_{\bot}+\frac{C_{z}}{\sqrt{\eta}}}{4}=-\frac{\beta+\gamma}{4},
\end{eqnarray}
where
\begin{eqnarray}
\beta\equiv\sqrt{\eta}C_{\bot},\qquad
\gamma\equiv\frac{C_{z}}{\sqrt{\eta}}.
\end{eqnarray}
In subsection~\ref{SubSec:OnlyCoulomb}, we have showed that
\begin{eqnarray}
\beta\rightarrow\frac{1}{2},\qquad \gamma\rightarrow0,
\end{eqnarray}
in the lowest energy limit. Then, the RG equation of $g$ in the low-energy regime can be
asymptotically approximated as
\begin{eqnarray}
\frac{dg}{d\ell}\sim-\frac{1}{8}.
\end{eqnarray}
It is easy to find that $g$ satisfies the asymptotical form
\begin{eqnarray}
g\sim e^{-\frac{1}{8}\ell}.
\end{eqnarray}
The Coulomb strength is defined as
\begin{eqnarray}
\alpha=\frac{g^{2}}{4\pi v}.
\end{eqnarray}
In figure~\ref{Fig:VRGClean}, we have showed that $v$ flows to a constant in the lowest energy limit.
Thus, $\alpha$ takes the asymptotical form
\begin{eqnarray}
\alpha\sim e^{-\frac{1}{4}\ell},
\end{eqnarray}
which flows to zero quickly in the lowest energy limit. We can find that $\eta_{\phi}$
satisfies
\begin{eqnarray}
\eta_{\phi}\rightarrow \frac{1}{8},
\end{eqnarray}
in the limit $\ell\rightarrow\infty$. It indicates that the boson field acquires a finite
anomalous dimension.
From the above analysis, we can find that Coulomb interaction in 3D AWSM becomes irrelevant, is due to that the scaling of boson field $\phi$ field acquires a finite nontrivial correction.

For 3D WSM, the boson self-energy is given by
\begin{eqnarray}
\Pi(\mathbf{q})=\frac{g^{2}}{6\pi^{2}v}\ell\left|\mathbf{q}\right|^{2}.
\end{eqnarray}
Considering the correction of $\Pi(\mathbf{q})$, the scaling of boson field $\phi$ can be
written as
\begin{eqnarray}
\phi=\phi'e^{\left(3-\eta_{\phi}\right)\ell}.
\end{eqnarray}
The term
\begin{eqnarray}
\eta_{\phi}=\frac{g^{2}}{12\pi^{2}v},
\end{eqnarray}
results from $\Pi(\mathbf{q})$.
Accordingly, the scaling for fermion-boson coupling parameter $g$ satisfies
\begin{eqnarray}
g=g'e^{\eta_{\phi}\ell}=g'
e^{\frac{g^{2}}{12\pi^{2}v}\ell}.
\end{eqnarray}
Then, the RG equation for $g$ takes the form
\begin{eqnarray}
\frac{dg}{d\ell}&=&-\frac{g^{2}}{12\pi^{2}v}g.
\end{eqnarray}
In 3D WSM, the fermion self-energy induced by long-range Coulomb interaction is
\begin{eqnarray}
\Sigma_{C}(\omega,\mathbf{k})&=&v\mathbf{k}\cdot\mathbf{\sigma}\frac{g^{2}}{6\pi^{2}v}\ell.
\end{eqnarray}
Considering the correction of fermion self-energy, we  obtain the
RG equation for the fermion velocity
\begin{eqnarray}
\frac{dv}{d\ell}&=&\frac{g^{2}}{6\pi^{2}v}v.
\end{eqnarray}
The Coulomb strength takes the form
\begin{eqnarray}
\alpha&=&\frac{g^{2}}{4\pi v}.
\end{eqnarray}
Thus, we can get the RG equation for $\alpha$
\begin{eqnarray}
\frac{d\alpha}{d\ell}
&=&-\frac{4}{3\pi}\alpha^{2}.
\end{eqnarray}
The solution of $\alpha$ is given by
\begin{eqnarray}
\alpha=\frac{\alpha_{0}}{1+\alpha_{0}\frac{4}{3\pi}\ell}.
\end{eqnarray}
It takes the asymptotical form
\begin{eqnarray}
\alpha\sim\frac{1}{\frac{3}{4\pi}\ell},
\end{eqnarray}
which flows to zero slowly in the limit $\ell\rightarrow\infty$. We
find that the term $\eta_{\phi}=\frac{\alpha}{3}$ satisfies
\begin{eqnarray}
\eta_{\phi}\rightarrow0,
\end{eqnarray}
in the limit $\ell\rightarrow\infty$.  It represents that the anomalous dimension of boson
field vanishes in the lowest energy limit.

From above comparison, we can clearly find that the obviously different roles of long-range Coulomb interaction in 3D AWSM and 3D DSM/WSM is due to that the boson field $\phi$ acquires finite anomalous dimension in 3D AWSM but
has vanishing anomalous dimension in 3D DSM/WSM.

\section{Calculation of the correlation length exponent \label{App:CLEDerivation}}

In this section, we show the detailed calculation of correlation length exponent at the fixed point
\begin{eqnarray}
\left(\Delta_{0}^{*}, \Delta_{1}^{*}, \Delta_{2}^{*, }\Delta_{3}^{*}\right)&\approx&
\left(0.239358, 0.0307505, 0.0307505,\right.
\nonumber
\\
&&\left.0.0667869\right),
\end{eqnarray}
which is obtained in subsection~\ref{SubSec:OnlyDisorderSubLeading}. Expanding the RG equations in
the vicinity of this fixed point, we get
\begin{eqnarray}
\frac{d\Delta_{0}}{d\ell}
&=&-\frac{1}{2}\delta\Delta_{0}
+\left[\frac{5}{2}\Delta_{0}^{*}\delta\Delta_{0}
+\frac{5}{4}\left(\Delta_{1}^{*}\delta\Delta_{0}
+\Delta_{0}^{*}\delta\Delta_{1}\right)\right.\nonumber
\\
&&+\frac{5}{4}\left(\Delta_{2}^{*}\delta\Delta_{0}+
\Delta_{0}^{*}\delta\Delta_{2}\right)\nonumber
\\
&&+\frac{33}{20}\left(\Delta_{3}^{*}\delta\Delta_{0}
+\Delta_{0}^{*}\delta\Delta_{3}\right)\nonumber
\\
&&+\frac{4}{5}\left(\Delta_{3}^{*}\delta\Delta_{1}
+\Delta_{1}^{*}\delta\Delta_{3}\right)\nonumber
\\
&&\left.
+\frac{4}{5}\left(\Delta_{3}^{*}\delta\Delta_{2}
+\Delta_{2}^{*}\delta\Delta_{3}\right)\right], \label{Eq:ExpandAppDelta0StepA}
\\
\frac{d\Delta_{1}}{d\ell}
&=&-\frac{1}{2}\delta\Delta_{1}
+\left[\frac{1}{20}\left(\Delta_{0}^{*}\delta\Delta_{1}+\Delta_{1}^{*}\delta\Delta_{0}\right)
\right.\nonumber
\\
&&+\frac{1}{10}\Delta_{1}^{*}\delta\Delta_{1}+\frac{9}{20}\left(\Delta_{2}^{*}\delta\Delta_{1}+\Delta_{1}^{*}\delta\Delta_{2}\right)
\nonumber
\\
&&+\frac{17}{20}\left(\Delta_{3}^{*}\delta\Delta_{1}+\Delta_{1}^{*}
\delta\Delta_{3}\right)\nonumber
\\
&&\left.+\frac{4}{5}\left(\Delta_{3}^{*}\delta\Delta_{0}+\Delta_{0}^{*}
\delta\Delta_{3}\right)\right], \label{Eq:ExpandAppDelta1StepA}
\\
\frac{d\Delta_{2}}{d\ell}
&=&-\frac{1}{2}\delta\Delta_{2}
+\left[\frac{1}{20}\left(\Delta_{0}^{*}\delta\Delta_{2}
+\Delta_{2}^{*}\delta\Delta_{0}\right)\right.\nonumber
\\
&&+\frac{9}{20}\left(\Delta_{1}^{*}\delta\Delta_{2}+\Delta_{2}^{*}\delta\Delta_{1}
\right)+\frac{1}{10}\Delta_{2}^{*}\delta\Delta_{2}\nonumber
\\
&&+\frac{17}{20}\left(\Delta_{3}^{*}\delta\Delta_{2}+\Delta_{2}^{*}\delta\Delta_{3}\right)\nonumber
\\
&&\left.+\frac{4}{5}\left(\Delta_{3}^{*}\delta\Delta_{0}+\Delta_{0}^{*}\delta\Delta_{3}\right)\right], \label{Eq:ExpandAppDelta2StepA}
\\
\frac{d\Delta_{3}}{d\ell}
&=&-\frac{1}{2}\delta\Delta_{3}
+\left[-\frac{7}{20}\left(\Delta_{0}^{*}\delta\Delta_{3}+\Delta_{3}^{*}\delta\Delta_{0}
\right)\right.\nonumber
\\
&&+\frac{17}{20}\left(\Delta_{1}^{*}\delta\Delta_{3}
+\Delta_{3}^{*}\delta\Delta_{1}\right)\nonumber
\\
&&+\frac{17}{20}\left(\Delta_{2}^{*}\delta\Delta_{3}+\Delta_{3}^{*}\delta\Delta_{2}
\right)
+\frac{1}{10}\Delta_{3}^{*}\delta\Delta_{3}\nonumber
\\
&&+\frac{4}{5}\Delta_{0}^{*}\delta\Delta_{0}
+\frac{4}{5}\Delta_{1}^{*}\delta\Delta_{1}
+\frac{4}{5}\Delta_{2}^{*}\delta\Delta_{2}\nonumber
\\
&&+\frac{4}{5}\left(\Delta_{1}^{*}\delta\Delta_{0}+\Delta_{0}^{*}\delta\Delta_{1}\right)\nonumber
\\
&&\left.
+\frac{4}{5}\left(\Delta_{2}^{*}\delta\Delta_{0}+\Delta_{0}^{*}\delta\Delta_{2}\right)\right], \label{Eq:ExpandAppDelta3StepA}
\end{eqnarray}
where $\delta\Delta_{j}$ stands for $\delta\Delta_{j}=\Delta_{j}-\Delta_{j}^{*}$.
Substituting the values of $\Delta_{j}^{*}$ into equations~(\ref{Eq:ExpandAppDelta0StepA})-(\ref{Eq:ExpandAppDelta3StepA}) and using
$\frac{d\delta\Delta_{j}}{d\ell}=\frac{d\Delta_{j}}{d\ell}$, we arrive at
\begin{eqnarray}
\frac{d\delta\Delta_{0}}{d\ell}
&=&0.285469\delta\Delta_{0}
+0.352627\delta\Delta_{1}+0.352627\delta\Delta_{2}\nonumber
\\
&&+0.444141\delta\Delta_{3}, \label{Eq:ExpandAppDelta0}
\\
\frac{d\delta\Delta_{1}}{d\ell}
&=&0.054967\delta\Delta_{0}-0.415888\delta\Delta_{1}+0.0138377\delta\Delta_{2}\nonumber
\\
&&+0.217624\delta\Delta_{3}, \label{Eq:ExpandAppDelta1}
\\
\frac{d\delta\Delta_{2}}{d\ell}
&=&0.054967\delta\Delta_{0}+0.0138377\delta\Delta_{1}-0.415888\delta\Delta_{2}\nonumber
\\
&&+0.217624\delta\Delta_{3}, \label{Eq:ExpandAppDelta2}
\\
\frac{d\delta\Delta_{3}}{d\ell}
&=&0.217312\delta\Delta_{0}+0.272855\delta\Delta_{1}+0.272855\delta\Delta_{2}\nonumber
\\
&&-0.524821\delta\Delta_{3}. \label{Eq:ExpandAppDelta3}
\end{eqnarray}
From equations~(\ref{Eq:ExpandAppDelta0})-(\ref{Eq:ExpandAppDelta3}), we find that
\begin{eqnarray}
&&\frac{d\left(\delta\Delta_{0}+c_{1}\delta\Delta_{1}+c_{2}\delta\Delta_{2}+c_{3}\delta\Delta_{3}\right])}{d\ell}\nonumber
\\
&=&c_{4}\left(\delta\Delta_{0}+c_{1}\delta\Delta_{1}+c_{2}\delta\Delta_{2}+c_{3}\delta\Delta_{3}\right), \label{Eq:VRGExpansionSumcjDeltaj}
\end{eqnarray}
where $c_{1}$, $c_{2}$, $c_{3}$, $c_{4}$ are determined by
\begin{eqnarray}
&&0.285469+0.054967c_{1}+0.054967c_{2}\nonumber
\\
&&+0.217312c_{3}=c_{4},
\\
&&0.352627-0.415888c_{1}+0.0138377c_{2}\nonumber
\\
&&+0.272855c_{3}=c_{4}c_{1},
\\
&&0.352627+0.0138377c_{1}-0.415888c_{2}\nonumber
\\
&&+0.272855c_{3}=c_{4}c_{2},
\\
&&0.444141+0.217624c_{1}+0.217624c_{2}\nonumber
\\
&&-0.524821c_{3}=c_{4}c_{3}.
\end{eqnarray}
Solving these equations yields
\begin{eqnarray}
&&c_{1}\approx0.598383, \qquad c_{2}\approx0.598383, \nonumber
\\
&&c_{3}\approx0.687136, \qquad c_{4}\approx0.500575.
\end{eqnarray}
From equation~(\ref{Eq:VRGExpansionSumcjDeltaj}), one could get
\begin{eqnarray}
&&\left(\Delta_{0}-\Delta_{0}^{*}\right)+c_{1}\left(\Delta_{1}-\Delta_{1}^{*}\right)
+c_{2}\left(\Delta_{2}-\Delta_{2}^{*}\right)\nonumber
\\
&&+c_{3}\left(\Delta_{3}-\Delta_{3}^{*}\right)\nonumber
\\
&=&\left[\left(\Delta_{0,0}-\Delta_{0}^{*}\right)+c_{1}\left(\Delta_{1,0}-\Delta_{1}^{*}\right)
\right.\nonumber
\\
&&\left.+c_{2}\left(\Delta_{2,0}-\Delta_{2}^{*}\right)+c_{3}\left(\Delta_{3,0}-\Delta_{3}^{*}\right)\right]e^{c_{4}\ell}\nonumber
\\
&\propto& e^{\nu^{-1}\ell}.
\end{eqnarray}
Thus, the correlation length exponent is given by
\begin{eqnarray}
\nu=\frac{1}{c_{4}}\approx2.
\end{eqnarray}


\begin{thebibliography}{99}


\bibitem{Vafek14}
Vafek O and Vishwanath A 2014 \emph{Annu. Rev. Condens. Matter Phys.} {\bf 5} 83

\bibitem{Wehling14}
Wehling T O, Black-Schaffer A M and Balatsky A V 2014 \emph{Adv. Phys.} {\bf 63} 1

\bibitem{Armitage17}
Armitage N P, Mele E J and Vishwanath A 2018 \emph{Rev. Mod. Phys.} {\bf 90} 015001

\bibitem{Wan11}
Wan X, Turner A M, Vishwanath A and Savrasov S Y 2011 \emph{Phys. Rev.} B {\bf 83} 205101

\bibitem{Huang15}
Huang S-M, Xu S-Y, Belopolski I, Lee C-C, Chang G, Wang B, Alidoust N,
Bian G, Neupane M, Zhang C, Jia S, Bansil A, Lin H and Hasan M Z 2015
\emph{Nat. Commun.} {\bf 6}  7373

\bibitem{WengHM15}
Weng H, Fang C, Fang Z, Bernevig B A and Dai X 2015 \emph{Phys. Rev.} X
{\bf 5} 011029

\bibitem{Xu15A}
Xu S-Y, Belopolski I, Alidoust N, Neupane M, Bian G, Zhang C, Sankar R,
Chang G, Yuan Z, Lee C-C, Huang S-M, Zheng H, Ma J, Sanchez D S, Wang B,
Bansil A, Chou F, Shibayev P P, Lin H, Jia S and Hasan M Z 2015 \emph{Science} {\bf 349} 613

\bibitem{Lv15A}
Lv B Q, Weng H M, Fu B B, Wang X P, Miao H, Ma J, Richard P,
Huang X C, Zhao L X, Chen G F, Fang Z, Dai X, Qian T and Ding H 2015
\emph{Phys. Rev.} X {\bf 5} 031013

\bibitem{Burkov16}
Burkov A A  2016 \emph{Nat. Mater.} {\bf 15} 1145

\bibitem{Yan17}
Yan B and Felser C 2017 \emph{Annu. Rev. Condens. Matter Phys.} {\bf 8} 337

\bibitem{Hasan17}
Hasan M Z, Xu S-Y, Belopolski I and Huang S-M 2017 \emph{Annu. Rev. Condens. Matter Phys.}
{\bf 8} 289

\bibitem{Weng16}
Weng H, Dai X and Fang Z 2016 \emph{J. Phys.: Condens. Matter} {\bf 28} 303001

\bibitem{FangChen16}
Fang C, Weng H, Dai X and Fang Z 2016
\emph{Chin. Phys.} B {\bf 25}, 117106

\bibitem{HuangChenGFGroup15}
Huang X, Zhao L, Long Y, Wang P, Chen D, Yang Z, Liang H,
Xue M, Weng H, Fang Z, Dai X and Chen G 2015 \emph{Phys. Rev.} X {\bf 5}
031023

\bibitem{ZhangChengLong16}
Zhang C-L, Xu S-Y, Belopolski I, Yuan Z, Lin Z, Tong B, Bian G, Alidoust N,
Lee C-C, Huang S-M, Chang T-R, Chang G, Hsu C-H, Jeng H-T, Neupane M, Sanchez D S, Zheng H,
Wang J, Lin H, Zhang C, Lu H-Z, Shen S-Q, Neupert T, Hasan M Z and Jia S 2016  \emph{Nat. Commun.} {\bf 7} 10735

\bibitem{Burkov17}
Burkov A A 2017 \emph{Phys. Rev.} B {\bf 96} 041110(R)

\bibitem{Nandy17}
Nandy S, Sharma G, Taraphder A and Tewari S 2017 \emph{Phys. Rev. Lett.} {\bf 119} 176804

\bibitem{Panfilov14}
Panfilov I, Burkov A A and Pesin D A 2014
\emph{Phys. Rev.} B {\bf 89} 245103

\bibitem{Shankar94}
Shankar R 1994 \emph{Rev. Mod. Phys.} {\bf 66} 129

\bibitem{ColemanBook}
Coleman P 2015 \emph{Introduction to Many-Body Physics} (Cambridge: Cambridge
University Press).

\bibitem{Kotov12}
Kotov V N, Uchoa B, Pereira V M, Guinea F and Castro Neto A H 2012
\emph{Rev. Mod. Phys.} {\bf 84} 1067

\bibitem{Gonzalez94}
Gonz\'{a}lez J, Guinea F and Vozmediano M A H 1994
\emph{Nucl. Phys.} B {\bf 424} 595

\bibitem{Son07}
Son D T 2007 \emph{Phys. Rev.} B {\bf 75} 235423

\bibitem{Hofmann14}
Hofmann J, Barnes E and Das Sarma S 2014 \emph{Phys. Rev. Lett.} {\bf 113} 105502

\bibitem{Goswami11}
Goswami P and Chakravarty S 2011  \emph{Phys. Rev. Lett.} {\bf 107} 196803

\bibitem{Hosur12}
Hosur P, Parameswaran S A and Vishwanath A 2012
\emph{Phys. Rev. Lett.} {\bf 108} 046602

\bibitem{Gonzalez14}
Gonz\'{a}lez J 2014  \emph{Phys. Rev.} B {\bf 90} 121107(R)

\bibitem{Hofmann15}
Hofmann J, Barnes E and Das Sarma S 2015 \emph{Phys. Rev.} B {\bf 92} 045104

\bibitem{Throckmorton15}
Throckmorton R E, Hofmann J, Barnes E and Das Sarma S 2015 \emph{Phys. Rev.} B {\bf 92} 115101

\bibitem{Lai15}
Lai H-H 2015 \emph{Phys. Rev.} B {\bf 91} 235131

\bibitem{Jian15}
Jian S-K and Yao H 2015 \emph{Phys. Rev.} B {\bf 92} 045121

\bibitem{ZhangShiXin17}
Zhang S-X, Jian S-K and Yao H 2017 \emph{Phys. Rev.} B {\bf 96} 241111(R)

\bibitem{Cho16}
Cho G Y and Moon E-G 2016 \emph{Sci. Rep.} {\bf 6} 19198

\bibitem{Isobe16}
Isobe H, Yang B-J, Chubukov A, Schmalian J and Nagaosa N 2016
\emph{Phys. Rev. Lett.} {\bf 116} 076803

\bibitem{WangLiuZhang17A}
Wang J-R, Liu G-Z and Zhang C-J 2017
\emph{Phys. Rev.} B {\bf 95} 075129

\bibitem{Abrikosov72}
Abrikosov A A  1972 \emph{J. Low. Temp. Phys.} {\bf 8} 315

\bibitem{Yang14A}
Yang B-J, Moon E-G, Isobe H and Nagaosa N 2014 \emph{Nat. Phys.} {\bf10} 774

\bibitem{Huh16}
Huh Y, Moon E-G and Kim Y B 2016  \emph{Phys. Rev.} B {\bf 93} 035138

\bibitem{Abrikosov74}
Abrikosov A A 1974 \emph{Sov. Phys. JETP} {\bf 39} 709

\bibitem{Moon13}
Moon E-G, Xu C, Kim Y B and Balents L 2013 \emph{Phys. Rev. Lett.}
{\bf 111} 206401

\bibitem{Herbut14}
Herbut I F and Janssen L 2014 \emph{Phys. Rev. Lett.} {\bf 113} 106401

\bibitem{Janssen15}
Janssen L and Herbut I F 2015 \emph{Phys. Rev.} B {\bf 92} 045117

\bibitem{Dumitrescu15}
Dumitrescu P T 2015  \emph{Phys. Rev.} B {\bf 92} 121102(R)

\bibitem{Janssen16}
Janssen L and Herbut I F 2016 \emph{Phys. Rev.} B {\bf 93} 165109

\bibitem{Janssen17}
Janssen L and Herbut I F 2017 \emph{Phys. Rev.} B {\bf 95} 075101

\bibitem{PALee85}
Lee P A and Ramakrishnan T V 1985  \emph{Rev. Mod. Phys.}
{\bf 57} 287

\bibitem{SyzranovReview}
Syzranov S V and Radzihovsky L 2018
\emph{Annu. Rev. Condens. Matter Phys.} {\bf 9} 35

\bibitem{Evers08}
Evers F and Mirlin A D 2008  \emph{Rev. Mod. Phys.} {\bf 80},
1355

\bibitem{DasSarma11}
Das Sarma S, Adam S, Hwang E H and Rossi E 2011 \emph{Rev. Mod. Phys.} {\bf 83} 407

\bibitem{Finkelstein84}
Finkel'stein A M  1984 \emph{Z. Phys.} B {\bf 56} 189

\bibitem{Castellani84}
Castellani C, Di Castro C, Lee P A and Ma M 1984
\emph{Phys. Rev.} B {\bf 30} 527

\bibitem{PunnooseFinkelstein05}
Punnoose A and Finkel'stein A M 2005 \emph{Science} {\bf 310} 289

\bibitem{Abrahams01}
Abrahams E, Kravchenko S V and Sarachik M P 2001 \emph{Rev. Mod. Phys.} {\bf 73} 251

\bibitem{Kravchenko04}
Kravchenko S V and Sarachik M P 2004 \emph{Rep. Prog. Phys.} {\bf 67} 1

\bibitem{Spivak10}
Spivak B, Kravchenko S V, Kivelson S A and Gao X P A  2010
\emph{Rev. Mod. Phys.} {\bf 82} 1743

\bibitem{WangJing11}
Wang J, Liu G-Z and Kleinert H 2011 \emph{Phys. Rev.} B {\bf 83}
214503

\bibitem{WangLiuZhang16}
Wang J-R, Liu G-Z and Zhang C-J 2016 \emph{New J. Phys.} {\bf 18} 073023

\bibitem{WangJing17}
Wang J, Zhao P-L, Wang J-R, and Liu G-Z 2017 \emph{Phys. Rev.} B {\bf 95}, 054507 (2017).

\bibitem{Ye98}
Ye J and Sachdev S 1998 \emph{Phys. Rev. Lett.} {\bf 80} 5409

\bibitem{Ye99}
Ye J 1999 \emph{Phys. Rev.} B {\bf 60} 8290

\bibitem{Stauber05}
Stauber T, Guinea F and Vozmediano M A H 2005 \emph{Phys. Rev.} B {\bf 71}
041406(R)

\bibitem{Herbut08}
Herbut I F, Juri\v{c}i\'{c} V and Vafek O 2008  \emph{Phys. Rev. Lett.}
{\bf 100} 046403

\bibitem{Vafek08}
Vafek O and Case M J 2008
\emph{Phys. Rev.} B {\bf 77} 033410

\bibitem{Foster08}
Foster M S and Aleiner I L 2008 \emph{Phys. Rev.} B {\bf 77} 195413

\bibitem{WangLiu14}
Wang J-R and Liu G-Z 2014 \emph{Phys. Rev.} B {\bf 89} 195404

\bibitem{Moon14}
Moon E-G and Kim Y B 2014 arXiv:1409.0573.

\bibitem{Roy16}
Roy B and Das Sarma S 2016 \emph{Phys. Rev.} B {\bf 94} 115137

\bibitem{Gonzalez17}
Gonz\'{a}lez J 2017 \emph{Phys. Rev.} B {\bf 96} 081104(R)

\bibitem{Zhao16}
Zhao P-L, Wang J-R, Wang A-M and Liu G-Z 2016
\emph{Phys. Rev.} B {\bf 94} 195114

\bibitem{Nandkishore17}
Nandkishore R M and Parameswaran S A 2017
\emph{Phys. Rev.} B {\bf 95} 205106

\bibitem{Mandal18}
Mandal I and Nandkshore R M 2018
\emph{Phys. Rev.} B {\bf 97} 125121

\bibitem{WangYuXuan17}
Wang Y and Nandkishore R M 2017
\emph{Phys. Rev.} B {\bf 96} 115130

\bibitem{WangLiuZhang17B}
Wang J-R, Liu G-Z and Zhang C-J 2017
\emph{Phys. Rev.} B  {\bf 96} 165142

\bibitem{YangZK18}
Yang Z-K, Wang J-R and Liu G-Z 2018 \emph{Phys. Rev.} B {\bf 98} 195123

\bibitem{ZhaoPL19}
Zhao P-L and Wang A-M 2019 \emph{Phys. Rev.} B {\bf 100} 125138

\bibitem{Sikkenk19}
Sikkenk T S and Fritz L 2019
\emph{Phys. Rev.} B {\bf 100} 085121

\bibitem{Yang13}
Yang B-J, Bahramy M S, Arita R, Isobe H, Moon E-G and
Nagaosa N 2013 \emph{Phys. Rev. Lett.} {\bf 110} 086402

\bibitem{Yang14B}
Yang B-J and Nagaosa N 2014  \emph{Nat. Commun.} {\bf 5} 4898

\bibitem{Roy18}
Roy B, Slager R-J and Juri\v{c}i\'{c} V 2018  \emph{Phys. Rev.} X {\bf 8}
031076

\bibitem{Luo18A}
Luo X, Xu B, Ohtsuki T and Shindou R 2018 \emph{Phys. Rev.} B {\bf 97} 045129

\bibitem{Luo18B}
Luo X, Ohtsuki T and Shindou R 2018 \emph{Phys. Rev.} B {\bf 98} 020201(R)

\bibitem{LiXin18}
Li X, Wang J-R and Liu G-Z 2018  \emph{Phys. Rev.} B {\bf 97} 184508

\bibitem{Copetti19}
Copetti C and Landsteiner K 2019  \emph{Phys. Rev.} B {\bf 99} 195146

\bibitem{Roy14}
Roy B and Das Sarma S 2014 \emph{Phys. Rev.} B {\bf 90} 241112(R)

\bibitem{Sbierski16}
Sbierski B, Decker K S C and Brouwer P W 2016 \emph{Phys. Rev.} B {\bf 94} 220202(R)

\bibitem{Roy17A}
Roy B, Goswami P and Juri\v{c}i\'{c} V 2017 \emph{Phys. Rev.} B {\bf 95} 201102(R)

\bibitem{RoyFoster18}
Roy B and Foster M S 2018 \emph{Phys. Rev.} X {\bf 8} 011049

\bibitem{Ostrovsky06}
Ostrovsky P M, Gornyi I V and Mirlin A D 2006  \emph{Phys. Rev.} B {\bf 74} 235443

\bibitem{Foster12}
Foster M S 2012 \emph{Phys. Rev.} B {\bf 85} 085122

\bibitem{Syzranov16}
Syzranov S V, Ostrovsky P M, Gurarie V and Radzihovsky L 2016
\emph{Phys. Rev.} B {\bf 93} 155113

\bibitem{Pixley16}
Pixley J H, Huse D A and Das Sarma S 2016  \emph{Phys. Rev.} X
{\bf 6} 021042

\bibitem{FuBo17}
Fu B, Zhu W, Shi Q, Li Q, Yang J and Zhang Z 2017 \emph{Phys. Rev.
Lett.} {\bf 118} 14601

\bibitem{Bahramy12}
Bahramy M S, Yang B-J, Arita R and Nagaosa N
2012 \emph{Nat. Commun.} {\bf 3} 679

\bibitem{Xi13}
Xi X, Ma C, Liu Z, Chen Z, Ku W, Berger H, Martin C,
Tanner D B and Carr G L  2013 \emph{Phys. Rev. Lett.} {\bf 111} 155701

\bibitem{Park15}
Park J, Jin K-H, Jo Y J, Choi E S, Kang W, Kampert E, Rhyee J-S,
Jhi S-H and Kim J S 2015  \emph{Sci. Rep.} {\bf 5} 15973

\bibitem{Yuan17}
Yuan X, Zhang C, Liu Y, Narayan A, Song C, Shen S, Sui X,
Xu J, Yu H, An Z, Zhao J, Sanvito S, Yan H and Xiu F 2016 \emph{NPG Asia
Mater.} {\bf 8} e325

\bibitem{ZhangJingLei17}
Zhang J L, Guo C Y, Zhu X D, Ma L, Zheng G L, Wang Y Q, Pi L,
Chen Y, Yuan H Q and Tian M L 2017 \emph{Phys. Rev. Lett.} {\bf 118}
206601

\bibitem{ZhangCengXiuFaXian17}
Zhang C, Sun J, Liu F, Narayan A, Li N, Yuan X, Liu Y, Dai J,
Long Y, Uwatoko Y, Shen J, Sanvito S, Yang W, Cheng J and Xiu F 2017
\emph{Phys. Rev.} B {\bf 96} 155205

\bibitem{Ochiai09}
Ochiai T and Onoda M 2009 \emph{Phys. Rev.} B {\bf 80} 155103

\bibitem{Bittner10}
Bittner S, Dietz B, Miski-Oglu M, Oria Iriarte P, Richter A and
Sch\"{a}fer F 2010 \emph{Phys. Rev.} B {\bf 82} 014301

\bibitem{LuLing14}
Lu L, Joannopoulos J D and Solja\v{c}i\'{c} M 2014
\emph{Nat. Photon.} {\bf 8} 821

\bibitem{LuLing15}
Lu L, Wang Z, Ye D, Ran L, Fu L, Joannopoulos J D and Solja\v{c}i\'{c} M 2015
\emph{Science} {\bf 349} 622

\bibitem{ChenCTChanGroup16}
Chen W-J, Xiao M and Chan C T 2016 \emph{Nat. Commun.} {\bf 7} 13038

\bibitem{WangQiang17}
Wang Q, Xiao M, Liu H, Zhu S and Chan C T 2017 \emph{Phys. Rev.} X {\bf 7} 031032

\bibitem{YanLuLi18}
Yan Q, Liu R, Yan Z, Liu B, Chen H, Wang Z and Lu L 2018 \emph{Nat. Phys.} {\bf 14}
461

\bibitem{Schwartz17}
Schwartz T, Bartal G, Fishman S and Segev M 2007  \emph{Nature} {\bf 446} 52

\bibitem{Billy08}
Billy J, Josse V, Zuo Z, Bernard A, Hambrecht B, Lugan P, Cl\'{e}ment D,
Sanchez-Palencia L, Bouyer P and Aspect A 2008 \emph{Nature} {\bf 453} 891

\bibitem{Roati08}
Roati G, D'Errico C, Fallani L, Fattori M, Fort C, Zaccanti M, Modugno G
Modugno M and Inguscio M 2008  \emph{Nature} {\bf 453} 895

\bibitem{Huh08}
Huh Y and Sachdev S 2008 \emph{Phys. Rev.} B {\bf 78} 064512




\end{thebibliography}
\end{document}